\documentclass[11pt]{article}
\usepackage[mathscr]{eucal}
\usepackage{epsfig,amsfonts}
\usepackage{amsmath}
\usepackage{amsthm,amssymb}
\usepackage{bm}
\usepackage{graphicx}
\usepackage{relsize}
\usepackage{hhline}
\usepackage{comment}
\usepackage{cite}
\usepackage{diagbox}
\usepackage{psfrag}
\usepackage{mathrsfs} 
\usepackage{hyperref}
\usepackage{bm}
\usepackage{array}
\usepackage{tikz}
\usepackage{enumerate}
\usepackage{textcomp}
\usepackage{hyperref}
\usepackage{mathrsfs} 

\usepackage{lettrine}

\usepackage{multirow}
\DeclareMathAlphabet{\mathpzc}{OT1}{pzc}{m}{it}
\DeclareMathAlphabet\mathbfcal{OMS}{cmsy}{b}{n} 
\makeatletter
\@addtoreset{equation}{section}
\makeatother

\def\bea{\begin{eqnarray}}
\def\eea{\end{eqnarray}}
\def\be{\begin{equation}}
\def\ee{\end{equation}}

\def\be{\begin{equation}}
\def\ee{\end{equation}}
\def\bdm{\begin{displaymath}}
\def\edm{\end{displaymath}}
\def\bea{\begin{eqnarray}}
\def\eea{\end{eqnarray}}

\def\ri{{\rm i}}

\def\XXint#1#2#3{{\setbox0=\hbox{$#1{#2#3}{\int}$}
    \vcenter{\hbox{$#2#3$}}\kern-.5\wd0}}

\newcommand{\rd}{\mbox{d}}
\newcommand{\re}{\mbox{e}}

\DeclareMathAlphabet{\mathpzc}{OT1}{pzc}{m}{it}
\DeclareMathOperator*{\slim}{slim}

\begin{document}

\begin{titlepage}
$\phantom{I}$
\vspace{2.8cm}

\begin{center}
\begin{LARGE}

{\bf  Scaling limit of the ground state Bethe roots for the
inhomogeneous XXZ spin\,-\,$\frac{1}{2}$ chain}

\end{LARGE}

\vspace{1.3cm}
\begin{large}

{\bf 

Sascha Gehrmann$^{1}$, Gleb A. Kotousov$^{1}$ and Sergei  L. Lukyanov$^{2}$}

\end{large}

\vspace{1.cm}
${}^{1}$Institut f$\ddot{{\rm u}}$r Theoretische Physik, 
Leibniz Universit$\ddot{{\rm a}}$t Hannover\\
Appelstra\ss e 2, 30167 Hannover, Germany\\
\vspace{.4cm}
${}^{2}$NHETC, Department of Physics and Astronomy\\
     Rutgers University\\
     Piscataway, NJ 08855-0849, USA\\
\vspace{1.0cm}

\end{center}

\begin{center}

\parbox{13cm}{%
\centerline{\bf Abstract} \vspace{.8cm}
It is  known that for the Heisenberg XXZ spin\,-\,$\frac{1}{2}$ chain in the critical regime,
the scaling limit of the  vacuum Bethe roots
yields an infinite set of numbers that coincide with the energy spectrum  of the quantum mechanical 
3D anharmonic oscillator. 
The discovery of this curious relation, among others,
 gave rise to the approach  referred to as the ODE/IQFT (or ODE/IM) correspondence.
Here we consider a multiparametric generalization of the Heisenberg spin chain, which is
associated with the inhomogeneous  six-vertex model. 
When quasi-periodic boundary conditions are imposed the lattice 
system may be explored within the Bethe Ansatz technique.
We argue that for the critical spin chain,
with a properly formulated scaling limit,
the scaled Bethe roots for the ground state are  described
by second order differential equations, which are multi-parametric generalizations of the Schr\"{o}dinger equation for
the anharmonic oscillator.
}
\end{center}

\vfill

\end{titlepage}

\tableofcontents

\setcounter{page}{2}

\section{Introduction and summary}
Since the seminal work of Polyakov \cite{Polyakov:1970xd}, it is believed that the isotropic and homogeneous scaling behaviour of a critical
statistical system can be adequately described within the framework of Conformal Field Theory (CFT). While 
this fundamental insight remains unproven in general, 
there exists an enormous amount of  supporting evidence.
Much of it comes from the study of   2D classical statistical models. In many cases, 
such models admit an equivalent description  in terms of 1D quantum spin chains. Then the problem of identification
of the underlying CFT  can be approached by studying
the scaling limit of the spectrum of a spin chain Hamiltonian \cite{Cardy}. This provides a 
convenient way for a numerical investigation of the critical behaviour. For the so-called   integrable statistical systems,
a detailed analytical study is also possible. By integrable, what is usually
meant is that the model can be explored
via the Bethe Ansatz approach or some variant thereof  \cite{Baxter:1982zz}.
\medskip

In the case of integrable systems at criticality, apart from the conformal structure of the 
underlying continuous theory, it is natural to investigate also the  integrable structure of the corresponding CFT \cite{Bazhanov:1994ft,Bazhanov:1996dr,Bazhanov:1998dq}. 
The latter is encoded in the
scaling limit of the Bethe roots -- the solutions of the Bethe Ansatz equations. The most powerful approach for 
studying the integrable structure of CFT or, more generally, QFT is the so-called ODE/IM correspondence \cite{Voros:1994,Dorey:1998pt,Bazhanov:1998wj,Bazhanov:2003ni}.
The abbreviation IM stands for Integrable Model.  For the most
interesting  examples these
are 2D Integrable Quantum Field Theories (IQFT) and, as such, we prefer to use the term 
ODE/IQFT correspondence rather than ODE/IM.
\medskip

One of the simplest manifestations of the ODE/IQFT correspondence, which is directly relevant 
to this work, occurs in the scaling limit of the XXZ spin\,-\,$\frac{1}{2}$ chain.
The latter is the archetypical example of an integrable model and, in fact, the
Bethe Ansatz approach was originally introduced  for its special (isotropic) case  \cite{Bethe:1931}.
The Hamiltonian of the spin-$\frac{1}{2}$ XXZ chain of length $N$ is given by
\bea\label{asiisaias}
\mathbb{H}_{\rm XXZ}=-
\sum_{m=1}^N\Big( \sigma_m^x\sigma_{m+1}^x+\sigma_m^y\sigma_{m+1}^y+\Delta
\, \sigma_m^z\sigma_{m+1}^z\Big)\ .
\eea
With quasi-periodic boundary conditions imposed \cite{Barber},
\be\label{BC1b}
\sigma^x_{N+m}\pm\ri\,\sigma^y_{N+m}=\omega^{\pm 2}\,\big(\,\sigma^x_{m}\pm\ri\,\sigma^y_{m}\,\big)\,,
\qquad \qquad \sigma_{N+m}^z=\sigma^z_m \,,
\ee
$\mathbb{H}_{\rm XXZ}$ commutes with the $z$\,-\,component of the total spin operator and the
eigenstates are labelled by the  (half-)integer  numbers $S^z$.
The energy spectrum is described in terms of solutions of the Bethe Ansatz equations:
\be\label{bae}
\bigg(
\frac{1+q^{+1} \,\,\zeta_j}
{1+q^{-1}\,\zeta_j }
\bigg)^{N}
=-\omega^2\,q^{2S^z}\,
\prod_{i=1}^{N/2-S^z}\,
\frac{\zeta_i-q^{+2}\,\zeta_j }
{\zeta_i-q^{-2}\,\zeta_j }\  ,
\ee
where  $q$ parameterizes the anisotropy as $\Delta=\frac{1}{2}(q+q^{-1})$.
It turns out that the system is critical when $-1\le \Delta <1$ \cite{Luther,Kadanoff,Barber}
 or, equivalently, $q$ is a unimodular number, i.e.,
\be\label{ask7823hj}
q=\re^{\ri \gamma}\ .
\ee
In this case, for
an even number of sites, the ground state of $\mathbb{H}_{\rm XXZ}$, i.e., the state with the lowest energy, 
is non-degenerate and occurs in the sector $S^z=0$.
It was proven in the work of Yang and Yang \cite{Yang} that for $\gamma\in(0,\pi]$ and periodic boundary conditions $(\omega=1)$, 
 the solution set to the Bethe Ansatz equations corresponding to the ground state
has $\zeta_j$ all real, distinct positive numbers. 
The proof may be  extended to quasi-periodic boundary conditions when
\be\label{askj823hj}
\omega=\re^{\ri\pi{\tt k}}
\ee
with sufficiently small real ${\tt k}$, namely, 
\be
|{\tt k}|<\frac{\gamma}{2\pi}\ .
\ee
The vacuum Bethe roots can be naturally ordered as
\be
0<\zeta_1<\zeta_2<\ldots<\zeta_{N/2}\ .
\ee
In the limit when the number of sites $N$ goes to infinity,
most of the roots  become densely distributed
 such that $(\zeta_j-\zeta_{j+1})/\zeta_j\sim 1/N$. However, at the edges of
the distribution the Bethe roots develop a a certain scaling behaviour. In particular,  keeping $j$ fixed as $N\to\infty$
the following limits exist
\be\label{as832hjd}
\lim_{N\to\infty\atop
j-{\rm fixed}} N^{2(1-\gamma/\pi)}\zeta_j
\ee
and form an infinite  set of non-vanishing numbers.
These turn out to admit a remarkable description in terms of 
the Schr\"{o}dinger equation for the 3D anharmonic oscillator \cite{Bazhanov:1998wj},
\bea\label{askj8923jh}
\bigg(- {\frac{{\rd}^2}{{\rd} x^{2}}} +\frac{l(l+1)}{x^2}+x^{2\alpha}-E\bigg)\, \Psi=0\ .
\eea
Namely, the differential 
equation possesses only a discrete component of the spectrum and the corresponding energy levels $E_j$,
coincide with the limits \eqref{as832hjd} up to an overall factor provided the parameters of the ODE are taken to be
 \bea\label{sosaiisa}
\alpha=\frac{\pi}{\gamma}-1\ ,\ \ \ \ \ \ \ \ \ \ \ \ l+\frac{1}{2}=\frac{\pi {\tt k}}{\gamma}\ .
 \eea
A similar relation between the scaling behaviour of the Bethe roots of not only the ground state,
but also the class of so-called low energy excited states, holds true  (see, e.g., ref.\cite{Kotousov:2019ygw}). 
Moreover  such a phenomenon, i.e., the ODE/IQFT correspondence, has been observed
for a  number of other integrable, critical statistical models as well 
\cite{Dorey:2007zx,Bazhanov:2008yc,Bazhanov:2017nzh,Bazhanov:2019xvy,Kotousov:2021vih}.
\medskip

The XXZ spin chain belongs to the integrability class of a 2D classical statistical system known as the 
six-vertex model. As was shown by Baxter in the work \cite{Baxter:1971}, the latter admits a multiparametric generalization,
which is solvable within the Bethe Ansatz approach. 
In this case, the
equations \eqref{bae} are modified to 
\be\label{baekasdba}
\prod_{J=1}^{N}
\frac{\eta_J+q^{+1} \,\,\zeta_j}
{\eta_J+q^{-1}\,\zeta_j }
=-\omega^2\,q^{2S^z}\,
\prod_{i=1}^{N/2-S^z}\,
\frac{\zeta_i-q^{+2}\,\zeta_j }
{\zeta_i-q^{-2}\,\zeta_j }
\,\qquad\qquad (\,j=1,2,\ldots,\tfrac{N}{2}-S^z\,)
\ee
and involve the complex parameters $\{\eta_J\}_{J=1}^N$,  the so-called inhomogeneities.
 The corresponding statistical system is known as the inhomogeneous
six-vertex model. With the parameters $q$ and $\eta_J$ obeying certain conditions, the
lattice system develops critical behaviour.
 We'll assume that the number
of sites $N$ is divisible by $r$ and 
 the inhomogeneities  obey the $r$-site periodicity condition
\be\label{mn21bszz}
\eta_{J+r}=\eta_J\qquad\qquad\qquad (\,J=1,2,\ldots, N-r\,;\ N/r\in\mathbb{Z}\,)\, .
\ee
In the scaling limit, $N\to\infty$ while    the integer $r$ is kept fixed.
This ensures the presence of translational invariance for the continuous theory.

\medskip

The critical behaviour of the $r$\,-\,site periodic inhomogeneous six-vertex model can be explored
by studying the spectrum of the Hamiltonian $\mathbb{H}$ of a certain spin chain. For the
case $r=1$ it coincides, up to an overall multiplicative factor and additive constant, with
$\mathbb{H}_{\rm XXZ}$ \eqref{asiisaias}. The explicit form of $\mathbb{H}$ with $r\ge 2$
is rather cumbersome (see, e.g., \cite{Ikhlef:2008zz,Frahm:2013cma} and the work \cite{Bazhanov:2020new},
where the same notation as in this paper is used). 
What is important is  that   $\mathbb{H}$ is given by a sum over terms that describe the interactions of
up to $r+1$ adjacent $\frac{1}{2}$ spins, i.e., the corresponding Hamiltonian density is local.
 The quasi-periodic boundary conditions \eqref{BC1b}
are still assumed and, similar to the homogeneous case,
the Hamiltonians commute with the $z$\,-\,projection of the total spin operator.
\medskip

A numerical investigation shows that 
the  spin chain governed by the Hamiltonian $\mathbb{H}$ exhibits critical behaviour 
when the anisotropy parameter $q$ is a unimodular number, $q=\re^{\ri \gamma}$. 
 It turns out  that the phase can be restricted to the domain $\gamma\in(0,\pi)$ and
the critical behaviour is described differently as
$\frac{\pi}{r}\,A<\gamma<\frac{\pi}{r}\,(A+1)$. 
Inside each segment labelled by  the integer $A$  we will parameterize $\gamma$ 
by real positive $n$ such that
\be\label{asio8732uy}
\gamma=\frac{\pi }{r}\,A+\frac{\pi}{n+r}\ \qquad\qquad \qquad\qquad  (A=0,1,\ldots,r-1\,,\; n>0)\  .
\ee
\medskip

 Mapping out the critical surfaces in the  space of the parameters $\big(\gamma,\{\eta_\ell\}\big)$
is a complicated problem that, apart from some special cases, has not been fully achieved.
A useful starting point is  when the inhomogeneities
are given by
\be\label{ajk8932h}
\eta_\ell=(-1)^r\,\re^{\frac{\ri\pi}{r}\,(2\ell-1)}\ \ \ \qquad\qquad (\ell=1,\ldots,r)
\ee
for which the model possesses  ${\cal Z}_r$ invariance 
(see, e.g., sec.\,7 in ref.\cite{Bazhanov:2020new}).
In this case the description of the scaling limit for the ground state is especially simple. 
For  $N$ divisible by $2r$ and $S^z=0$ the corresponding Bethe roots are split into
$r$ groups of equal size. The Bethe roots within each group lie on the rays
\bea
\arg(\zeta)=-\frac{\pi }{r}\, A,\ \frac{\pi }{r} \,(2-A),\ \ldots,\,  \frac{\pi }{r}\,\big( 2\, (r-1)-A\big) \ \ \ \ \ \ \ \ \ \ \ ({\rm mod}\,2\pi)\ .
\eea
These can be labelled by the integer 
 $a=1,2,\ldots,r$ and we denote by 
$\zeta_{m}^{(a)}$   the roots lying on the $a^{\rm th}$ ray ordered by their absolute value.
Then  the real positive numbers $\big|\zeta_{m}^{(a)}\big|^r$ with $m=1,2,\ldots, N/(2r)$  are the same for any $a$. 
They turn out to 
coincide with the solution to
\eqref{bae} with $S^z=0$ and the substitutions
$q\mapsto \re^{\frac{\ri \pi r}{n+r}}$, $N\mapsto N/r$ which corresponds to the ground state of
the XXZ spin chain. It follows that the limits
\be\label{asjk732j}
E_{m}^{(a)}=\lim_{N\to\infty\atop m-{\rm fixed}}\,\bigg(\frac{N}{r N_0}\bigg)^{\frac{2n}{r(n+r)}}\zeta_m^{(a)}
\ee 
exist and are non-vanishing.
Here for later convenience, an extra $n$ dependent factor has been introduced with
\be\label{asm8932hg21}
N_0=\frac{\sqrt{\pi}\Gamma(1+\frac{r}{2n})}{r\Gamma(\frac{3}{2}+\frac{r}{2n})}\ .
\ee
The numbers $E_m^{(a)}$ would be expressed in terms of the  spectrum of the ODE \eqref{askj8923jh}.
For the purpose of this work, it is  convenient to re-write the differential equation using the variables 
$y=\frac{2}{r}\log(x)+\frac{2}{n+r}\log(\frac{r}{2})$
 and $\psi=x^{-\frac{1}{2}}\,\Psi$,  so that it becomes\footnote{%
With some abuse of notation we use the same symbol $E$ in eqs.\,\eqref{askj8923jh} and \eqref{hasyssstsat}.
In fact, $E$ from \eqref{askj8923jh} coincides with 
the combination $(-1)^A\,(2/r)^{\frac{2n}{n+r}}
 E^r$ with $E$ from \eqref{hasyssstsat}. Also, $\alpha=\frac{n}{r}$ and $l+\frac{1}{2}=\frac{2p}{r}$.}
\bea\label{hasyssstsat}
 \bigg[-\partial_y^2+p^2+ \re^{(n+r)y}-(-1)^A\,E^r\ \re^{ry}\bigg]\,\psi=0
\eea
with
\bea\label{jashgh12b1b}
p=\frac{n+r}{2}\ {\tt k}\ .
\eea
The above differential equation admits the Jost solution $\psi_p(y)$, which is uniquely defined through the asymptotic
\be\label{9823hdf12121}
\psi_p(y)=\re^{py}\qquad\qquad {\rm as}\qquad\qquad y\to-\infty\qquad\qquad\qquad\qquad\big(\Re e(p)\ge 0\big)\ .
\ee
For generic $E$ the solution  grows unboundedly as $y\to+\infty$.
However, at the special values $E=E_m^{(a)}$
\eqref{asjk732j}, one has
\be\label{as9832jaaahdf}
\psi_p(y)\big|_{E=E_m^{(a)}}\to 0 \qquad {\rm as}\qquad y\to +\infty\ .
\ee
\medskip

In this work we  consider the spin chain, where the ${\cal Z}_r$ symmetry is ``softly'' broken for
finite system size.  This is carried out by treating
the inhomogeneities $\eta_\ell$   as $N$\,-\,dependent bare couplings of the spin chain Hamiltonian
which tend to the values \eqref{ajk8932h} as $N\to\infty$.
The quantities
\bea\label{saaaakju32hjsd}
{\mathfrak a}_{s}=\frac{1}{s}\ \bigg(\frac{N}{r N_0}\bigg)^{d_s}\  \frac{1}{r}\ \sum_{\ell=1}^r(\eta_\ell)^{-s}
\eea
will be taken to be independent of $N$, i.e., they play the r\^{o}le of the RG invariants. With their proper specialization (which,
among other things, involves fixing the exponent $d_s>0$)
the Bethe roots possess a scaling behaviour similar to that as for the ${\cal Z}_r$ invariant case. 
In particular, for the ground state in the sector with $S^z=0$,
the roots  may  still be split into $r$ groups with roughly equal phases.
Then there exist the limits \eqref{asjk732j}, where the integer $m$ labels the roots according to their
absolute values. We will argue that  
$E_m^{(a)}$ can be described as above, but with 
the differential equation \eqref{hasyssstsat}  replaced by  the ODE of the form
\bea\label{c}
 \bigg[-\partial_y^2+p^2+ \re^{(n+r)y}-(-1)^A\,E^r\ \re^{ry}-\sum_{(\mu,j)\in\,\bm{\Xi}_{r,A}}
 c_{\mu,j}\, E^{\mu}\ \re^{\big(  (A\mu-rj)\,\frac{n+r}{r}+\mu\big) y}\,\bigg]\,\psi=0\ .
\eea
The  coefficients $c_{\mu,j}$ are related to the values of the RG invariants 
${\mathfrak a}_s$ from eq.\,\eqref{saaaakju32hjsd}, while
the summation indices $\mu$ and $j$ run over a certain  integer set. The latter 
  is given by\footnote{The number of  admissible pairs $(\mu,j)$ \eqref{cv}  coincides  with the   total number of  solutions of the Diophantine equations  $A\mu-rj=L$
subject to the constraints  $1\leq\mu<r-L\, \&\ j\geq 0$ for $L=1,2,\ldots$\ .}
\bea\label{cv}
\bm{\Xi}_{r,A}&=&\big\{(\mu,j)\, :\ \  
\tfrac{rj}{A}<\mu<
\tfrac{  r}{A+1}\  
(j+1)  \ \ \& \ \  j\ge  0\big\}\qquad{\rm for}\qquad A=1,2,\ldots,r-2\,,
 \eea
while for $A=0$,
\bea\label{cu}
\bm{\Xi}_{r,0}&=&\big\{(\mu,j)\, :\ \  
\mu=1,2,\ldots,r-1\ \ \& \ \  j=  0\big\}\,.
\eea
 As $0<A<r-1$ we expect that it is  possible to organize the scaling limit
in such a way that the differential equation \eqref{c}
appears for any given values of the coefficients $c_{\mu,j}$
(some restrictions on the domain of $n$  may be required). In particular, 
the specification of the RG invariants such that there is only 
 one non-vanishing term occurring in the sum in
\eqref{c} can be found in Appendix \ref{AppD2} for 
$r=3,4,\ldots,10$.
When $A=0$ only the coefficient
$c_{\mu,0}$ with $\mu\ge \frac{r}{2}$ can be chosen at will, while the others turn out to be not independent. 
The case  $A=r-1$ was already discussed in ref.\cite{Kotousov:2021vih}.
Its peculiar feature is that all of the exponents $d_s$ in \eqref{saaaakju32hjsd} turn out to be zero.
The scaling of the Bethe roots for the ground state is still described by means of
the differential equation  of the form \eqref{c} with
\bea
\bm{\Xi}_{r,r-1}=\big\{(\mu,j)\,:\ \ \mu=j+1\ \ \&\  \ j=0,1,\ldots,r-2\big\}\qquad\qquad\qquad (A=r-1)\,.
\eea

\section{The case odd $r$ and $A=\frac{r-1}{2}$\label{sec2}}
An important source of intuition in the study of the  spin\,-\,$\frac{1}{2}$ Heisenberg  chain
comes from the XX spin chain --- the special case when the anisotropy parameter $\Delta$
in \eqref{asiisaias} is set to zero or, equivalently, $q=\ri$. By means of the Jordan-Wigner transformation,
the lattice model can then be mapped to a system of free fermions, which dramatically simplifies
the analysis. A similar phenomenon occurs 
for the spin chain associated with the $r$-site periodic inhomogeneous six-vertex model for any odd $r$,
\bea\label{sakj8932a}
r=3,5,7,\ldots\,,
\eea 
when the integer $A$ in formula \eqref{asio8732uy} for the argument of $q$  is chosen to be
\bea\label{sakj8932b}
A=\frac{r-1}{2}\ .
\eea
In this case
\bea
q=\ri\, \re^{\frac{\ri\pi(r-n)}{2r (n+r)}}
\eea
so that as $n=r$ one has $q=\ri$. Below we discuss the scaling limit of the ground state for the spin chain
in the regime \eqref{sakj8932b}  ($r$ odd) using the ``free fermion'' point as a launch pad.

\medskip
\subsection{Scaling limit of the Bethe roots for the free fermion point \label{sec21}}
At  $q=\ri$ the Bethe Ansatz equations \eqref{baekasdba}  for the $r$-site periodic model \eqref{mn21bszz} become
\be
\bigg(\prod_{\ell=1}^{r}
\frac{\eta_\ell+\ri\zeta_j}
{\eta_\ell-\ri\zeta_j}\bigg)^{\frac{N}{r}}
=-\re^{\ri\pi (2{\tt k}+S^z)}\ .
\ee
Taking the logarithm of both sides, one obtains
\be
\frac{N}{\ri\pi r }\  \log\Bigg(\prod_{\ell=1}^r\frac{1+\ri \zeta_j/\eta_\ell}{%
1-\ri\zeta_j/\eta_\ell}\Bigg)=2n_j-1 +2{\tt k}+S^z
\ee
with $n_j$ being certain integers. For the vacuum state
of the ${\cal Z}_r$ invariant case, where the inhomogeneities are given by \eqref{ajk8932h},
the Bethe roots are split into the groups $\zeta_m^{(a)}$  with $a=1,\ldots,r$. Then it turns out that with a proper
choice of the branch of the logarithm, 
\be\label{baekasdbaaaa}
\frac{N}{\ri\pi r }\  \log\Bigg(\prod_{\ell=1}^r\frac{1+\ri \zeta_m^{(a)}/\eta_\ell}{%
1-\ri\zeta_m^{(a)}/\eta_\ell}\Bigg)=2m-1+2{\tt k}\qquad\qquad \big(m=1,2,\ldots,N/(2r)\big)\ ,
\ee
where  $S^z$ has been set to zero. One may expect that the above formula remains true
when the inhomogeneities $\eta_\ell$ differ slightly  from the values
$(-1)^r\,\re^{\frac{\ri\pi}{r}\,(2\ell-1)}$. 
\bigskip

To perform the scaling limit, we write $\zeta$ in the form
\bea
 \zeta=
\bigg(\frac{\pi}{2N}\bigg)^{\frac{1}{r}}\ E
\eea
and keep $E$ fixed as $N\to\infty$. Further, the inhomogeneities are assigned an $N$\,-\,dependence in such a way that the quantities
\bea\label{asn3hg27821}
{\mathfrak a}_{2j+1}&=&\frac{1}{2j+1}\  \bigg(\frac{2N}{\pi}\bigg)^{1-\frac{2j+1}{r}}\ \frac{1}{r}\ \sum_{\ell=1}^r(\eta_\ell)^{-2j-1}
\ \ \ \  \ \qquad \big(j=1,\ldots,\tfrac{r-1}{2}\big)
\eea
remain independent of the length of the spin chain. For $j=A=\frac{r-1}{2}$, the exponent of the
$N$\,-\,dependent factor in the
r.h.s. vanishes and, 
 without loss of generality,
${\mathfrak a}_r$ can be set to be any positive number. We take its value to be the same as for the ${\cal Z}_r$
invariant case \eqref{ajk8932h}, 
i.e., ${\mathfrak a}_r=\frac{1}{r}$.
This way, in the scaling limit, the Bethe Ansatz equations \eqref{baekasdbaaaa} simplify to
\bea\label{jas7832hjs}
\frac{(-1)^A}{r}\,E^r+\sum_{j=0}^{A-1}(-1)^j\, {\mathfrak a}_{2j+1}\ E^{2j+1}=2m-1+2{\tt k}\ \ \ \qquad \ \ \ \ (m=1,2,\ldots)\ .
\eea
For given $m$, the above algebraic equation  possesses exactly $r$ complex solutions, which we denote
as $E_m^{(a)}$. Assuming that the absolute value of ${\mathfrak a}_{2j+1}$ is sufficiently small, the roots may be specified by the condition
that
\be
E_m^{(a)}\sim \re^{\frac{\ri\pi}{r}(2a-A)}\,  \big((2m-1+2{\tt k})\,r\big)^{\frac{1}{r}}
\qquad\qquad {\rm for}\qquad\qquad m\gg 1
\ee
and $E_m^{(a)}$ are split into $r$ groups of roots which have approximately the same phase
\be
\arg\big(E_m^{(a)}\big)\approx\frac{\pi}{r}\ \big(2 a-A\big)\ \ \ \ \ \ \qquad (a=1,\ldots,r)\ .
\ee
As such, the scaling limit of the Bethe roots for the ground state is described by the formula 
\be \label{jsa8723hjsds}
E_{m}^{(a)}=\lim_{N\to\infty\atop m-{\rm fixed}}\,\bigg(\frac{2N}{\pi}\bigg)^{\frac{1}{r}}\zeta_m^{(a)}\,.
\ee 
This   provides an illustration of the scaling relation
\eqref{asjk732j} discussed in the Introduction.
\medskip

It is clear how to come up with a differential equation, whose spectrum is given by $E_m^{(a)}$.
One just takes \eqref{hasyssstsat} with $n=r$ and replaces $(-1)^A E^r$ by 
$r\lambda(E)$, where 
\be\label{3iubewbv12vb}
\lambda(E)=\frac{(-1)^A}{r} E^r+ \sum_{j=0}^{\frac{r-3}{2}}(-1)^j\,{\mathfrak  a}_{2j+1}\  E^{2j+1}\ .
\ee 
The ODE becomes
\bea\label{asoi8923jhd}
 \Big[-\partial_y^2+p^2+\re^{2ry}-
r\lambda(E)\ \re^{ry}\Big]\ \psi=0\ ,
 \eea
which is a form of the  confluent hypergeometric equation. The solution defined via the asymptotic condition \eqref{9823hdf12121}
reads explicitly as
 \be
 \psi_p(y)=\re^{py}\ \exp\Big(-\tfrac{1}{r}\,\re^{ry}\Big)\ {}_1F_1\Big(\tfrac{1}{2}+\tfrac{p}{r}-
\tfrac{\lambda}{2},1+\tfrac{2p}{r},\tfrac{2}{r}\,\re^{ry}\Big)\, .
 \ee
For $y\to+\infty$, it develops the asymptotic behaviour
\be
 \psi_p(y) \asymp 
(2r)^{\frac{p}{r}+\frac{1}{2}}\  
\frac{\Gamma(1+\tfrac{p}{r})}{2\sqrt{\pi}}\  \ D_+\  \Big(\frac{r}{2}\Big)^{\frac{1}{2}\lambda} \exp\Big(\tfrac{1}{r}\,\re^{ry}-\big(\lambda+1\big)\, \tfrac{ry}{2}+o(1)\Big)\ ,
\ee
where
\be\label{askj8723hjdsas}
D_+=  \frac{\Gamma(\frac{1}{2}+\tfrac{p}{r})}{\Gamma\big(\tfrac{1}{2}+\tfrac{p}{r}-
\tfrac{1}{2} \lambda(E)\big)}\ .
\ee
The Jost solution $\psi_p(y)$ vanishes at large $y$ when $E$ is a zero of  $D_+=D_+(E)$.
Hence, the spectrum is defined through the equation $\lambda(E)=2m-1+\frac{2p}{r}$ with $m=1,2,3,\ldots\ $.
This coincides with  \eqref{jas7832hjs}, provided  $\frac{p}{r}={\tt k}$ as in \eqref{jashgh12b1b} with $n=r$.

\subsection{Differential equation \label{sec22}}
For the case  $q=\ri\, \re^{\frac{\ri\pi(r-n)}{2r (n+r)}}$ with arbitrary $n>0$ we propose that the
differential equation describing the  scaling limit
of the Bethe roots for the ground state is given by
\be\label{hasytsat}
 \bigg[-\partial_y^2+p^2+ \re^{(n+r)y}-(-1)^A\,E^r\ \re^{ry}-\sum_{j=0}^{A-1}
 c_{2j+1}\, E^{2j+1}\ \re^{\big(\frac{n+r}{2}-\frac{n-r}{2r}\,(2j+1)\big) y}\,\bigg]\,\psi=0\ \  \ \ \ \big(A=\tfrac{r-1}{2}\big)\,.
\ee
When all the coefficients $c_{2j+1}$  vanish, the ODE  becomes that for the ${\cal Z}_r$ invariant case \eqref{hasyssstsat}. 
Also, as $n=r$, it reduces to the  equation \eqref{asoi8923jhd} where $\lambda(E)$ is given by \eqref{3iubewbv12vb}, 
provided 
$c_{2j+1}$ are set to $(-1)^j\,r\,{\mathfrak a}_{2j+1}$.  In general, these coefficients
would be some  functions of the RG invariants of the form \eqref{saaaakju32hjsd}.
\medskip

The ODE may be advocated for via the same types of arguments that
were originally developed in ref.\cite{Bazhanov:1998wj} for the Schr\"{o}dinger
equation for the anharmonic oscillator (which is equivalent to \eqref{hasyssstsat}).
They are based on the invariance of
\eqref{hasytsat}   w.r.t. the transformation
\be\label{asji8732jhds}
\hat{\Omega}\ :\ y\mapsto y+\frac{2\pi\ri}{n+r}\,,\qquad\qquad E\mapsto q^{-2}\,E
\ee
that holds true not only for $c_{2j+1}=0$, but any values of these coefficients.
Using this symmetry, one can derive the set of functional relations for the connection coefficients of the
ODE.
The most fundamental one is
\be\label{as8931hjds}
\re^{\frac{2\pi\ri p}{n+r}}\,D_+(q E)\,D_{-}(q^{-1}E)-\re^{-\frac{2\pi\ri p}{n+r}}\,D_+(q^{-1}E)\,D_{-}(qE)=
2\ri\sin\big(\tfrac{2\pi p}{n+r}\big)\ ,
\ee
 which is obeyed by the spectral determinants
$D_{\pm}(E)$. In order to define the latter, apart from $\psi_p(y)$ \eqref{9823hdf12121}, one should consider 
other solutions of the differential equation, $\psi_{-p}(y)$ and $\chi(y)$.
\medskip

For large negative $y$, the potential in \eqref{hasytsat} approaches $p^2$. 
As a result, the ODE possesses a   solution  satisfying the asymptotic condition
\bea\label{dksdjksdjsx}
\psi_{-p}(y)\to \re^{-py}\ \ \ \  \ \ \qquad {\rm as}\ \ \ \qquad y\to-\infty\qquad\qquad \big(\Re e(p)\le 0\big)\, .
\eea
It turns out that the analytic continuation of $\psi_{-p}(y)$  to the full complex $p$\,-\,plane 
results in  a meromorphic function of $p$. This allows one to specify the solution
$\psi_{-p}(y)$
for any complex $p$, except for the  discrete set of real, positive values
$p=\frac{1}{2}\,(n+r),\,(n+r),\,\frac{3}{2}\,(n+r), \ldots\ $. Similarly $\psi_p(y)$,
defined in the Introduction in formula  \eqref{9823hdf12121},  makes sense
for any $p\ne -\frac{1}{2}\,(n+r),\,-(n+r),\ldots\ $. The Wronskian  
$W[\psi_{-p},\psi_p]\equiv \psi_{-p}\,\partial_y\psi_{p}-\psi_{p}\,\partial_y\psi_{-p}$ is given by
\be\label{askjaaa2389jsd}
W[\psi_{-p},\psi_p]=2p
\ee
so that the two functions  form a basis in the space of solutions of the ODE \eqref{hasytsat} if
$\frac{2p}{n+r}\notin\mathbb{Z}$.
\medskip

The solution $\chi(y)$ is unambiguously defined through its asymptotic
behaviour
\bea\label{as8723hjsdaa}
\chi(y)\asymp\exp\Big(-\tfrac{n+r}{4}\ y-\tfrac{2}{n+r}\ \re^{\frac{n+r}{2} y}+o(1)\Big) \qquad\qquad{\rm as}\qquad \qquad
y\to+\infty\qquad\qquad (n>r)\ .
\eea
Note that, for now, we make the technical assumption $n>r$, while the case $0<n<r$ will be discussed below.
The spectral determinants are expressed via the Wronskians of $\psi_{\pm p}(y)$ and $\chi(y)$ as
\bea\label{jassysa}
D_{\pm }(E)=\frac{\sqrt{\pi}}{\Gamma(1\pm\frac{2p}{n+r})}\ (n+r)^{-\frac{1}{2}\mp\frac{2p}{n+r}}\ 
W[\chi,\psi_{\pm p}]\ .
\eea
They are entire functions of $E$ and the overall multiplicative factor has been chosen in such a way that
\be
D_{\pm }(0)=1\ .
\ee
For the proof of the  relation \eqref{as8931hjds} one should expand $\chi$ in the basis
of solutions $\psi_{\pm p}$. It follows from \eqref{jassysa}  that
\be
\chi(y)=
\frac{1}{\sqrt{\pi(n+r)}}\ \Big(\Gamma\big(-\tfrac{2p}{n+r}\big)\ (n+r)^{-\frac{2p}{n+r}}\,
D_{-}(E)\,
\psi_{+p}(y)+
\Gamma\big(\tfrac{2p}{n+r}\big)\ (n+r)^{\frac{2p}{n+r}}\,
D_{+}(E)\,
\psi_{-p}(y)\Big)\,.
\ee
Then one applies the symmetry transformation $\hat{\Omega}$ \eqref{asji8732jhds} to both sides of this equation
and computes the  Wronskian of $\chi(y)$ and $\hat{\Omega}\chi(y)$.
Taking into account that $W[\hat{\Omega}\chi,\chi]=2\ri$, 
which is a consequence of \eqref{as8723hjsdaa}, one  easily arrives at
 \eqref{as8931hjds}. 
\medskip

To make a link to the lattice system, recall that the inhomogeneous six-vertex model
possesses a commuting family of operators, a prominent member of which is the 
Baxter $Q$\,-\,operator \cite{Baxter}. In fact, there are two of them  \cite{Bazhanov:1996dr,Bazhanov:1998dq}. We will use the notation
$\mathbb{A}_\pm(\zeta)$ borrowed from ref.\cite{Bazhanov:2020new} and  the
reader may consult sec.\,3 of that work for their construction. Important properties of $\mathbb{A}_\pm(\zeta)$ is that they 
commute  amongst themselves for different values of the
spectral parameter,
\bea\label{tqcomm}
\big[\mathbb{A}_\pm(\zeta_1),\,\mathbb{A}_\pm(\zeta_2)\big]=\big[\mathbb{A}_\pm(\zeta_1),\,\mathbb{A}_\mp(\zeta_2)\big]=0
\ ,
\eea
and satisfy the  quantum Wronskian relation
\be\label{qwron}
q^{+2\mathbb P}\ 
\mathbb{A}_+\big(q^{+1}\zeta\big)\,\mathbb{A}_-\big(q^{-1}\zeta \big)-
q^{-2\mathbb P}\ 
  \mathbb{A}_-\big(q^{+1}\zeta\big)\,\mathbb{A}_+\big(q^{-1}\zeta
  \big)=\big(q^{+2\mathbb P}-q^{-2\mathbb P}\,\big)\, f(\zeta)\ .
 \ee
Here 
\be
f(\zeta)=\prod_{J=1}^N \big(1+\zeta/\eta_J\big)\,,
\ee
while
\be
q^{2{\mathbb P}}=\re^{\ri\pi{\tt k}}\, q^{\mathbb{S}^z}
\ee
with $\mathbb{S}^z$ being the $z$\,-\,projection of the total spin operator. In addition, in
the sector with given value of $S^z\ge 0$, the eigenvalues of $\mathbb{A}_+(\zeta)$
are polynomials in $\zeta$ of order $N/2-S^z$. The above immediately
 implies that the zeroes of these polynomials coincide with the roots $\zeta_j$ of the
Bethe Ansatz equations \eqref{baekasdba}. The eigenvalues of $\mathbb{A}_-(\zeta)$
are likewise polynomials in $\zeta$ of order $N/2+S^z$, whose zeroes obey the
equations similar to \eqref{baekasdba}. 
The relation \eqref{qwron} holds true for arbitrary inhomogeneities, though we focus on the case
where $\eta_J$ obey the $r$\,-\,site periodicity condition \eqref{ajk8932h}. 

\medskip

The Hamiltonian
$\mathbb{H}$ associated with the inhomogeneous six-vertex model belongs to the commuting family, i.e.,
\be
\big[\mathbb{H},\mathbb{A}_\pm(\zeta)\big]=0\, .
\ee
Let $A_\pm(\zeta)$ be the  eigenvalues
of $\mathbb{A}_\pm(\zeta)$ corresponding to the ground state of the spin chain.
In the ${\cal Z}_r$ invariant case  the scaling relation for the Bethe roots \eqref{asjk732j}
can be equivalently written as 
\be\label{asj9823hjs}
\lim_{N\to\infty} A_\pm\Big(\big(N/(rN_0)\big)^{-\frac{2n}{r(n+r)}}\,E\Big)=D_\pm(E)
\qquad\qquad \qquad\qquad (n>r)
\ee
with the constant $N_0$ defined in \eqref{asm8932hg21} and $D_\pm(E)$
are the spectral determinants for the ODE \eqref{hasyssstsat}. In the scaling limit,
the quantum Wronskian relation for
$A_\pm(\zeta)$, which follows from \eqref{qwron}, 
 becomes the   functional relation \eqref{as8931hjds}.
Motivated by the analysis of the free fermion point, we propose to extend  \eqref{asj9823hjs}
to the case when $D_\pm$ are the spectral determinants for the ODE \eqref{hasytsat}.
In taking the limit, the inhomogeneities become $N$\,-\,dependent such that the values of
\bea\label{sajk89732hjsdA}
{\mathfrak a}_{2j+1}=\frac{1}{2j+1}\  \bigg(\frac{N}{rN_0}\bigg)^{1-\frac{2j+1}{r}}\ \frac{1}{r}\ \sum_{\ell=1}^r(\eta_\ell)^{-2j-1}
\ \ \ \  \ \qquad \big(j=0,\ldots,\tfrac{r-3}{2}\big)
\eea
are kept fixed. Like in the free fermion case, ${\mathfrak a}_r$  can be specified to be $\frac{1}{r}$ or, equivalently,\footnote{%
In this section, $r$ is always taken to be odd so that the sign factor in the r.h.s. 
of \eqref{sajk89732hjsdC} is one. Nevertheless, keeping it there  makes the formula applicable for any integer $r$, 
which will be explored later.}
\be\label{sajk89732hjsdC} 
\frac{1}{r}\sum_{\ell=1}^r (\eta_\ell)^{-r}=(-1)^{r-1}\, .
\ee
As for the quantities ${\mathfrak a}_s$ \eqref{saaaakju32hjsd}  with $s$ even, we set them to be zero:
\be\label{sajk89732hjsdB} 
\sum_{\ell=1}^r(\eta_\ell)^{-2j}=0\qquad\qquad\qquad \big(j=1,\ldots,\tfrac{r-1}{2}\big)\, .
\ee
This can be thought of as taking the exponent $d_s$ in  \eqref{saaaakju32hjsd}  to be $+\infty$ for even $s$.
\medskip

Our numerical work in support of the relation \eqref{asj9823hjs} went along the following line.
First of all we note that for given values of ${\mathfrak a}_{2j+1}$,  formulae \eqref{sajk89732hjsdA}-\eqref{sajk89732hjsdB}
may be regarded as a system of equations for the determination of the inhomogeneities. It is 
easy to show that $(\eta_\ell)^{-1}$ are zeroes of a single polynomial equation of
order $r$, whose coefficients are expressed in terms of ${\mathfrak a}_{2j+1}$. 
This defines the set $\{\eta_\ell\}_{\ell=1}^r$ 
up to
permutations, which has no effect on the Bethe Ansatz equations. The  roots corresponding to the ground
state of the Hamiltonian $\mathbb{H}$   can be obtained by
 continuously deforming the
solution to the Bethe Ansatz equations
from the ${\cal Z}_r$ invariant case with ${\mathfrak a}_{2j+1}=0$ to the given values of the RG invariants.
Then one may consider the sums
\bea\label{sa982321uuy}
h_s^{(N)}= s^{-1} \sum_{j=1}^{\frac{N}{2}} \big(\zeta_j\big)^{-s}\qquad\qquad (s=1,2,\ldots)\ .
\eea
Taking into account that
\be\label{ais87237632}
A_+(\zeta)=\prod_{j=1}^{\frac{N}{2}}\bigg(1-\frac{\zeta}{\zeta_j}\bigg)\,,
\ee
the convergence of the  l.h.s. of \eqref{asj9823hjs}  implies the existence of the limits
\be\label{askj329812}
h_s^{(\infty)}=\lim_{N\to\infty}\bigg(\frac{rN_0}{N}\bigg)^{\frac{2sn}{r(n+r)}}h_s^{(N)}\, .
\ee
For $n>r$ this was indeed observed numerically.
The spectral determinant  $D_+(E)$ occurring in the scaling limit of $A_+(\zeta)$ is an entire function of $E$ which
satisfies the normalization condition $D_+(0)=1$. Hence the series
\bea\label{hasytsaaaat}
\log D_+(E)=-\sum_{s=1}^\infty J_s\ E^s
\eea
has a finite radius of convergence.  This way, the scaling relation \eqref{asj9823hjs} is equivalent to the infinite set of so-called
sum rules:
\be\label{sakj2873jsd}
h_s^{(\infty)}=J_s\qquad\qquad \qquad (s=1,2,\ldots)\, .
\ee
They can be used to provide numerical support for the ODE/IQFT correspondence as was originally done
in ref.\cite{Bazhanov:2017nzh} in the study of the scaling limit of the Fateev-Zamolodchikov spin chain 
\cite{Fateev,Albertini,Ray}.

\medskip

It is possible to calculate
the first few coefficients $J_s$  analytically by applying 
 perturbation theory in $E$  to the ODE \eqref{hasytsat}. From the results of the work \cite{Bazhanov:1998za}
one can show that
\bea\label{aissuaus}
J_1&=&c_1\rho_1\ f_1(h,g_0)\ \nonumber\\[0.2cm]
J_2&=&\big(c_1\rho_1\big)^2\ f_2(h,g_0)\\[0.2cm]
J_3&=&\big(c_1\rho_1\big)^3\, f_3(h,g_0)+c_3\rho_3\, f_1(h,g_1)\ .\nonumber
\eea
The functions $f_s$ are given in the Appendix \ref{AppA}, while
\bea
\rho_{2j+1}=
\frac{(n+r)^{2g_j-2}}{\Gamma^2(1-g_j)}
\eea
and
\bea
h=\frac{p}{n+r}\ ,\ \ \ \ \ \ \  \ \ \ \ g_j=\frac{1}{2}-\frac{(2j+1)(n-r)}{2r(n+r)}\ .
\eea
Since $h=\frac{\tt k}{2}$, see \eqref{jashgh12b1b},  formula \eqref{sakj2873jsd} yields a  prediction for the dependence of $h_s^{(\infty)}$ on the twist parameter.   This can be checked 
even without the explicit knowledge of the relation between the RG invariants ${\mathfrak a}_{2j+1}$ and the coefficients $c_{2j+1}$
of the differential equation. Note that the expressions for $J_s$ for larger $s$ are not available in analytical form.
 Nevertheless, it may be verified that
the  ${\tt k}$\,-\,dependence of both sides of \eqref{sakj2873jsd} agrees for $s\ge 4$
by computing $J_s$ via a numerical integration of the
ODE \eqref{hasytsat}.
\medskip

Through the study of the cases $r=3,5,7$ we found that the relation between the coefficients
of the differential equation and the RG invariants takes the general form
\bea\label{aks322091}
c_{2j+1}=C_{j}^{(j)}\,{\mathfrak a}_{2j+1}+\sum_{k=2}^{A-j}\ \sum_{j_1+\ldots+j_k=j+(k-1)A\atop
0\le j_1,\ldots, j_k\le A-1}\ C^{(j)}_{j_1\cdots j_k}\ {\mathfrak a}_{2j_1+1}\cdots {\mathfrak a}_{2j_k+1}\ .
\eea
Here, for given $r$,  $C^{(j)}_{j_1\cdots j_k}$ are some $n$\,-\,dependent constants.
In the case $r=n$, all  $C^{(j)}_{j_1\cdots j_k}$ with $k\ge 2$ vanish, while $C_{j}^{(j)}=(-1)^{j} r$. 
In what follows, we present a conjectured formula that expresses ${\mathfrak a}_{2j+1}$ in terms of $c_{2j+1}$,
i.e., the inversion of \eqref{aks322091}. However, this first requires a discussion of the case $0<n<r$.

\medskip
\subsection{Relation between the RG invariants and the coefficients of the ODE \label{sec23}}
For $0<n\le r$ the asymptotic condition \eqref{as8723hjsdaa}, which is used to specify the subdominant as 
$y\to+\infty$
solution of the differential equation, is not  literally applicable. The function $\chi(y)$ can be instead defined by means of
its WKB asymptotic, and the spectral determinants are then calculated, as before, by means of formula \eqref{jassysa}.
This turns out to be most efficient for the numerical computation of $D_\pm(E)$.
However, for our purposes, it would be useful
to introduce them for generic $n>0$ via analytic continuation. 
 Assuming $n>r$, we first 
derive the large\,-\,$E$ asymptotic behaviour of $D_+(E)$. 
The latter has zeroes  at 
$E_m^{(a)}$ defined through the condition \eqref{as9832jaaahdf}:
\be\label{asi23hdshashg}
D_+\big(E_m^{(a)}\big)=0\,.
\ee
These accumulate along the rays 
\bea
\arg(E)=-\frac{\pi }{r}\, A,\ \frac{\pi }{r} \,(2-A),\ \ldots,\  \frac{\pi }{r}\,\big( 2\, (r-1)-A\big) \ \ \ \ \ \ \ \ \ \ \ ({\rm mod}\,2\pi)\, .
\eea
The rays are the Stokes lines that divide the complex $E$\,-\,plane into $r$ wedges, in which the large\,-\,$E$ asymptotic
behaviour of the spectral determinant is described differently.
Inside each wedge, labelled by $a=1,2,\ldots, r$, we can parameterize $E$ in terms of $\theta$ as\footnote{%
For odd $r$, which we consider in this section,
one can neglect the factor $(-1)^{r-1}$. It is included in eq.\,\eqref{saj7832hsdA}
since it will be applied for any $r$, $A=0,1,\ldots,r-2$
 and to make the notation consistent with that of the work \cite{Kotousov:2023zps}, devoted to the case $A=0$.
The remark also carries over to eq.\,\eqref{832hj7832h}, below.}
 \bea\label{saj7832hsdA}
E=(-1)^{r-1}\ 
 \re^{\frac{\ri \pi   }{r} (2a-1-A)}\ \re^{\frac{2 n\theta}{r(n+r)}}
\eea
with
\be\label{saj7832hsdB}
 \big|\Im m(\theta)\big|<\frac{\pi(n+r)}{2n}\, .
\ee
Then the standard WKB analysis yields
\be\label{asui87273hjaasd}
\log D_+\asymp
\frac{N_0\,\re^{\theta}}{\cos(\frac{\pi r}{2n})}-\sum_{j=0}^{A-1}\ \frac{D_{2j+1}}{\sin\big(\frac{\pi (2j+1) (n-r) }{2nr}\big)}\ 
\re^{\frac{(2j+1)}{r}\,\big(\theta+\ri\pi(2a-1-A)\big)}
-\frac{2np \,\theta}{r(n+r)}+\log C_p
+o(1)
\ee
as $\Re e (\theta)\to+\infty$.
Here
\bea\label{9832jh43hb1}
D_{2j+1}&=&\frac{\sqrt{\pi}
\Gamma\big(\tfrac{1}{2}-\tfrac{n-r}{2n r}(2j+1)\big)}{2n \Gamma\big(1-\tfrac{n-r}{2n r}(2j+1)\big) }
\sum_{k=1}^A\re^{-\frac{\ri\pi}{2}(r+1) (k-1)}\ 
\frac{\Gamma\big(m-1+\tfrac{n-r}{2n r}(2j+1)\big)}{m!\,\Gamma\big(\tfrac{n-r}{2n r}(2j+1)\big)}
\nonumber\\[0.2in]
&\times&
  \sum_{j_1+\ldots+j_k=j+(k-1)A\atop
 j_1,\ldots, j_k\geq 0}c_{2j_1+1}\cdots c_{2j_k+1}
\eea
and
\bea\label{jasays}
C_p=\sqrt{\frac{r}{n+r}}\ r^{\frac{2p}{r}}\,(n+r)^{-\frac{2p}{n+r}}\,
\frac{\Gamma(1+\frac{2p}{r})}{\Gamma(1+\frac{2p}{n+r})}\, .
\eea
 For $0<n<r$
the spectral determinant can be defined to be the entire function of $E$, whose zeroes coincide with $E_m^{(a)}$ 
\eqref{asi23hdshashg}    and which obeys the asymptotic formula \eqref{asui87273hjaasd}. This specifies 
$D_+(E)$   for any $n>0$ except for a discrete set of values when the coefficients
in the expansion \eqref{asui87273hjaasd} possess simple poles:
\be\label{as3872h21}
n=\frac{r\, (2j+1)}{2j+1+2 w r}\qquad\qquad {\rm with}\qquad\qquad j=0,\ldots,A-1;\  w=0,1,2,\ldots\ .
\ee
 The spectral determinant
$D_-(E)$ may be introduced analogously. Together, $D_\pm(E)$ satisfy 
the quantum Wronskian relation \eqref{as8931hjds} for any $n>0$ except \eqref{as3872h21}.\footnote{%
If we require the spectral determinants to be entire functions of $E$,
the quantum Wronskian relation \eqref{as8931hjds} must be modified when $n$ takes the values \eqref{as3872h21}.  
The simplest illustration is provided by the case $r=n$, where
$D_+(E)$ is given by \eqref{askj8723hjdsas}, while $D_-(E)$ is obtained from the former via the substitution
$p\mapsto -p$.} We will exclude such exceptional $n$ from our analysis
to avoid  making the discussion too technical 
(see, e.g., ref.\cite{Bazhanov:2019xvyA} for a treatment of the cases $r=1,2$). 
\medskip

The scaling limit of the individual Bethe roots, 
\be 
\lim_{N\to\infty\atop m-{\rm fixed}}\,\bigg(\frac{N}{r N_0}\bigg)^{\frac{2n}{r(n+r)}}\zeta_m^{(a)}\,,
\ee 
exists for any positive $n$.
However as $0<n\le r$, it turns out there are issues with the existence of the limit   \eqref{askj329812}.
To properly define $h_s^{(\infty)}$  
certain subtractions need to be made   so that the large\,-\,$N$ limit can
be taken.  Namely, introduce the ``regularized'' version of the sum $h_s^{(N)}$:
\bea
h_s^{(N,{\rm reg})}=s^{-1}\sum_{m=1}^M (\zeta_m)^{-s}+\frac{(-1)^{s-1} N }{2s r\cos(s\gamma)}\ \sum_{\ell=1}^r(\eta_\ell)^{-s}\ ,
\eea
where $M$ stands for the total number of Bethe roots and in the case of the ground state $M=\frac{N}{2}$.
Without going into details, we mention that the form of the counterterm is motivated by the 
Bethe Ansatz equations \eqref{baekasdba} 
with the inhomogeneities subject to the $r$\,-\,site periodicity condition $\eta_{J+r}=\eta_J$.
Taking  into account the definition of ${\mathfrak a}_s$ \eqref{saaaakju32hjsd}, 
the above formula can be re-written as 
\bea\label{asosaiisa}
h_s^{(N,{\rm reg})}=s^{-1}\sum_{m=1}^M (\zeta_m)^{-s}+\frac{(-1)^{s-1}\ rN_0 {\mathfrak a}_s}{2\cos(s\gamma)}\ 
\bigg(\frac{N}{ rN_0}\bigg)^{1-d_s}\ .
\eea
This 
will be used to define $h_s^{(N,{\rm reg})}$ for any value of $A=0,1,2,\ldots,r-2$.
In the case we are currently considering, where  $r$ is odd and $A=\frac{r-1}{2}$,  all ${\mathfrak a}_{2j}$ with
$j=1,2,\ldots,A$ are set to zero, so that
\be\label{sa8921uy32}
h_{2j}^{(N,{\rm reg})}=h_{2j}^{(N)}\,.
\ee
At the same time, the values of ${\mathfrak a}_{2j+1}$ with $j=0,1,\ldots,A-1$ are generic complex numbers, which 
are held fixed in taking the scaling limit,
and the exponent $d_{2j+1}$ in \eqref{asosaiisa} is given by
\bea
d_{2j+1}=1-\frac{2j+1}{r}\ \ \ \ \ \ \qquad\ \ \  \ (j=0,1,\ldots,A-1)\, .
\eea
Hence,
\bea\label{askj8923h21}
 \bigg(\frac{rN_0}{N}\bigg)^{\frac{2(2j+1)n}{r(n+r)}}\,h_{2j+1}^{(N,{\rm reg})}=
 \bigg(\frac{rN_0}{N}\bigg)^{\frac{2(2j+1)n}{r(n+r)}}\,h_{2j+1}^{(N)}+\frac{(-1)^{j}\, rN_0\,   \mathfrak{a}_{2j+1}}{2
 \sin\big(\frac{\pi (2j+1) (n-r)}{2r(n+r)}\big)}\ 
\bigg(\frac{ rN_0}{N}\bigg)^{\frac{(2j+1)(n-r)}{r(n+r)}}\  .
\eea
For $n>r$ the second term in the r.h.s. tends to zero as $N\to\infty$ and
can be neglected. However, as $0<n<r$ it grows and cancels the divergent behaviour of the
first term, ensuring the existence of the limit
\be\label{akjs871323221hg}
h_s^{(\infty)}=\lim_{N\to\infty}\bigg(\frac{rN_0}{N}\bigg)^{\frac{2sn}{r(n+r)}}h_s^{(N,{\rm reg})}\, .
\ee
Note that even when $n>r$, the last formula is more efficient than \eqref {askj329812} for numerical purposes since
the inclusion of the counterterm greatly improves the convergence. For instance,
 we observed 
\bea\label{iasisisa}
 \bigg(\frac{rN_0}{N}\bigg)^{\frac{2n}{r(n+r)}}\,h_{1}^{(N,{\rm reg})}=h_1^{(\infty)}+O\big(N^{-2}\big)\ \quad  \ \ \ 
 (n>r-2)\,,
\eea
while for $0<n< r-2$ the correction term goes to zero  as $N\to\infty$ with a certain exponent which depends on $n$.
\medskip

Closely related to the divergent behaviour of $h_s^{(N)}$ for $0<n\le r$ is that some of the  coefficients
in the expansion of the spectral determinant \eqref{hasytsaaaat} possess a simple pole
at $n=r$. It is worth  discussing this in some detail since it turns out to give a hint of how to obtain the relations between
the RG invariants and the coefficients of the differential equation.
 Since $E_m^{(a)}$ are zeroes
of $D_+(E)$, it follows that $J_s$  can be written in the form
\bea\label{aks712jh213}
J_s=\frac{1}{s}\ \sum_{m=1}^{\Lambda}\sum_{a=1}^r\big(E_m^{(a)}\big)^{-s}+\Xi_s(\Lambda)\, .
\eea
Here the remainder $\Xi_s(\Lambda)\to 0$ as the cutoff $\Lambda\to\infty$ for $n>r$. 
Obtaining the large\,-\,$\Lambda$ behaviour of $\Xi_s(\Lambda)$  requires knowledge of
the asymptotics of the zeroes $E_m^{(a)}$ as $m\gg1$. The latter can be extracted from the ODE \eqref{hasytsat}
using the WKB approximation.
\medskip

The asymptotic formula  \eqref{asui87273hjaasd} for $D_\pm(E)$  holds true inside the wedge 
in the complex $E$\,-\,plane described by 
eqs.\,\eqref{saj7832hsdA},\,\eqref{saj7832hsdB} and can not be applied  along its boundary.
At the boundary of the  $a^{\rm th}$  and $(a+1)^{\rm th}$   (mod $r$)   wedges, $E$ can be parameterized by real positive $\theta$ as
\bea\label{832hj7832h}
E=
(-1)^{r-1}\  \re^{\frac{\ri\pi  }{r}(2a-A)}\ \re^{\frac{2 n\theta}{r(n+r)}}\ \ \ \ \qquad\qquad (\theta>0)\, .
\eea
Along this ray, the asymptotic behaviour of $D_+(E)$ is given by
\bea
D_+\asymp 2C_p\  \re^{-\frac{2np \theta}{r(n+r)}}\  \re^{-\chi_a(\theta)} \cos\big( \phi_a(\theta)\big)\ \ \ \ \ \ \ \
\qquad (\theta\to+\infty)\ ,
\eea
where
\bea
\chi_a(\theta)= 
N_0\tan\big(\tfrac{\pi r}{2n}\big)\, \re^\theta+\sum_{j=0}^{A-1}\ D_{2j+1}\ \cot\big(\tfrac{\pi (2j+1)(n-r)}{2nr}\big)\ 
\re^{\frac{(2j+1)}{r}\,\big(\theta+\ri\pi(2a-A)\big)}+o(1)
\eea
and
\bea\label{sa98i23ujhsds}
\phi_a(\theta)= N_0\re^\theta+\sum_{j=0}^{A-1}\re^{\frac{\ri\pi}{r}(2a-A) (2j+1)}\ D_{2j+1}
 \ \re^{\frac{2 j+1 }{r}\, \theta}-\tfrac{\pi p}{r}+o(1)\, .
\eea
In view of eq.\,\eqref{832hj7832h} we write the zeroes $E_m^{(a)}$  in the form
\be\label{asui781hew}
E_m^{(a)}= (-1)^{r-1}\  
\re^{\frac{\ri \pi  }{r}(2a-A)}\ 
\exp\Big(\tfrac{2 n}{r(n+r)}\, 
\theta_m^{(a)}\Big)\ 
.
\ee
Then $\theta_m^{(a)}$   solves the equation
\bea\label{sja7832h12}
\phi_a\big(\theta_m^{(a)}\big)=\pi\big(m-\tfrac{1}{2}\big)\ \ \ \ \ \ \ \ \ \  (m=1,2,\ldots)\ ,
\eea
which is similar to the usual Bohr-Sommerfeld quantization condition. It is convenient to re-write the latter in
a different way. Introduce
 $X$ and $Y$ such that
\bea\label{sa7823gh12}
X=-\ri\,\re^{\frac{\ri\pi}{r}(2a-A)}\ \re^{\frac{1}{r}\,\theta_m^{(a)}}\ ,\ \ \ \    Y=-\ri \,\re^{\frac{\ri\pi}{r}(2a-A)}\ \bigg(\frac{\pi}{N_0}\ \Big(m-\frac{1}{2}+\frac{p}{r}\,\Big)\bigg)^{\frac{1}{r}}
\eea
as well as
\bea\label{zxmnh32SA}
G_{r-2j-1}= (-1)^j\ \frac{D_{2j+1}}{N_0}\ .
\eea
Formula \eqref{sja7832h12}, upon dropping the term $o(1)$ in \eqref{sa98i23ujhsds}, then takes the form
\bea\label{98asui32112}
X^r+
 \sum_{m=1}^{A}  G_{2m} X^{r-2m}={ Y}^r\ ,
\eea
which can be considered as a polynomial equation to determine $X$ given $Y$.
Since $m$ is assumed to be large it follows from the definition \eqref{sa7823gh12} that 
$|Y|\gg1$. Moreover, eq.\,\eqref{98asui32112} has $r$ roots and we should focus on the one,
$X(Y)$, such that
\bea
X(Y)=Y+o(1)\ \ \ \ \ {\rm as}\ \ \ \ Y\to\infty\ .
\eea
 The Lagrange formula \cite{Lagrange,Belardinelli} (see also Appendix \ref{AppB}) gives  this root as a  series in inverse
powers of $Y$:
\bea
X(Y)=Y+\sum_{j=0\atop 
2j+1\not=0\ {\rm mod}\, r}^\infty R_{2j+1}\ Y^{-2j-1}
\eea
with
\be\label{sakj87321}
R_{2j+1}=\frac{1}{r}\sum_{
\alpha_1+2\alpha_2+\ldots+(j+1) \alpha_{j+1}=j+1\atop
\alpha_1,\ldots,\alpha_{j+1}\geq 0}
\frac{(-1)^{\alpha_1+\ldots+\alpha_{j+1}}}{\alpha_1! \alpha_2!\ldots \alpha_{j+1}!}\ 
\frac{\Gamma(\alpha_1+\ldots+\alpha_{j+1}-\frac{2j+1}{r})}{\Gamma(1-\frac{2j+1}{r})}\ 
G_2^{\alpha_1}G_4^{\alpha_2}\dots G_{2j+2}^{\alpha_{j+1}}
\ee
In the case at hand, as it follows from \eqref{zxmnh32SA} and \eqref{9832jh43hb1}, $G_{2m}$ is given by 
\bea\label{sakj87321AAA}
G_{2 m}&=& (-1)^{A-m}\  \frac{
\Gamma\big(\frac{3}{2}+
\frac{r}{2n}\big)\Gamma\big(\frac{1}{2}-
\frac{(r-2m)(n-r)}{2nr}\big)}{\Gamma\big(
\frac{r}{2n}\big)
\Gamma\big(1-\frac{(r-2m)(n-r)}{2nr}\big)}
\ 
\sum_{k=1}^m(-1)^{(A+1)(k-1)}\ 
\frac{\Gamma\big(k-1+\tfrac{n-r}{2n r}(r-2m)\big)}{m!\,\Gamma\big(\tfrac{n-r}{2n r}(r-2m)\big)}
\nonumber\\[0.2in]
&\times&
\sum_{j_1+\ldots+j_k= kA-m\atop
 j_1,\ldots,j_k\geq 0}c_{2j_1+1}\cdots c_{2j_k+1}\ .
\eea
Combining the above with \eqref{asui781hew} yields the asymptotic formula for the zeroes:
\be
E_m^{(a)}\asymp \re^{\frac{\ri \pi}{r}(2a-A)}\ 
 |Y_m|^{\frac{2n}{n+r}}\ 
\bigg(1+\sum_{k=1} ^AR_{2k-1}\ \re^{-\frac{\ri\pi}{r}(4a+1) k}\ |Y_m|^{-2k}+O\big(m^{-1}
\big)\bigg)^{\frac{2n}{n+r}}\qquad  (m\gg 1)\ ,
\ee
where
\be
|Y_m|=\bigg(\frac{\pi}{N_0}\, \Big(m-\frac{1}{2}+\frac{p}{r}\Big)\bigg)^{\frac{1}{r}}\, .
\ee
The remainder $\Xi_s(\Lambda)$ in \eqref{aks712jh213} involves a sum over
 negative integer powers of $E_m^{(a)}$.
They are expressed through
 $Q_{2k-1}(z)$, which are the coefficients in the formal expansion
\bea\label{askj87322h1}
\bigg(1+\sum_{k=1\atop 
2k-1\not=0\ {\rm mod}\, r} ^\infty R_{2k-1}\ Y^{-2k}\bigg)^{z}=1+\sum_{k=1}^\infty Q_{2k-1}(z)\ Y^{-2k}\ .
\eea
Then
\bea
\big(E_m^{(a)}\big)^{-s}\asymp\re^{-\frac{\ri \pi }{r} (2a-2-A) s}\  
 |Y_m|^{-s\nu }\ \bigg(1+\sum_{k=1}^A Q_{2k-1}(-s\nu)\  \re^{-\frac{\ri\pi}{r} (4a+1) k}\ |Y_m|^{-2k}
 +O\big(m^{-1}\big)\bigg)\,,
\eea
where  we use
\bea
\nu=\frac{2n}{n+r}\ .
\eea
Taking $s$ to be odd and performing the sum over $a$  leads to
\be
\sum_{a=1}^r\big(E_m^{(a)}\big)^{-s}\asymp r\, \re^{\frac{\ri\pi}{2} (s -1)}\ Q_{r-s-1}(-s\nu)\ 
 |Y_m|^{-r+s(1-\nu) }+O\Big(m^{-2+(1-\nu)\frac{s}{r}}\Big)\qquad (s=2j+1)\ .
\ee
It follows that
 the large $\Lambda$ behaviour for the remainder  in \eqref{aks712jh213} is given by
\bea\label{98saj3ads12}
\Xi_{2j+1}(\Lambda)=
\frac{  (-1)^j\, r N_0}{\big(\frac{\pi (2j+1) (n-r) }{r(n+r)}\big)}\ \,\frac{Q_{r-2j-2}\big(- (2j+1)\nu\big)}{2j+1}
\  \bigg(\frac{N_0}{\pi \Lambda}\bigg)^{\frac{(2j+1)(n-r)}{r(n+r)}}+
O\bigg(\Lambda^{-1-\frac{(2j+1)(n-r)}{r(n+r)}}\bigg)\, .
\eea
\medskip

Let's compare the residues of \eqref{98saj3ads12} and \eqref{askj8923h21} at the pole $n=r$. It leads to the relation
\be
 \mathfrak{a}_{2j+1}=\frac{1}{2j+1}\ Q_{r-2j-2}\big(- (2j+1)\nu \big)\ ,
\ee
or, explicitly, taking into account the definition \eqref{askj87322h1},
\bea\label{hassy}
\mathfrak{a}_{2j+1}=\frac{1}{2j+1}\ \sum_{k=1}^{A-j}(-1)^k\ \frac{\Gamma\big( (2 j+1)\nu+k \big) }{k! \Gamma\big( (2 j+1)\nu \big) }\ \sum_{j_1+j_2\ldots +j_k=A-j-k\atop
 j_1,\ldots,j_k\geq 0}R_{2j_1+1}\ldots R_{2j_k+1}\ ,
\eea
where $R_{2j+1}$ is given by eqs.\,\eqref{sakj87321} and \eqref{sakj87321AAA}.
Of course, the above is expected to be valid only for $n=r$.
Nevertheless, we found from the numerical work that it holds true for any $n>0$. 
The relation can be inverted, whereupon
 it takes the form \eqref{aks322091}. The coefficients  $ C^{(j)}_{j_1\cdots j_k}$ are rather cumbersome in general
and the simplest of them is given by
\be\label{as983hjds21aaaa}
C_j^{(j)}=(-1)^j\,\frac{r\Gamma\big(
\frac{r}{2n}\big)
\Gamma\big(1-\frac{(2j+1)(n-r)}{2r n}\big)}{
\Gamma\big(\frac{1}{2}+
\frac{r}{2n}\big)\Gamma\big(\frac{1}{2}-
\frac{(2j+1)(n-r)}{2r n}\big)}\ .
\ee
Formula \eqref{aks322091} is written out explicitly in the Appendix \ref{AppC} for the
cases $r=3,5,7$. Numerical work confirms its validity.

\medskip
\subsection{Some remarks about the scaling limit\label{sec24}}

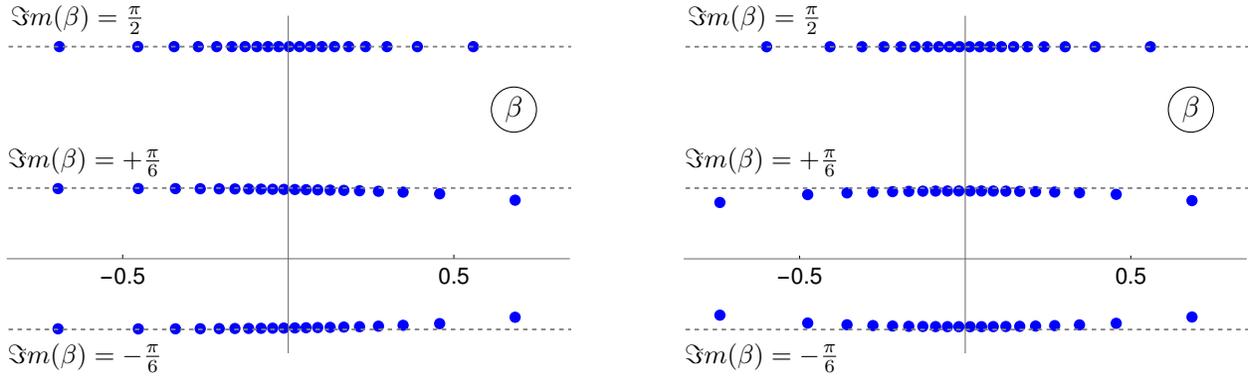
\begin{figure}
\begin{center}
\begin{tikzpicture}
\draw (3,1) circle (0.3cm);
\node at (3,1) {$\beta$};
\draw (12,1) circle (0.3cm);
\node at (12,1) {$\beta$};
\node at (-2.8,2.2) {\small $\Im m(\beta)=\frac{\pi}{2}$};
\node at (-2.7,0.3) {\small $\Im m(\beta)=+\frac{\pi}{6}$};
\node at (-2.7,-2.3) {\small $\Im m(\beta)=-\frac{\pi}{6}$};
\node at (6.2,2.2) {\small $\Im m(\beta)=\frac{\pi}{2}$};
\node at (6.3,0.3) {\small $\Im m(\beta)=+\frac{\pi}{6}$};
\node at (6.3,-2.3) {\small $\Im m(\beta)=-\frac{\pi}{6}$};
\node at (0,0) {\includegraphics[width=7.5cm]{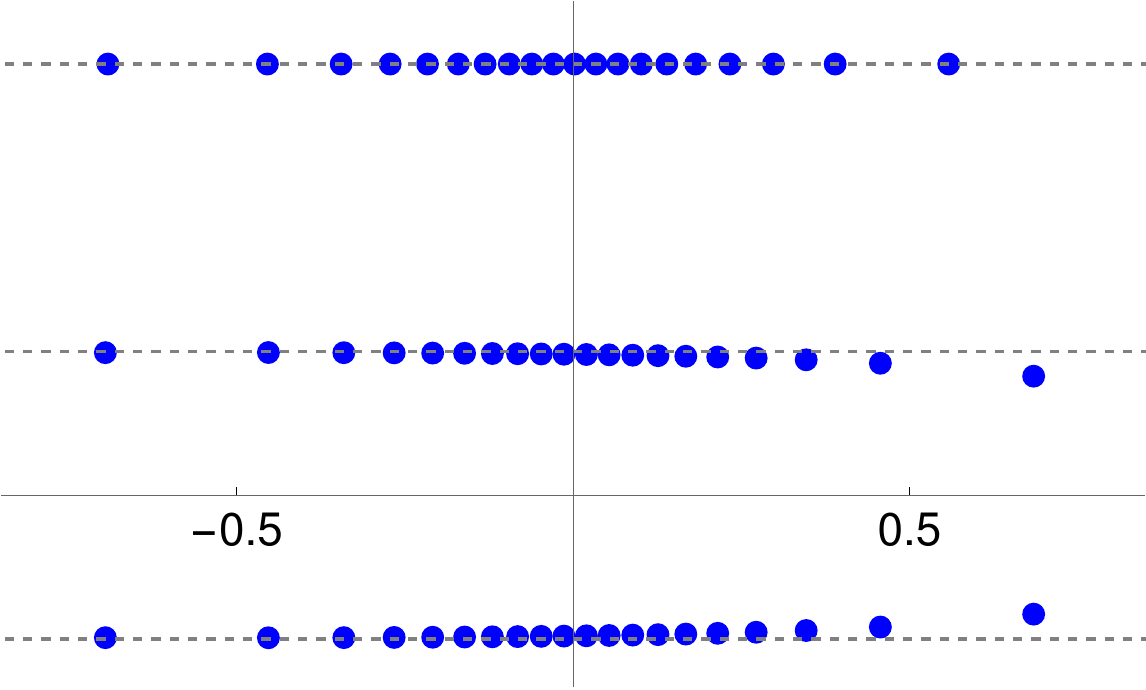}};
\node at (9,0) {\includegraphics[width=7.5cm]{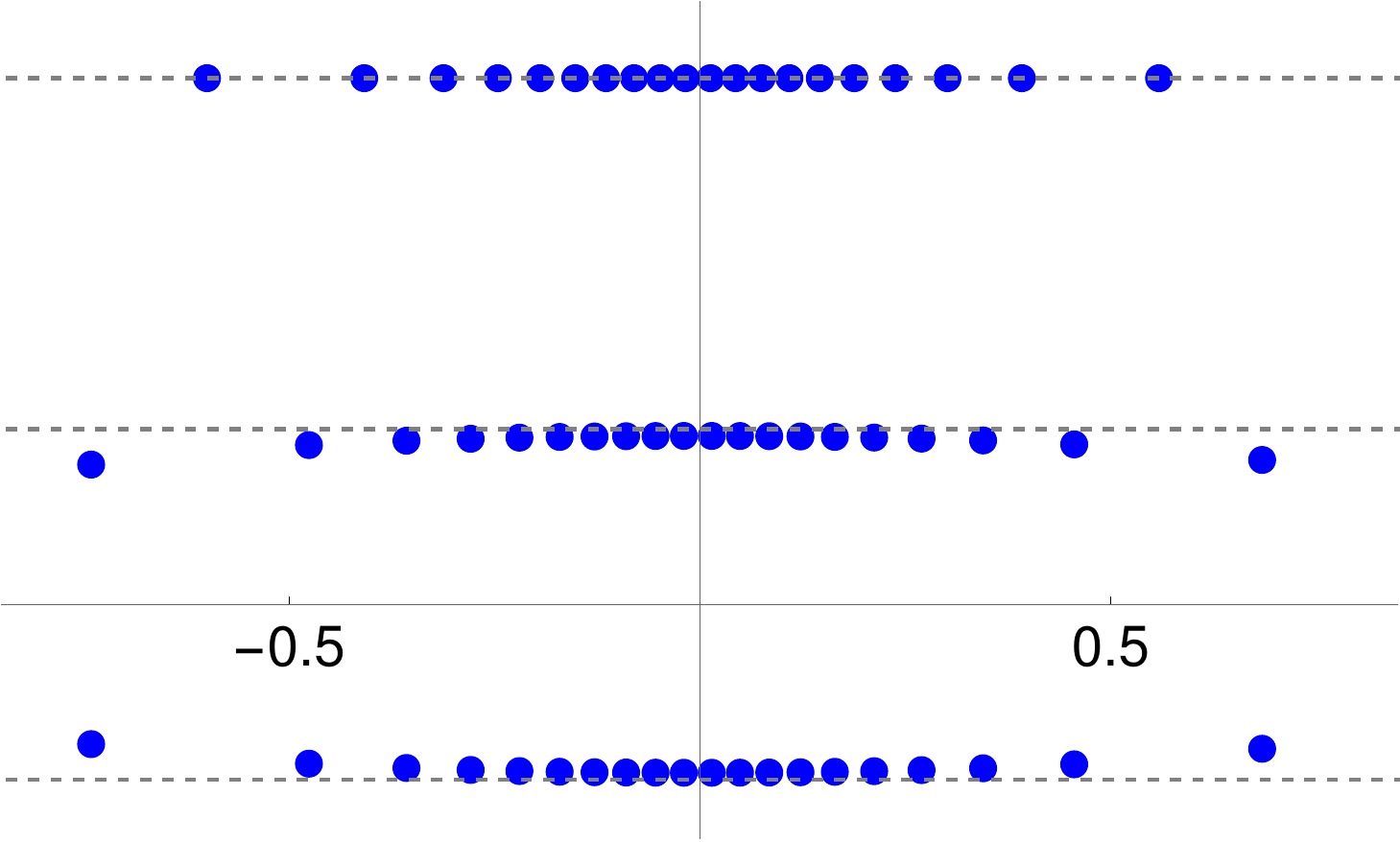}};
\end{tikzpicture}
\end{center}
\caption{\label{fig3}\small%
Depicted are plots of the Bethe roots in the complex
 plane $\beta=-\frac{1}{2}\log(\zeta)$ for $r=3$, $N=120$,
 $n=5$ and ${\tt k}=0.05$.  For the left panel, the inhomogeneities are taken to satisfy the conditions
\eqref{sajk89732hjsdA}\,-\,\eqref{sajk89732hjsdB} with $\mathfrak{a}_1=0.4$. The right panel corresponds to the
case when the spin chain possesses ${\cal C}{\cal P}$ and ${\cal T}$ symmetry. The inhomogeneities are 
specified according to \eqref{ask912ihjsdbiuasjAAA}\,-\,\eqref{as8732g21} with $\mathfrak{b}_1=0.4$.
}
\end{figure}

In our discussion we focused on describing the large\,-\,$N$ behaviour of the
 Bethe roots  in the vicinity of  $\zeta=0$. A similar analysis can be performed for the Bethe roots located near $\zeta=\infty$.
In this case, \eqref{asjk732j} is replaced by
\be\label{983jjhwqew}
\bar{E}_{m}^{(a)}=\lim_{N\to\infty\atop m-{\rm fixed}}\,\bigg(\frac{N}{r N_0}\bigg)^{\frac{2n}{r(n+r)}}
\Big(\zeta_{M_a-m}^{(a)}\Big)^{-1}\,,
\ee 
where for the ground state in the sector $S^z=0$ all $M_a$ are equal and coincide with $N/(2r)$.
If the scaling limit is taken with $\mathfrak{a}_{2j+1}$ \eqref{sajk89732hjsdA}
kept fixed and the conditions \eqref{sajk89732hjsdC},\,\eqref{sajk89732hjsdB} are imposed on the inhomogeneities,
 it turns out that $\bar{E}_{m}^{(a)}$ are described  by the differential equation
corresponding to the ${\cal Z}_r$ invariant case similar to \eqref{hasyssstsat}. 
This can be observed by plotting the  Bethe roots in the complex $\beta$\,-\,plane
with $\beta=-\frac{1}{2}\log(\zeta)$, which is done in the left panel of fig.\,\ref{fig3} for the case $r=3$. 
One sees that for the roots on the right side of the distribution, for which $|\zeta_j|\ll 1$, 
there is significant deviation from the lines $\Im m(\beta)=\pm\frac{\pi}{6},\,\frac{\pi}{2}$
so that their scaling behaviour is described via the differential equation \eqref{hasytsat} with $r=3$.
The Bethe roots depicted on the left side of the figure have $|\zeta_j|\gg 1$ and
are visibly located along the lines with  $\Im m(\beta)=\pm\frac{\pi}{6},\,\frac{\pi}{2}$.
Their scaling limit turns out to be same as that of the Bethe roots for the ${\cal Z}_3$ invariant case.
\medskip

Keeping in mind the invariance of the Bethe Ansatz equations \eqref{baekasdba}
w.r.t. the simultaneous inversion
$\zeta_j\mapsto \zeta_j^{-1}$, $\eta_\ell\mapsto (\eta_\ell)^{-1}$ and
$\omega\mapsto \omega^{-1}$ (${\tt k}\mapsto -{\tt k}$), it 
is clear how to organize the scaling limit such that the scaling behaviour of the roots near $\zeta=0$
and $\zeta=\infty$ is reversed.
One should treat
\be
\bar{{\mathfrak a}}_{s}=\frac{1}{s}\ \bigg(\frac{N}{r N_0}\bigg)^{\bar{d}_s}\  \frac{1}{r}\ \sum_{\ell=1}^r(\eta_\ell)^{s}
\ee
with $s=2j+1$ and  $\bar{d}_{2j+1}=1-\frac{2j+1}{r}$ as the RG invariants and impose the
additional  restrictions on the inhomogeneities:
\be
\sum_{\ell=1}^r (\eta_\ell)^{2j}=0\qquad\qquad (j=1,2,\ldots,A)
\ee
and
\be
\frac{1}{r}\sum_{\ell=1}^r (\eta_\ell)^{r}=(-1)^{r-1}\,.
\ee
The numbers $\bar{E}_{m}^{(a)}$ \eqref{983jjhwqew} would then
be described by the ``barred'' version of the ODE \eqref{hasytsat}:
\be\label{oias9813hjh}
 \bigg[-\partial_{\bar{y}}^2+\bar{p}^2+ \re^{(n+r)\bar{y}}-(-1)^A\,\bar{E}^r\ \re^{r\bar{y}}-\sum_{j=0}^{A-1}
 \bar{c}_{2j+1}\, \bar{E}^{2j+1}\ \re^{\big(\frac{n+r}{2}-\frac{n-r}{2r}\,(2j+1)\big)\bar{y}}\,\bigg]\,\bar{\psi}=0\ \  \ \ \ \big(A=\tfrac{r-1}{2}\big)\,.
\ee
Here $\bar{p}=-p=-\frac{n+r}{2}\ {\tt k}$,
while formula \eqref{aks322091}  with the replacements
$(\mathfrak{a}_{2j+1},c_{2j+1})\mapsto (\bar{\mathfrak{a}}_{2j+1},\bar{c}_{2j+1})$ 
gives the relation between the RG invariants and the coefficients of the differential equation. 
\bigskip

In view of applications to QFT, 
of special interest are
the spin chains admitting extra global symmetries ---
charge conjugation ${\cal C}$, parity  ${\cal P}$
and time reversal ${\cal T}.$
If the inhomogeneities
obey the condition
\be\label{cont1}
\eta_\ell=\big(\eta_{r+1-\ell}\big)^{-1}\qquad\qquad (\ell=1,2,\ldots,r)
\ee
then the system possesses ${\cal CP}$ invariance, i.e.,
\be
\hat{{\cal C}}\hat{{\cal P}}\,\mathbb{H}\,\hat{{\cal C}}\hat{{\cal P}}=\mathbb{H}\, .
\ee
For the action of the involutory operators $\hat{{\cal C}}$ and 
$\hat{{\cal P}}$ on the space of states of the spin chain  we refer the reader to
sec.\,4 of the paper \cite{Bazhanov:2020new}. In the latter, it is also described how the
${\cal C}{\cal P}$
 transformation acts on the Baxter $Q$\,-\,operators:
\bea\label{CP}
{ \hat {\cal C}}{\hat  {\cal P}}
\,{\mathbb A}_\pm(\zeta)\,{ \hat {\cal C}}{\hat  {\cal P}}
&=&\zeta^{\frac{N}{2}\mp\,\mathbb{S}^z}\ {\mathbb A}_\mp\big(\zeta^{-1}\big)\, \big[\, \mathbb{A}^{(\infty)}_\pm\,\big]^{-1}\, ,
\eea
where
\be
 { \mathbb{A}}^{(\infty)}_\pm=\lim_{\zeta\to\infty}\zeta^{-\frac{N}{2}\pm\mathbb{S}^z}\mathbb{A}_\pm(\zeta)\, .
\ee
Note that $\hat{\cal C}\hat{\cal P}$ intertwines the sectors with $S^z$ and $-S^z$.  
If, in addition to \eqref{cont1},
the inhomogeneities obey the reality condition
\be\label{cont2}
\eta_\ell^*=\eta_\ell^{-1}\,,
\ee
then the spin chain possesses time reversal symmetry ${\cal T}$ (again
 see ref.\cite{Bazhanov:2020new} for details). This way, 
the requirements \eqref{cont1} and \eqref{cont2} will guarantee that
the field theory underlying the critical behaviour
of the spin chain is ${\cal CP}$ and ${\cal T}$ invariant.
Also note that, without loss of generality,
the inhomogeneities can be chosen in such a way to satisfy the condition
\be\label{cont3}
\prod_{\ell=1}^r\eta_\ell=1
\ee
even for the spin chain, where no other constraints on $\eta_\ell$ are being assumed. 
\medskip

For the ${\cal Z}_r$ invariant case, $\eta_\ell=(-1)^r\,\re^{\frac{\ri\pi}{r}\,(2\ell-1)}$ 
and the conditions
 \eqref{cont1}, \eqref{cont2} and \eqref{cont3} are satisfied.
For the  model, which possesses ${\cal C}{\cal P}$ and
${\cal T}$ symmetries but ${\cal Z}_r$ invariance is broken, the inhomogeneities can be parameterized
through the phases $\delta_\ell$ as
\be\label{ask912ihjsdbiuasjAAA}
\eta_\ell=(-1)^r\re^{\frac{\ri\pi}{r}(2\ell-1)+\ri\delta_\ell}
\ee
with
\be
\delta_\ell=\delta_\ell^*=-\delta_{r+1-\ell}\,,\qquad\qquad \delta_1+\ldots+\delta_r=0\,.
\ee
In the case when $r$ is odd and $A=\frac{r-1}{2}$, one can consider the limit
with $\delta_\ell\to0$ as $N\to\infty$  such that
\bea\label{as8732g21}
\delta_{\ell}=-\delta_{r+1-\ell}=2\, \sum_{j=0}^{A-1}\ \sin\big(\tfrac{\pi}{r}(2\ell-1)(2j+1)\big)\ {\mathfrak b}_{2j+1}\  \bigg(\frac{r N_0}{N}\bigg)^{1-\frac{2j+1}{r}}\ \ \ (\ell=1,\ldots, A+1)
\eea
keeping the real numbers ${\mathfrak b}_{2j+1}$ fixed. 
The  Bethe roots for the ground state of the spin chain with  $N=120$ 
and $\mathfrak{b}_{1}=0.4$ are  depicted in the right panel of fig.\,\ref{fig3}, which may be compared with
the plot in the left panel.
Our numerical work for the cases $r=3,5,7$ shows that
the Bethe roots still possess the scaling behaviour \eqref{asjk732j} and \eqref{983jjhwqew} with 
$E_m^{(a)}$ and $\bar{E}_m^{(a)}$ described in terms of the differential equations 
\eqref{hasytsat} and \eqref{oias9813hjh}, respectively. If we assume that the scaling limit exists then
formula \eqref{CP} implies that the coefficients $c_{2j+1}$ must coincide with their barred counterparts:
\bea
\bar{c}_{2j+1}={c}_{2j+1}\ \ \ \ \qquad\ (j=0,\ldots,A-1)\ .
\eea
 The relation between  $\mathfrak{b}_{2j+1}$ and $c_{2j+1}$ turns out to take the form
\bea
c_{2j+1}={C}_{j}^{(j)}\,{\mathfrak b}_{2j+1}+\ldots\,,
\eea
where the ellipses involve a sum over monomials in ${\mathfrak b}_{2j'+1}$ with $j'<j$, each of which is quadratic
or of higher order in the RG invariants.
The coefficient for the linear term coincides with ${C}_{j}^{(j)}$ from \eqref{aks322091} and is given by  \eqref{as983hjds21aaaa}.
It is worthing noting that though the $N\to\infty$ limit \eqref{akjs871323221hg} involving  the regularized sum over the Bethe roots
$h_s^{(N,{\rm reg})}$ still exists and is described by \eqref{sakj2873jsd}, the rate of convergence is considerably slower.
Moreover, the definition of $h_s^{(N,{\rm reg})}$ requires
different subtractions, depending on the RG invariant $\mathfrak{b}_{2j+1}$ but not on ${\tt k}$,  in order   
to ensure the existence of the  limit for any $n>0$.
In particular, when the scaling limit was defined as in sec.\,\ref{sec22}, $h_{2j}^{(N,{\rm reg})}=h_{2j}^{(N)}$, 
see \eqref{sa8921uy32}, while for the case \eqref{as8732g21} this is no longer true.
\medskip

Thus we see that there is  freedom in the  way in which one takes the scaling limit ---
different prescriptions give rise to the ODEs of the same form 
\eqref{hasytsat} and \eqref{oias9813hjh} which describe 
 the scaling of the Bethe roots for the vacua. 
The
coefficients of the differential equations depend on the set of RG invariants 
that specify the scaling procedure. Note that the number of non-vanishing
RG invariants may not necessarily coincide with the number of coefficients $c_{2j+1}$.
For instance,  we discussed the scaling limit, where $\sum_{\ell=1}^r(\eta_\ell)^{-2j}$
were set to zero for $j=1,2,\ldots,A$. As was mentioned before, this formally corresponds to taking the exponent
$d_{2j}$ in  \eqref{saaaakju32hjsd}  to be $+\infty$. However, if the condition is  relaxed
and one considers the limit with non-vanishing RG invariants $\mathfrak{a}_{2j}$ but
$d_{2j}$  sufficiently large positive numbers,  the ODEs describing the scaling limit
remain the same. In particular for the cases $r=3,5$ we found  that if the
exponent $d_{2j}>\frac{1}{r}$, the presence of the non-zero RG invariants $\mathfrak{a}_{2j}$
does not essentially affect the scaling behaviour.

\medskip
\subsection{Low energy spectrum in the scaling limit}
In this paper, we focus on
describing the scaling of the Bethe roots for the ground state of the spin chain,
which may be regarded as a rather formal mathematical
problem.
From the physical point of view the most interesting question would be the study of the
field theory underlying the critical behaviour of the lattice system. While this goes beyond
the scope of our work, here we nevertheless give a brief discussion of the
spectrum of the Hamiltonian $\mathbb{H}$ in the scaling limit
in the regime with 
$\frac{\pi}{2}(1-\frac{1}{r})<\gamma<\frac{\pi}{2}(1+\frac{1}{r})$ and $r$ odd.
The analysis is mostly informed by a study of the free fermion point
with $\gamma=\frac{\pi}{2}$.
\medskip

\subsubsection{Low energy spectrum for the ${\cal Z}_r$ invariant spin chain\label{sec251}}

For a given solution of the Bethe Ansatz equations \eqref{baekasdba},
with the inhomogeneities obeying the $r$\,-\,site periodicity condition
\eqref{mn21bszz}, the corresponding eigenvalue of the Hamiltonian $\mathbb{H}$ reads as
\be\label{aoisd9819821A}
{\cal E}=2\ri\sum_{\ell=1}^r\sum_{m=1}^{N/2-S^z}\,\bigg(\frac{1}{1+\zeta_m\,q^{-1}\,/\eta_\ell}
-\frac{1}{1+\zeta_m\,q/\eta_\ell}\bigg)\ .
\ee
The Hamiltonian commutes with the $r$\,-\,site translation operator $\mathbb{K}$, whose
eigenvalue ${\cal K}$ is computed from the Bethe roots as (see, e.g., sec.\,6 of ref.\cite{Bazhanov:2020new} for details)
\be\label{asj21gh}
{\cal K}=\re^{\ri r\pi{\tt k}}\,q^{-r(\frac{N}{2}-S^z)}\,\prod_{\ell=1}^r
\prod_{m=1}^{N/2-S^z}
\frac{\zeta_m+\eta_{\ell}\,q^{+1}}{\zeta_m+\eta_{\ell}\,q^{-1}}\ .
\ee
The lowest energy state in the sector $S^z=0$ is translationally invariant, i.e., ${\cal K}_{\rm GS}=1$. 
As was mentioned in the Introduction, in the ${\cal Z}_r$ invariant case with 
\be
\eta_\ell=(-1)^r\,\re^{\frac{\ri\pi}{r}\,(2\ell-1)}\ \ \ \qquad\qquad (\ell=1,\ldots,r)
\ee 
 the corresponding Bethe roots
are simply related to those for the ground state
of the XXZ spin chain. Using well known facts about the latter, it is straightforward to deduce 
the large\,-\,$N$ behaviour for the ground state energy:
\be\label{aiosd9012221}
{\cal E}^{(0)}_{\rm GS}=Ne_{\infty}+\frac{2\pi r v_{{\rm F}}}{N}\,\Big(\,\frac{1}{2}\,(n+r)\,{\tt k}^2
-\frac{r}{12}\Big)+o(N^{-1}) \, ,
\ee
where 
the specific bulk energy
$e_\infty$ and the Fermi velocity $v_{\rm F}$ 
are  given by 
 \bea\label{uassaysa}
e_{\infty}= -\frac{2v_{{\rm F}}}{\pi}\,\int_0^\infty{\rm d}t\ \frac{\sinh\big(\frac{r t}{n}\big)}
{\sinh\big(\frac{(n+r)\,t}{n}\big)\,\cosh(t)}\,,\qquad \qquad\qquad
v_{{\rm F}}=\frac{r\,(n+r)}{n}\ .
\eea
The superscript ``$\scriptstyle (0)$'' has been introduced in order to emphasize that the above formulae
are valid for the ${\cal Z}_r$ invariant spin chain.
Eq.\,\eqref{aiosd9012221}
 leads one to expect that in the scaling limit $\mathbb{H}^{(0)}$ can be written as
\be\label{kjas8923hjds}
\mathbb{H}^{(0)}\asymp N e_{\infty}+\frac{2\pi r v_{{\rm F}}}{N}\,\hat{H}_{\rm CFT}+o(N^{-1}) \, .
\ee
Here $\hat{H}_{\rm CFT}$ stands for the Hamiltonian of the underlying CFT, 
\be\label{932jd8932jds}
\hat{H}_{\rm CFT}=\int_0^{2\pi}\frac{\rd x}{2\pi}\,\big(T+\overline{T}\,\big)\,,
\ee 
with  $T$ and $\overline{T}$  being the chiral components ($\partial\overline{T}=\bar{\partial}T=0$)
of the 
canonically normalized energy-momentum tensor. 
\medskip

For the XXZ spin chain \eqref{asiisaias}, corresponding to $r=1$, the 
low energy states for any value of the anisotropy parameter
$-1<\Delta<1$ can be classified in 
the same way as for the free fermion point with $\Delta=0$. Namely,
one should choose the basis of eigenstates of the Hamiltonian to be the Bethe states $\Psi_N$,
which are labelled by the solution of the Bethe Ansatz equations. 
In the scaling limit  $\Psi_N$ takes the form of the tensor product of the chiral states
\be\label{askj8723jh2}
\slim_{N\to\infty}\Psi_N= |\alpha\rangle\otimes |\bar{\alpha}\rangle\,,
\ee
where the symbol ``$\slim $'' is used as a reminder that   the formula applies 
for the class of low energy states only. Also, strictly speaking, the  existence of the limit requires that the Bethe state
$\Psi_N$ be properly normalized (for details see ref.\cite{Kotousov:2019ygw}).
As was discussed in \cite{Bazhanov:2003ni} (see also \cite{Masoero}),
in a given sector with fixed $S^z$ the chiral state $|\alpha\rangle$ can be labelled by two sets of non-negative integers
 \be\label{s8932hj2187A}
  1\leq n_{1}^{\pm}<  n_{2}^{\pm}<\ldots<   n_{M^{\pm}}^{\pm}\, .
\ee
Similarly $|\bar{\alpha}\rangle$ is labelled by
\be\label{s8932hj2187B}
  1\leq \bar{n}_{1}^{\pm}<  \bar{n}_{2}^{\pm}<\ldots<   \bar{n}_{\bar{M}^{\pm}}^{\pm}
\ee
and the difference $\bar{M}^+-\bar{M}^-$ should coincide with $M^--M^+$.
The latter can be identified with the so-called winding number:
\be\label{ask2187hjad}
{\tt w}=M^--M^+=\bar{M}^+-\bar{M}^-\ .
\ee
Formula \eqref{askj8723jh2} implies that the eigenvalues of the 
CFT Hamiltonian $\hat{H}_{\rm CFT}$ split into two terms,
$I_1+\bar{I}_1$\,,
where $I_1$ ($\bar{I}_1$) stands for the contribution of the chiral state $|\alpha\rangle$ ($|\bar{\alpha}\rangle$).
In the case of the scaling limit of the XXZ spin chain,
\be
I_1=\frac{p^2}{n+1}-\frac{1}{24}+{\tt L}
\,,\qquad\qquad \bar{I}_1=\frac{\bar{p}^2}{n+1}-\frac{1}{24}+\bar{\tt L}
\qquad\qquad (r=1)\ .
\ee
Here 
\be
p=\frac{S^z}{2}+\frac{n+1}{2}\ ({\tt k}+{\tt w})\,,\qquad\qquad 
\bar{p}=\frac{S^z}{2}-\frac{n+1}{2}\ ({\tt k}+{\tt w})\ ,
\ee
while the non-negative integers ${\tt L}$, $\bar{\tt L}$ are the conformal levels. 
In terms of the sets \eqref{s8932hj2187A},\,\eqref{s8932hj2187B} they are given by 
\bea
{\tt L}&=&\sum_{j=1}^{M^{-}}\big(n_{j}^{-}-\tfrac{1}{2}\big)+
\sum_{j=1}^{M^{+}} \big(n_{j}^{+}-\tfrac{1}{2}\big)-\sum_{j=1}^{|{\tt w}|}\,(j-\tfrac{1}{2})\nonumber\\[0.0cm]
&&\qquad\qquad\qquad\qquad\qquad\qquad\qquad\qquad\qquad\qquad\ . \\[0.0cm]
\bar{\tt L}&=&
\sum_{j=1}^{\bar{M}^{-}}\big(\bar{n}_{j}^{-}-\tfrac{1}{2}\big)+
\sum_{j=1}^{\bar{M}^{+}} \big(\bar{n}_{j}^{+}-\tfrac{1}{2}\big)-\sum_{j=1}^{|{\tt w}|}\,(j-\tfrac{1}{2}) \nonumber
\eea 
As for the
 eigenvalue of the lattice translation operator for the low energy state $\Psi_N$, one has
\be\label{asjk3872yt2h1g3a}
{\cal K}=(-1)^{{\tt w}}\,\exp\bigg(\frac{2\pi\ri r }{N}\ \big(\,I_1-\bar{I}_1\,\big)\bigg)\qquad\qquad (r=1)\, .
\ee
\medskip

In Appendix \ref{AppD} we briefly discuss the  spectrum of the Baxter $Q$\,-\,operator $\mathbb{A}_+(\zeta)$ \eqref{tqcomm}
in the scaling limit
for the ${\cal Z}_r$ invariant  spin chain  at the free fermion point $\gamma=\frac{\pi}{2}$. 
For a low energy Bethe state
in the sector with given $S^z$, the
chiral states $|\alpha\rangle$ and $|\bar{\alpha}\rangle$ appearing in the r.h.s. of  \eqref{askj8723jh2}  can be labelled  by the positive integers
 \bea
  1\leq n_{1,a}^{\pm}<  n_{2,a}^{\pm}<\ldots<   n_{M^{\pm}_a,a}^{\pm}
  \eea
and
 \bea
  1\leq \bar{n}_{1,a}^{\pm}<  \bar{n}_{2,a}^{\pm}<\ldots<   \bar{n}_{\bar{M}^{\pm}_a,a}^{\pm}\ ,
  \eea
respectively, where $a=1,2,\ldots,r$. The number of elements in the sets
$\{ n_{j,a}^{\pm}\}$ and $\{ \bar{n}_{j,a}^{\pm}\}$, i.e.,
$M^\pm_a$ and $\bar{M}^\pm_a$, are subject to the constraint
\be
\sum_{a=1}^r\big(M^-_a-M^+_a\big)+\sum_{a=1}^r\big(\bar{M}^-_a-\bar{M}^+_a\big)=0\,.
\ee
For $r>1$, the relation \eqref{ask2187hjad} is replaced by
\be
\frac{1}{r}\,\sum_{a=1}^r \big(M_a^{-}-M_a^{+}\big)=
\frac{1}{r}\,\sum_{a=1}^r\big(\bar{M}_a^{+}-\bar{M}_a^{-}\big)={\tt w}+\frac{{\tt m}}{r}\ ,
\ee
which, together with the winding number ${\tt w}\in\mathbb{Z}$,  involves
\be
{\tt m}\ : \ \ \ \ \ {\tt m}=0,1,2,\ldots,r-1\ .
\ee
Recall that $N$ is always assumed to be divisible by the odd integer $r$.
For this reason,  in taking the limit \eqref{askj8723jh2}
the size of the lattice should be increased by $2r$ in order to remain in the sector with given $S^z$. As such, one can introduce
an extra quantum number
\bea\label{asjh7812hgsa}
{\tt s}=\frac{N}{2}-S^z\ \ \ \ \ \ \  ({\rm mod }\ r)\,,
\eea
which  takes the values
\be
{\tt s}=0,1,2,\ldots,r-1\ .
\ee
It is expected that, away  from the free fermion point with $\frac{\pi}{2}(1-\frac{1}{r})<\gamma<\frac{\pi}{2}(1+\frac{1}{r})$,
 the chiral states 
can be characterized in the same way and, in particular, are labelled by  $S^z,{\tt w}\in\mathbb{Z}$ and ${\tt m},{\tt s}=0,1,\ldots,r-1$.
\medskip

In order to describe the eigenvalues  of the conformal Hamiltonian  \eqref{kjas8923hjds}, one should distinguish between the following four cases:
\be\arraycolsep=0.8cm
\begin{array}{ll}
{\rm (i)}\ \ \, \frac{{\tt s}+S^z}{r}-S^z\ {\rm even\ and} \ {\tt k}\in(-\frac{1}{2},0)\,, &   {\rm (ii)}\ \ \frac{{\tt s}+S^z}{r}-S^z\  {\rm even\  and}\  {\tt k}\in(0,\frac{1}{2}) \\[0.2cm]
{\rm (iii)}\ \frac{{\tt s}+S^z}{r}-S^z\ {\rm odd\ and} \  {\tt k}\in(-\frac{1}{2},0)\,, & {\rm (iv)}\  \frac{{\tt s}+S^z}{r}-S^z\ {\rm odd\ and} \ {\tt k}\in(0,\frac{1}{2})
\end{array}
\ee
(note that $\frac{{\tt s}+S^z}{r}-S^z$  is always an integer, as follows from the definition \eqref{asjh7812hgsa}
and that $r$ is odd). Also, introduce the integers $\tilde{\tt m}$ and $\tilde{\tt s}$ as
\be
{\tilde {\tt m}}=\begin{cases}
|r-{\tt s}- {\tt m}| & \qquad {\rm for\ cases\ (i)\ \& \ (iv)} \\
|{\tt s}- {\tt m}| & \qquad {\rm for\ cases\ (ii)\ \& \ (iii)} 
\end{cases}\ ,\qquad\qquad
{\tilde {\tt s}}=\begin{cases}
+{\tt s}& \qquad {\rm for\ case\ (i)}  \\
-{\tt s}& \qquad {\rm for\ case\ (ii)} \\
r-{\tt s}& \qquad {\rm for\ case\ (iii)}  \\
{\tt s}-r& \qquad {\rm for\ case\ (iv)}  
\end{cases}\ .
\ee
Based on an analysis of the free fermion point and  numerical work  we come to
\bigskip

\noindent
{\bf Conjecture:} 
The eigenvalues  of  $\hat{H}_{\rm CFT}$ \eqref{kjas8923hjds}
 are given by $I_1+\bar{I}_1$ with 
 \be\label{asjhyghdfvAAAA}
I_1=\frac{p^2}{n+r}-\frac{r}{24}+\frac{{\tt m}\,(r-{\tt m})}{2r}+{\tt L}\,,\qquad\qquad
\bar{I}_1=\frac{\bar{p}^2}{n+r}-\frac{r}{24}+\frac{{\tilde  {\tt m}}\,(r-{\tilde {\tt m}})}{2r}+\bar{{\tt L}}\ ,
\ee
in  the cases (i) and (iv) and
\be
I_1=\frac{p^2}{n+r}-\frac{r}{24}+\frac{{ {\tilde {\tt m}}}\,(r-{\tilde  {\tt m}})}{2r}+{\tt L}\,,\qquad\qquad
\bar{I}_1=\frac{\bar{p}^2}{n+r}-\frac{r}{24}+\frac{{ {\tt m}}\,(r-{ {\tt m}})}{2r}+\bar{{\tt L}}\ ,
\ee
 in the cases (ii) and (iii).
Here $p$ and $\bar{p}$ are expressed in terms of the quantum numbers $S^z$, ${\tt w}$, ${\tt m}$ and ${\tt s}$ according to
\be\label{asjhyghdfvBBBB}
p+\bar{p}=S^z\, ,\qquad\qquad \frac{p-\bar{p}}{n+r}= {\tt k}+{\tt w}+\frac{{\tt m}}{r}+\frac{\tilde{\tt s}}{2r}\ ,
\ee
while ${\tt L}$, $\bar{\tt L}$ are non-negative integers.
The  eigenvalues of the $r$\,-\,site lattice translation operator for the low energy states
are given by  
\be\label{asjhyghdfvCCCC}
{\cal K}=(-1)^{\frac{r-1}{2}(\frac{N}{2}-S^z)+{\tt m}+r{\tt w}}\ \re^{-\frac{\ri\pi}{2} {\tilde {\tt s}}}\ 
\,\exp\bigg(\frac{2\pi\ri r }{N}\ \big(\,I_1-\bar{I}_1\,\big)\bigg)\ .
\ee

\bigskip

\subsubsection{Some comments on the low energy spectrum away from the ${\cal Z}_r$ invariant point}
Let ${\cal E}_{\rm vac}$  (vacuum energy) be the lowest energy of the spin chain 
in the sector with given $S^z$  and $N/2-S^z$ is divisible by $r$, i.e.,
the quantum number ${\tt s}$ \eqref{asjh7812hgsa} is equal to zero. Note that by the ground state energy, ${\cal E}_{\rm GS}$,
what is meant is the vacuum energy  for $S^z=0$.
We studied the large\,-\,$N$ behaviour of ${\cal E}_{\rm vac}$, where
 the  inhomogeneities   were taken
to satisfy the conditions 
 \eqref{sajk89732hjsdA}-\eqref{sajk89732hjsdB} with $\mathfrak{a}_{2j+1}$ being  treated as fixed, $N$\,-\,independent
 parameters. It was found that the extensive part does not depend on the values of the
 RG invariants and, of course, $S^z$ so that the specific bulk energy
$e_\infty$ is still given by  eq.\,\eqref{uassaysa}. However, the difference between ${\cal E}_{\rm vac}$ and 
${\cal E}_{\rm vac}^{(0)}$  (the vacuum energy for the ${\cal Z}_r$ invariant spin chain)
goes at large $N$ as 
\be\label{sakj892hj12asas}
{\cal E}_{\rm vac}-{\cal E}^{(0)}_{\rm vac}=\frac{2\pi r v_{\rm F}}{N}\,f(\mathfrak{a}_1,\ldots,\mathfrak{a}_{r-2})+o(N^{-1})\ ,
\ee
 where $f$ is a certain polynomial in $\mathfrak{a}_{2j+1}$ whose coefficients depend on $n$.
Note that $f$ is independent of the value of  $S^z$ as well as the twist parameter ${\tt k}$. 
This motivated 
us to separate out the factor  $2\pi r v_{\rm F}$ in \eqref{sakj892hj12asas}, 
where the Fermi velocity is defined by the same formula as for the ${\cal Z}_r$ invariant case, i.e., 
\eqref{uassaysa}. For
$r=3$,
\bea
f=-C\,\mathfrak{a}_1^3
\eea
with $C$ being a positive  $n$\,-\,dependent coefficient which vanishes as $n^{-1}$ for $n\to\infty$
and diverges   $\sim n^{-2}$ as $n\to 0$. Also, at the free fermion point
$C\big|_{n=3}=\frac{1}{2}$.
\bigskip

It turns out that the form of the leading behaviour \eqref{sakj892hj12asas} is not universal in that it depends
on the way in which one performs the scaling limit. 
For example, as was discussed, the  limit can be organized differently
such that the inhomogeneities obey \eqref{cont1},\,\eqref{cont2}
 and the spin chain Hamiltonian possesses both ${\cal CP}$ and ${\cal T}$ symmetries for any number of lattice sites.
If $r=3$, there would be a single real valued RG invariant $\mathfrak{b}_1$, which is defined
by eqs.\,\eqref{ask912ihjsdbiuasjAAA}\,-\,\eqref{as8732g21}. Then, we observed  that
\bea
	{\cal E}_{\rm vac}-{\cal E}^{(0)}_{\rm vac}=
-\frac{C_1}{N^{\frac{1}{3}}}\, \mathfrak{b}^2_1-	\frac{2\pi r v_{{\rm F}}}{N}\,C_2\, {\mathfrak b}^3_1+o(N^{-1})\ \ \ \ \ \ \ \ \ \ \ \ \ \ \ \ \ (r=3)\ ,
\eea
where $C_1$, $C_2$ are some positive constants depending on $n$.
\bigskip

Regardless of the scheme in which the scaling limit is taken,   the operator
$N(\mathbb{H}-{\cal E}_{\rm GS})$, when restricted to the class of low energy states,
is expected to possess the $N\to\infty$ limit.
The low energy states can be chosen to be the Bethe states, which are specified by solutions to the Bethe
Ansatz equations.  It seems to be possible to 
 unambiguously and continuously deform 
the  corresponding Bethe roots away from the
${\cal Z}_r$ invariant case, at least in some domain of the values of the RG invariants. 
A preliminary numerical study shows that the spectrum of the operator
\bea\label{hastsaasuu}
\hat{H}\equiv\frac{1}{2}\,(n+r)\,{\tt k}^2-\frac{r}{12}+\frac{1}{2\pi r v_{\rm F}}\ 
\slim_{N\to\infty}\, N\big({\mathbb H}-{\cal E}_{\rm G S}\big)
\eea
depends on the RG invariants. For $n=r$, i.e., at the free fermion point,
this can  also be confirmed  via
 a straightforward extension of the analysis presented in Appendix \ref{AppD}.
A detailed investigation of the spectrum of $\hat{H}$ 
lies beyond the scope of the current paper.

\bigskip

\section{Basic constraints on ODEs for  spin chain with anisotropy $0<\gamma<\pi$}
The quantum  Wronskian relation(s), obeyed by the $Q$\,-\,operators, first appeared in the series of works 
\cite{Bazhanov:1994ft,Bazhanov:1996dr,Bazhanov:1998dq}
and has since then become a central notion in 
 integrability in Quantum Field Theory and Statistical Mechanics. In the case of the inhomogeneous
six-vertex model this would be the  relation \eqref{qwron} satisfied by 
 $\mathbb{A}_\pm(\zeta)$
and it holds true for any values of the anisotropy parameter $q$
and inhomogeneities $\eta_J$. It is expected that, with a properly defined scaling limit,
 the eigenvalues of $\mathbb{A}_\pm(\zeta)$ corresponding to the low energy  states would become the functions
$D_\pm(E)$, which obey the functional
relation inherited from the quantum Wronskian relation.   For the ground state 
of the spin chain this would be \eqref{as8931hjds} with $p=\frac{n+r}{2}\,{\tt k}$.
In developing the ODE/IQFT correspondence the main step 
is to identify $D_\pm(E)$ with the 
spectral determinants of a certain 2$^{\rm nd}$ order ODE. Unfortunately, 
we do not know of a systematic way to do this  at the moment.
While the previous section was devoted to the case with $r$ odd and $q=\re^{\ri\gamma}$, where
$\gamma\in\big(\frac{\pi}{2}(1-\frac{1}{r}),\frac{\pi}{2}(1+\frac{1}{r})\big)$,
it contains the essential hints for obtaining the differential equation  appearing 
in the scaling limit of the spin chain for the other domains 
\be
\gamma\in\big(\tfrac{\pi }{r} A,\tfrac{\pi }{r} (A+1)\big)
\ee
labelled
by the integer $A=0,1,\ldots,r-1$.
Among them is that the class of possible ODEs is greatly reduced by the requirement
that the differential equation be invariant w.r.t.  the transformation
\be\label{asn3mn21nbbn}
\hat{\Omega}\ :\ y\mapsto y+\frac{2\pi\ri}{n+r}\,,\qquad\qquad E\mapsto q^{-2}\,E\,,
\ee
which is used to derive the quantum Wronskian relation for the spectral determinants.
Our proposal is that the scaling limit can be defined in such a way that
the ground state of the spin chain is described in terms
of the differential equation \eqref{c}, i.e.,
\bea\label{a873wehg12} %
 \Big[-\partial_y^2+p^2+ \re^{(n+r)y}-(-1)^A\,E^r\ \re^{ry}-
 \delta U(y)
 \,\Big]\,\psi=0
\eea
with
\bea\label{akjsaussauy}
\delta U(y)=\sum_{(\mu,j)\in\bm{\Xi}_{r,A}}
 c_{\mu,j}\, E^{\mu}\ \re^{\big(  (A\mu-rj)\,\frac{n+r}{r}+\mu\big) y}
\eea
and its barred counterpart. It is easy to check that
the ODE   is indeed invariant w.r.t. the transformation $\hat{\Omega}$
with $q=\re^{\frac{\ri\pi }{r}A+\frac{\ri\pi}{n+r}}$. Formally this holds true for  
integer $j$ and
arbitrary values of $\mu$. However,
in order for the spectral determinant to be an entire function of $E$, $\mu$ should be a non-negative integer.
To specify the ODE completely one needs to describe the set  $\bm{\Xi}_{r,A}$, 
which occurs in the sum in \eqref{akjsaussauy}.
 For the case $r$ odd and $A=\frac{r-1}{2}$, this has already been carried out in the previous section, 
\be
\bm{\Xi}_{r,\frac{r-1}{2}}=\big\{(\mu,j)\, :\ \ \mu=2j+1\ \  \&\ \ j=0,\ldots,\tfrac{r-3}{2}\big\}\,\qquad\qquad (r\ {\rm odd})\, .
\ee
Then the differential equation \eqref{a873wehg12} becomes  \eqref{hasytsat}
with $c_{2j+1}\equiv c_{2j+1,j}$

\medskip
\subsection{The case $A=r-1$\label{subsec31}}
The ODE/IQFT correspondence in the regime $\pi(1-\frac{1}{r})<\gamma<\pi$ was  proposed in 
ref.\cite{Kotousov:2021vih}. Here we give a brief summary of the  results of that work 
relevant to our current discussion.
\medskip

For $A=r-1$ the integers $\mu$ and $j$ are specified as
\bea
\mu=j+1\ \ \ \ \ \  \ \quad {\rm with}\quad \ \ \ \ \ \ j=0,\ldots, A-1\,.
\eea
Then the differential equation \eqref{a873wehg12} takes the form
\bea\label{hasytsatd}
 \bigg[-\partial_y^2+p^2+  \re^{(n+r)y}+(- E)^r\,\re^{ry}-\sum_{\mu=1}^{r-1}
 c_{\mu}\, E^{\mu}\ \re^{ry+\frac{n}{r} (r-\mu) y}\,\bigg]\,\psi=0\,.
\eea
Performing  the change of variables
\bea
z=-E^{-1}\,\re^{\frac{n}{r}y}\ ,\ \ \ \ \Psi=\re^{\frac{n}{2r}y}\,\psi
\eea
one arrives at
\bea\label{sak23jhsdnba12}
\Big[-\partial^2_{z}+z^{-2}\,\Big(\big(\tfrac{rp}{n}\big)^2-\tfrac{1}{4}\Big)+\kappa^2 z^{-2+\xi r}\, P_r(z)\Big]\, \Psi=0\,.
\eea
Here
\bea
\kappa^2=(-1)^r\, \Big(\frac{rp}{n}\Big)^2\, E^{\frac{r(n+r)}{n}}\ , \qquad\qquad \xi=\frac{r}{n}\, ,
\eea
while  $P_r(z)$ is a polynomial of degree $r$ given by
\bea
 P_r(z)=z^r-\sum_{\mu=1}^{r-1}
 (-1)^{\mu}\, c_{\mu}\, z^{r-\mu}+1\ .
\eea
If one assumes that all the coefficients $c_{\mu}$ are sufficiently small, then the roots of this polynomial would be close to
$\re^{\frac{\ri\pi}{r}(2a+1)}$ and hence  simple. This way, $P_r(z)$ can be written
as
\bea\label{sak23jhsdnba12111}
P_r(z)=\prod_{a=1}^r(z-z_a)\qquad\qquad {\rm with}\qquad\qquad
\prod_{a=1}^r (-z_a)=1\, ,
\eea
where $z_a$ are distinct.
\medskip

It was proposed
in section 12 of ref.\cite{Kotousov:2021vih} that the differential equation \eqref{sak23jhsdnba12}
\eqref{sak23jhsdnba12111} describes the scaling limit of the Bethe roots for the ground state
of the spin chain with $q=-\re^{\frac{\ri\pi}{r}(\beta^2-1)}$, where $0<\beta^2\equiv\frac{r}{n+r}<1$.
 In taking the limit, the inhomogeneities are assumed to be fixed
and sufficiently close to their values for the ${\cal Z}_r$  invariant case. Also, 
without loss of generality, the normalization condition was imposed 
as
\be
\prod_{\ell=1}^r\eta_\ell=1\, .
\ee
In our notation, this prescription for the scaling limit corresponds to setting the  exponent $d_s$ in the definition of
the RG invariants $\mathfrak{a}_s$ \eqref{saaaakju32hjsd}
 to be zero, i.e.,
\bea\label{saaaakju3d}
{\mathfrak a}_{s}=  \frac{1}{s r}\ \sum_{\ell=1}^r(\eta_\ell)^{-s}\qquad\qquad (s=1,2,\ldots,r-1)\,.
\eea
The relation between the set $\{\mathfrak{a}_s\}_{s=1}^{r-1}$ 
and the coefficients $\{c_{\mu}\}_{\mu=1}^{r-1}$ of the differential equation \eqref{hasytsatd}
is rather complicated. It is described explicitly for the case $r=2$  in section 12.3 in  \cite{Kotousov:2021vih}.
\medskip

For $A=r-1$ 
 the full class of ODEs which appear in 
the scaling limit of the low energy excited states of the  spin chain  is known \cite{Kotousov:2021vih}.
It will not be presented  here. We just mention that though
the integrable structure underlying the CFT depends on the RG invariants,
they have no effect on the conformal structure. In particular, the 
eigenvalues and degeneracies of the  Hamiltonian $\hat{H}$ defined through \eqref{hastsaasuu}
does not depend on $\mathfrak{a}_s$   \eqref{saaaakju3d} provided that their absolute values are sufficiently small.

\subsection{General restrictions on  integers $\mu$ and $j$}
As was mentioned in the Introduction, our work is devoted to the study of the scaling limit of the spin chain
with ``softly'' broken ${\cal Z}_r$ symmetry.
At the level of the differential equation  \eqref{a873wehg12}, this can be formulated as the requirement that the term $\delta U$
\eqref{akjsaussauy}
may be treated as a perturbation. In view of this, we impose the following constraints:

\begin{enumerate}
\item[(a)] The function 
\bea\label{893qhjbnsdb2}
\delta U(y)=\sum_{(\mu,j)}
 c_{\mu,j}\, E^{\mu}\ \re^{\big(  (A\mu-rj)\,\frac{n+r}{r}+\mu\big) y}
\eea
vanishes as $y\to-\infty$:
\bea\label{auususa}
\lim_{y\to-\infty}
\delta U(y)= 0\ \ \ \ \ {\rm for}\ \ \ \forall n>0\ .
\eea
 Then one can still introduce the Jost solutions of the ODE,
which are involved in the derivation of the quantum Wronskian relation  \eqref{as8931hjds},
 through the asymptotic conditions
\bea\label{dksdjksdjs}
\psi_{\pm p}(y)\to \re^{\pm py}\ \ \ \  \ \ \qquad {\rm as}\ \ \ \qquad y\to-\infty\qquad\qquad \big(\Re e(\pm p)\geq 0\big)\,.
\eea
It is easy to see that \eqref{auususa}
implies that both the integers $\mu$ and $j$ can not be simultaneously zero  as well as the inequalities
\begin{subequations}\label{ksissuaALL}
\bea\label{ksissua}
&&\mu\geq  {\rm max}\Big(0,\frac{r j}{A}\Big) \qquad{\rm for}\qquad A\ge 1 \\[0.2cm]
&& \mu\ge 0\ \&\ j\le 0\qquad\ \ \  {\rm for}\qquad A=0\, .
\eea
Here we also took into account that  $\mu$ must be non-negative, which is required in order that the spectral determinant
admits a  power series expansion at $E=0$. 
\end{subequations}

\item[(b)]  The other constraint is that
$\delta U(y)$ should grow slower than  the term $\propto\re^{(n+r) y}$ 
in eq.\,\eqref{a873wehg12} as $y\to+\infty$, i.e.,
\bea\label{sak9823aaaakjds}
\lim_{y\to+\infty}
\re^{-(n+r) y}\,\delta U(y)= 0\ \ \ \ \ {\rm for}\ \ \ \forall n>0\ .
\eea
This implies
\bea\label{aiissaiias}
\mu<\frac{  r}{A+1}\  (j+1) \  \ \ \ \ \ \ \ \ \ \ \ \ \qquad (0\le A<r-1)\ .
\eea
It follows from \eqref{sak9823aaaakjds} that the zeroes of the spectral determinants  would
accumulate along the same  rays as for the ${\cal Z}_r$ invariant case. 
For $A=r-1$ it was discussed in the previous subsection that $\mu=j+1$  and the inequality \eqref{aiissaiias} is saturated.
In turn, the Stokes lines are considerably more cumbersome
to describe (see, e.g., Appendix B in ref.\cite{Kotousov:2021vih}). This makes the case $A=r-1$ 
somewhat special. 
\end{enumerate}

The  conditions \eqref{ksissuaALL} and \eqref{aiissaiias} have some immediate consequences.
In particular, one has for the integer $j$ that
\be
 0\leq j\leq A-1\qquad\qquad (A=1,\ldots,r-2)\, .
\ee
Since we exclude the case when both $\mu$ and $j$ are zero, it follows from \eqref{ksissua} that $\mu$
is a positive integer and, furthermore,
\be\label{8932issanbsadbv21}
1\leq \mu<\frac{A}{A+1}\ r\,,\qquad \qquad \qquad (A=1,\ldots,r-2)\, .
 \ee
 As  for $A=0$,
the possible values of the pair $(\mu,j)$ are described as
\be\label{as9812aaaasy32hsd}
\mu=1,2,\ldots,r-1\ \&\ j=0\qquad \qquad \qquad (A=0)\, .
\ee
\bigskip

If $\delta U(y)$ is given by \eqref{893qhjbnsdb2}, where the summation is taken over the pairs
$(\mu,j)$ satisfying the above conditions,
the spectral determinants for the ODE \eqref{a873wehg12} can be  introduced along the same lines as was done
in secs.\,\ref{sec22} and \ref{sec23}.\footnote{%
Strictly speaking formula \eqref{as8723hjsdaa} used to specify the subdominant at $y\to+\infty$ solution $\chi$, entering into the
definition \eqref{jassysa} of the spectral determinant, is not valid in general. Nevertheless, $\chi$ may still be
unambiguously introduced via its large\,-\,$y$ asymptotic obtained within the WKB approximation.} 
The proof that they satisfy the quantum Wronskian relation \eqref{as8931hjds},
which is based on the invariance of the differential equation w.r.t. the transformation \eqref{asn3mn21nbbn}, 
remains unchanged.  Further, $D_\pm(E)$ are entire functions of $E$.
\bigskip

\section{ODE with single coefficient $c_{\mu,j}$ for $A\mu>rj$  and $A=1,\ldots,r-2$\label{sec33}}
We now come up against the problem of giving a precise description of the scaling limit of the spin chain
that would result in the differential equation belonging to the class \eqref{a873wehg12} and \eqref{akjsaussauy}.
This involves specifying the exponents $d_s$ entering into the definition of the RG invariants $\mathfrak{a}_s$
\eqref{saaaakju32hjsd} as well as relating their values to the coefficients $c_{\mu,j}$ of the ODE. 
However, when carried out in full generality, one runs
into, besides the technical complexity, also conceptual issues. For this reason, we first focus our attention
on formulating the scaling limit such that the 
corresponding differential equation is given by  
\bea\label{3892jh12ssasa}
 \bigg[-\partial_y^2+p^2+ \re^{(n+r)y}-(-1)^A\,E^r\ \re^{ry}-
c\, E^{\mu}\ \re^{\big(  (A\mu-rj)\,\frac{n+r}{r}+\mu\big) y}
 \,\bigg]\,\psi=0  \qquad\qquad (c\equiv c_{\mu,j})\,,
\eea
where the pair of integers $(\mu,j)$ obeys the conditions 
\be\label{askjas8712jhsa}%
\frac{rj}{A}<\mu<
\frac{  r}{A+1}\  
(j+1)\ \&\ j\ge 0 \qquad\qquad (1\le A<r-1)\, .
\ee
Notice that we have excluded the case $A\mu=rj$ which, as will be discussed below, requires a separate treatment.

\medskip
\subsection{Bohr-Sommerfeld quantization condition\label{sec41}}
The experience gained in sec.\,\ref{sec23} suggests to start with the analysis of the large\,-\,$E$ 
asymptotic behaviour of the spectral determinant.
A computation, based on the WKB approximation, leads to the asymptotic formula
\bea\label{aaakjs712hdsas}
\log D_+&\asymp& \frac{N_0\re^\theta}{\cos(\frac{\pi r}{2n})}
-
\sum_{1\leq k\leq  \frac{r}{2M}}\re^{(r-2k M)\frac{\theta}{r}}\  \re^{\frac{\ri\pi}{r}(2a-1-A) k \mu}\ 
\big( (-1)^{(r-1)\mu} \,c\big)^k\\[0.2in]
&\times&\frac{\Gamma\big(\frac{k \mu}{r}-\frac{n+r}{2nr}\ (r-2kM) \big )
\Gamma\big(k-\frac{1}{2}-\frac{k \mu}{r}+\frac{n+r}{2nr}\ (r-2k M)\big)}{2n\sqrt{\pi}\,k!}-\frac{2np \,\theta}{r(n+r)}+\log C_p
+o(1)\,.
 \nonumber
 \eea
 Here $N_0$ and  $C_p$ are given by \eqref{asm8932hg21} and \eqref{jasays}, respectively, while
 \bea\label{3902jds892hjs}
M=(j+1)\,r-(A+1) \, \mu>0
\eea
 is a  positive integer (which follows from \eqref{askjas8712jhsa}).
Again, $\theta$ parameterizes $E$ in the wedge  labelled by the integer $a$ as
 \bea
E=(-1)^{r-1}\ 
 \re^{\frac{\ri \pi   }{r} (2a-1-A)}\ \re^{\frac{2 n\theta}{r(n+r)}}
\eea
and the asymptotic formula is valid for $\Re e(\theta)\to+\infty $ with
$ \big|\Im m(\theta)\big|<\frac{\pi(n+r)}{2n}$.
\medskip

One can also derive the Bohr-Sommerfeld quantization condition, from which the  location 
of the zeroes $E_m^{(a)}$ of the spectral determinant $D_+(E)$ for $m\gg1$ may be deduced.  
Writing the zeroes in terms of  $\theta_m^{(a)}$ according to \eqref{asui781hew}, one
finds that 
\bea
X=\re^{-\frac{\ri\pi \mu}{2rM} (2a-A)}\ \re^{\frac{1}{r}\theta_m^{(a)}}
\eea
with $a=1,2,\ldots,r$ and $m\gg 1$ solves the equation
\bea\label{asusauasu}
X^r+\sum_{k=1}^{ [ \frac{r}{2M}]}{ G}_{2kM}\ X^{r-2k M} +o(1)=Y^r\ .
\eea
Here $[\ldots]$ stands for the integer part and the notation
\bea
 Y=\re^{-\frac{\ri\pi \mu}{2rM} (2a-A)}\
\bigg(\frac{\pi}{N_0}\ \Big(m-\frac{1}{2}+\frac{p}{r}\,\Big)\bigg)^{\frac{1}{r}}\ ,\ \ \ \ \ { G}_{2kM}={\mathfrak g}_{2k}\ \Big((-1)^{(r-1) \mu}\  c\Big)^k  
\eea
as well as
\bea\label{j3aaaa2h87hj2saaa}
{\mathfrak g}_{2k}=\frac{\Gamma(\frac{3}{2}+\frac{r}{2n})}{\Gamma(\frac{r}{2n})}\ 
\frac{
\Gamma\big(\, \frac{k}{r}\, L+\frac{1}{2n}\ (r-2kM)\big)}{k!\,
\Gamma\big(\frac{3}{2}-k+\frac{k}{r}\, L+\frac{1}{2n}\ (r-2kM)\big )}
\eea
is being used.
Also, together with the positive integer $M$ \eqref{3902jds892hjs}, we introduce 
\be\label{jh7823iujhsdghas}
L=A\mu-rj> 0\,,
\ee
which is a positive integer 
due to the condition \eqref{askjas8712jhsa}.
Notice that the argument of the $\Gamma$\,-\,function occurring in the numerator of the second fraction
in $\mathfrak{g}_{2k}$ is always
positive since, as follows from the  upper limit in the sum  in \eqref{asusauasu}, $k\le\frac{r}{2M}$.
It vanishes when this bound is saturated and $L=0$, which is only possible if  both $A$ and  $r$ are
 divisible by $2K$ and
\be\label{nash1278as}
j=\frac{A}{2K}\, (2K-1)\,,\ \ \qquad
\qquad\ \ \mu=\frac{r}{2K}\, (2K-1)\, \qquad\qquad (K=1,2,\ldots) \ .
\ee
Then, the last term in the sum in \eqref{asusauasu} would be proportional to $\log(X/{\rm const})$.
This, among other things, makes the case $L=0$ somewhat special and is one of the reasons why it will be
excluded from the discussion of this section.
\begin{comment}
$\re^{\frac{1}{r}\theta_m^{(a)}}$ with $a=1,2,\ldots,r$ and $m\gg 1$ are the roots of the equation

\bea\label{asusauasu}
x^r+ \sum_{1\leq k\leq  \frac{r}{2\mu_j}}{\tilde G}_{2k\mu_j}\  \re^{\frac{\ri\pi}{r}(2a -A) k m_j} 
\ x^{r-2k \mu_j}+o(1)=\frac{\pi}{N_0}\  \Big(m-\frac{1}{2}+\frac{p}{r}\Big)\,,
\eea
where
\bea
{\tilde G}_{2k\mu_j}=
(-1)^{(r-1) km_j} \ 
\frac{\Gamma(\frac{3}{2}+\frac{r}{2n})}{\Gamma(\frac{r}{2n})}
\frac{
\Gamma\big(\, \frac{k}{r}\, (Am_j- r j)+\frac{1}{2n}\ (r-2k\mu_j)\big)}{k!\,
\Gamma\big(\frac{1}{2}+\frac{k}{r}\, (Am_j- r j)+\frac{1}{2n}\ (r-2k\mu_j)\big )}\  \big( c_{m_{j}}\big)^k\ .
\eea
\medskip
\end{comment}
\medskip

If the positive integer $M$  entering into
 the quantization condition \eqref{asusauasu}
exceeds $r/2$, the sum vanishes and hence, as $m\to+\infty$,
\bea\label{askj8723jhsaaad}
E_m^{(a)}=(-1)^{r-1}\,\re^{\frac{\ri \pi  }{r}\,(2a-A)}\ \bigg( \frac{\pi}{N_0}\  \Big(m-\frac{1}{2}+\frac{p}{r}\,\Big)\bigg)^{\frac{2n}{r(n+r)}}\ \Big(1+o(m^{-1})\Big)\ \ \ \ \  \ \ \ \ \ (M>r/2)\ .
\eea
If
  $0< M\leq r/2$, i.e., 
  \bea\label{ask782jhewbn12}
\frac{  r}{A+1}\  
\Big(j+\
\frac{1}{2}\Big) \leq  \mu <\frac{  r}{A+1}\  (j+1) \ ,  
  \eea
the sum  in \eqref{asusauasu}  is not trivial and
 the leading large\,-\,$m$ asymptotic formula for $E_m^{(a)}$ reads as
\bea\label{saj8723hg21}
E_m^{(a)}=(-1)^{r-1}\,\re^{\frac{\ri \pi  }{r}\,(2a-A)}\ \bigg(\frac{\pi}{N_0}\  \Big(m-\frac{1}{2}+\frac{p}{r}\,\Big)\bigg)^{\frac{2n }{r(n+r)}}\ \bigg(1+O\Big(m^{-\frac{2M}{r}}\Big)\bigg)\  \ \ \ \ \ \ 
(0<M\leq r/2)\,.
\eea
In both cases the zeroes accumulate along the same Stokes lines 
as  in the ${\cal Z}_r$ invariant spin chain.
However, 
 the subleading term in \eqref{saj8723hg21}, which evaluates to a complex number, 
decays considerably slower than $o(m^{-1})$ from  eq.\,\eqref{askj8723jhsaaad} so that
the zeroes approach to the rays less rapidly when the value of the  integer $\mu$ is
restricted to the interval \eqref{ask782jhewbn12}.
This, in turn, impacts the convergence of the series
\bea
J_s=\frac{1}{s}\ \sum_{m=1}^{\infty}\sum_{a=1}^r\big(E_m^{(a)}\big)^{-s}\,,
\eea
which would require, in general,  the introduction of counterterms.
Similar to what was discussed in sec.\,\ref{sec23}, one can make use of this to try to deduce the
relation between the RG invariants \eqref{saaaakju32hjsd}
 and the coefficient $c$ in \eqref{3892jh12ssasa}.
\medskip

\subsection{Specification of the scaling limit}
Let's first consider the case when  the pair $(\mu,j)$  satisfies the inequality
\be\label{903217jajsqw}
 \mu\ge\frac{  r}{A+1}\  
\Big(j+\
\frac{1}{2}\Big) 
\ee
in addition to \eqref{askjas8712jhsa}.
Numerical work shows that it is possible to 
organize the scaling limit  in such a way that
 the scaled  Bethe roots for the ground state of the spin chain are described by
the ODE of the form  \eqref{3892jh12ssasa}. 
The exponent 
$d_s$, which appears in the definition of the RG invariants,
\bea\label{ais78176326512assadas}
{\mathfrak a}_{s}=\frac{1}{s}\ \bigg(\frac{N}{r N_0}\bigg)^{d_s}\  \frac{1}{r}\ \sum_{\ell=1}^r(\eta_\ell)^{-s}\,,
\eea
should be taken to be
\be\label{saaaakju32hjsdB}
d_{s}=\frac{2M}{r}\,i_s\,,
\ee
where   $i_s$ denotes  the {smallest} positive integer such that
\be\label{sak7832jhsd}
s=\mu\,i_s\ \ \ ({\rm mod}\ r)
\ee
and, as before,  $M=(j+1)\,r-(A+1) \, \mu$. If $s$ 
is not divisible by $\mu$ modulo $r$, i.e., the integer $i_s$ 
does not exist, the corresponding RG invariant can be set to zero:
\be\label{askj7823h21*asas}
\mathfrak{a}_s=0\qquad{\rm if}\qquad s\nmid \mu \ ({\rm mod}\ r) \, . 
\ee
In addition, we assume the normalization condition for the inhomogeneities
\be\label{iaassauasas}
\frac{1}{r}\,\sum_{\ell=1}^r (\eta_\ell)^{-r}=(-1)^{r-1}\, .
\ee
\bigskip

The inequality \eqref{903217jajsqw}  guarantees that there are RG invariants $\mathfrak{a}_s$ for which  $0<d_s\le 1$.
In this case one can use the trick discussed in sec.\,\ref{sec23}
to deduce the relation
\bea\label{sajk8937hj21hg}
{\mathfrak a}_{ s}=(-1)^{s-(j+1) i_s+{\cal N}_s}\ Z_{s}\ c^{i_s}\ .
\eea
Here the integer ${\cal N}_s$ is defined by the condition
\bea
s=\mu\,i_s+r\,{\cal N}_s\ .
\eea
The coefficients $Z_{s}$ are   built from  $\mathfrak{g}_{2k}$ \eqref{j3aaaa2h87hj2saaa} as
\bea\label{saj32jhghsaaaaa}%
Z_{s}
= \frac{1}{s}\sum_{m=1}^{i_s}(-1)^m\ \frac{\Gamma(s\nu+m)}{m! \Gamma(s\nu) }
\sum_{
k_1,\ldots, k_m\ge 1\atop
k_1+k_2+\ldots +k_m={i_s} }{\mathfrak R}_{k_1}\ldots \mathfrak{R}_{k_m}
\eea
with
\be\label{asi7823hdjssa}
\mathfrak{R}_{k}=\frac{1}{r} \sum_{
\alpha_1+2\alpha_2+\ldots+k\alpha_{k}=k\atop
\alpha_1,\ldots,\,\alpha_{k}\geq 0}
\frac{(-1)^{\alpha_1+\ldots+\alpha_{k}}}{\alpha_1!\alpha_2!\ldots\alpha_{k}!}\ 
\frac{\Gamma(\alpha_1+\ldots+\alpha_{k}-\frac{2kM-1}{r})}{\Gamma(1-\frac{2kM-1}{r})}\ 
\ {\mathfrak  g}_{2}^{\alpha_1} {\mathfrak g} _{4}^{\alpha_2}\dots {\mathfrak g} _{2k}^{\alpha_{k}}
\ee
and 
\bea\label{3902jdsjh}
\nu=\frac{2n}{n+r}\, .
\eea
\medskip

It needs to be emphasized that the argument used to obtain \eqref{sajk8937hj21hg} 
can not be literally applied to the case when the exponent $d_s$ for the RG invariant
$\mathfrak{a}_s$ exceeds one. Nevertheless,
we found that under certain conditions
which will be clarified in the examples below,  the scaling limit of the Bethe roots is described by the differential equation \eqref{3892jh12ssasa} if the RG invariants are specified as in \eqref{askj7823h21*asas},\, 
\eqref{sajk8937hj21hg}\,-\,\eqref{3902jdsjh}   for any
$s$. 
Moreover, numerical work shows that the values of some of the RG invariants $\mathfrak{a}_s$, where
$s$ is divisible by $\mu$ modulo $r$ and
$d_s>1$ have no effect on the scaling limit and can be set to zero. 
In particular, for $r\le 10$ this was always observed  if the corresponding exponent $d_s\ge 2$.  
 Unfortunately, 
we do not have at hand a full  description of the minimal set of non-trivial RG invariants that works for
generic values of $r$ and  $A=1,\ldots,r-2$.\footnote{%
Here and below, by ``non-trivial RG invariants'' we mean those $\mathfrak{a}_s$,
whose values affect the scaling limit of the Bethe roots.} For this reason, we list them in
the tables in Appendix \ref{AppD2},  along with the corresponding exponents $d_s$ for all the cases with
 $r=3,4,\ldots,10$, which were deduced from a numerical study.
\medskip

\subsubsection{Example: $r=5$, $A=1$ and $(\mu,j)=(2,0)$}
In this case we found that there are three non-trivial RG invariants
 $\mathfrak{a}_2$, $\mathfrak{a}_4$ and
$\mathfrak{a}_1$, whose corresponding exponents are given by $d_2=\tfrac{2}{5}$, $d_4=\tfrac{4}{5}$ and
$d_1=\tfrac{6}{5}$, respectively.
A specialization of formulae \eqref{sajk8937hj21hg}\,-\,\eqref{3902jdsjh} yields
\begin{subequations}
\bea\label{iaiassausa}
{\mathfrak a}_2&=&- \tfrac{\nu}{5}\, \mathfrak{ g}_{2}\ c
\nonumber\\[-0.1in]
&& \\[0.0cm]
{\mathfrak a}_4&=&\tfrac{\nu}{5} \, \big(\mathfrak{ g}_{4}+\tfrac{1}{10}\, (4\nu-1)\ \mathfrak{ g}_{2}^2\big) \ c^2\nonumber\\[0.1in]\label{9832jhdssa}
{\mathfrak a}_1&=&-\tfrac{\nu}{5}\, \big( \mathfrak{ g}_{6}+
\tfrac{1}{5}\, (\nu+1)\, \mathfrak{ g}_{2}\mathfrak{ g}_{4}+\tfrac{1}{150}\,
(\nu+1)(\nu-4)\, \mathfrak{ g}_{2}^3\big)\ c^3\,,
\eea
\end{subequations}
where $\nu=\frac{2n}{n+5}$ and the  functions $\mathfrak{g}_{2k}$ are  defined in eq.\eqref{j3aaaa2h87hj2saaa}
with $M=1$ and $L=2$.
Then, according to the above discussion,  the ODE appearing in the scaling limit is expected to be
\bea\label{kj8293hdsjh}
 \bigg[-\partial_y^2+p^2+ \re^{(n+5)y}+E^5\ \re^{5 y}-
 c\, E^2\,\re^{\frac{2}{5}(n+10)\, y}
 \,\bigg]\,\psi=0\ ,
\eea
where recall that $p=\frac{1}{2}\,(n+5)\,{\tt k}$, see \eqref{jashgh12b1b}.
One may observe that both $\mathfrak{g}_2$ and $\mathfrak{g}_4$ are non-singular functions of $n$ for any $n>0$.
In contrast $\mathfrak{g}_6$, which reads explicitly as
\be
\mathfrak{g}_6=
\frac{\Gamma(\frac{3}{2}+\frac{5}{2n})}{\Gamma(\frac{5}{2n})}\,
\frac{\Gamma(\frac{6}{5}-\frac{1}{2n})}{
6\,\Gamma(-\frac{3}{10}-\frac{1}{2n})}\,,
\ee
possesses simple poles when  $n=\frac{5}{2(1+5k)}$ with $k=1,2,3,\ldots\,$, accumulating at $n=0$.
As a result, the proportionality coefficient $Z_1^{(\mu=2)}=-\mathfrak{a}_1/c^3$  
 also contains poles at the same values of $n$, see eq.\,\eqref{9832jhdssa}.
Moreover, $Z_1^{(\mu=2)}$ has zeroes with the largest one located at $n=\frac{5}{2}$.
From the numerical work, we were able to confirm that the scaling limit
of the Bethe roots for the ground state is governed by the ODE \eqref{kj8293hdsjh}
for $n> \frac{5}{2}$. As for the domain $0<n\le\frac{5}{2}$, it turns out to be difficult to
explore for reasons to be discussed below.
\medskip

\subsubsection{Example: $r=5$, $A=1$ and $(\mu,j)=(1,0)$}
So far, we have taken $\mu$ and $j$ to obey the inequality \eqref{903217jajsqw} in addition to  \eqref{askjas8712jhsa}.
For $\mu<
\frac{  r}{A+1}\  
(j+\
\frac{1}{2}) $ or, equivalently, the integer 
$M$  \eqref{3902jds892hjs} exceeds $\frac{r}{2}$, it follows from formula \eqref{saaaakju32hjsdB}
that all the exponents corresponding to the non-trivial RG invariants are  greater than one.
The  numerical study shows that 
the relations obtained assuming  $\mu$ satisfies  both
$\frac{  r}{A+1}\  
(j+\
\frac{1}{2}) \leq  \mu $ and $\frac{rj}{A}< \mu<\frac{  r}{A+1}\  (j+1)$
 remain valid if the former condition is dropped. This can be illustrated on the example
$r=5$,  $A=1$ and $(\mu,j)=(1,0)$.  Then, there is only one non-trivial
RG invariant $\mathfrak{a}_1$ with corresponding exponent $d_1=\frac{6}{5}>1$.
Formula \eqref{sajk8937hj21hg}, specialized to the case at hand, reads as
\be
\mathfrak{a}_1= Z_1^{(\mu=1)}\,c\qquad\qquad {\rm with}\qquad \qquad
Z_1^{(\mu=1)}=\frac{\Gamma(\frac{1}{2}+\frac{5}{2n})}{5\,\Gamma(\frac{5}{2n})}\,
\frac{\Gamma(\frac{1}{5}-\frac{1}{2n})}{
\Gamma(\frac{7}{10}-\frac{1}{2n})}\ .
\ee
Notice that the factor $Z_1^{(\mu=1)}$ remains finite and non-vanishing as $n>\frac{5}{2}$.
In this domain, it was numerically verified 
 that the corresponding differential equation is given by
\bea
 \bigg[-\partial_y^2+p^2+ \re^{(n+5)y}+E^5\ \re^{5 y}-
 c\, E\,\re^{\frac{1}{5}(n+10)\, y}
 \,\bigg]\,\psi=0\ .
\eea
\smallskip

\subsubsection{Example:  combining the cases $\mu=1$ and $\mu=2$ for $r=5$, $A=1$, $j=0$}
We found it rather surprising that in the prescription for taking the scaling limit,
relations \eqref{sajk8937hj21hg}-\eqref{3902jdsjh} hold true
for the RG invariants $\mathfrak{a}_s$ with  $d_s>1$ at least for sufficiently large $n$. For this reason, 
we performed an additional numerical study with a slightly different setup. The exponents $d_s$ for 
all the non-trivial  RG invariants  \eqref{ais78176326512assadas} were taken as in \eqref{saaaakju32hjsdB}.
Furthermore, the value of the RG invariants $\mathfrak{a}_s$ with $0<d_s\le 1$ were fixed
according to \eqref{sajk8937hj21hg}, while those with $d_s>1$
were kept as free parameters. It was observed that for sufficiently large $n$
the  scaling limit of the Bethe roots for the ground state is described in terms of the differential equation  
of the form
\bea \label{392821uyhsabbds}%
 \bigg[-\partial_y^2+p^2+ \re^{(n+r)y}-(-1)^A\,E^r\ \re^{ry}-
 c\, E^{\mu}\ \re^{\big(  (A\mu-rj)\,\frac{n+r}{r}+\mu\big) y}-\delta\tilde{U}(y)
 \,\bigg]\,\psi=0\,,
\eea
where 
\be\label{ask387jas}
\delta \tilde{U}(y)=\sum_{s\in{\Sigma}}\ \tilde{c}_s\, E^{s}\ \re^{\big( (A s-rj)\,\frac{n+r}{r}+s\big)y}
\ee
and
\be
\Sigma=\Big\{s:\   s=\mu\, i\  ({\rm mod}\, r)\ \ {\rm with}\ 
i=\big[\tfrac{r}{2M}\big]+1,\ldots,r-1\ \ \ \&\ \  \ \tfrac{rj}{A}\le  s<\tfrac{r}{A+1}\,(j+\tfrac{1}{2}) \Big\}\ .
\ee
The appearance of such an ODE may be expected since it belongs to the class 
 \eqref{a873wehg12},\,\eqref{akjsaussauy}, while the presence of the term
$\delta \tilde{U}$ does not affect the Bohr-Sommerfeld quantization condition \eqref{asusauasu},
which was derived for the case with $\delta \tilde{U}=0$. The coefficients $\tilde{c}_s$ depend on 
$c$ as well as  the values of  $\mathfrak{a}_s$ with $d_s>1$. 
They can be obtained numerically, for instance, from the study of the first few sum rules,
along the lines of sec.\,\ref{sec22}. 
\medskip

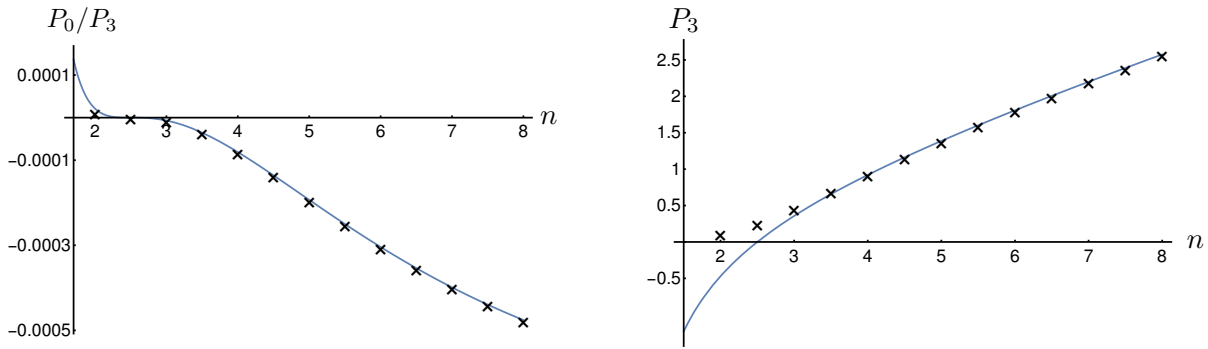
\begin{figure}
\begin{center}
\begin{tikzpicture}
\node at (0,0) {\includegraphics[width=7cm]{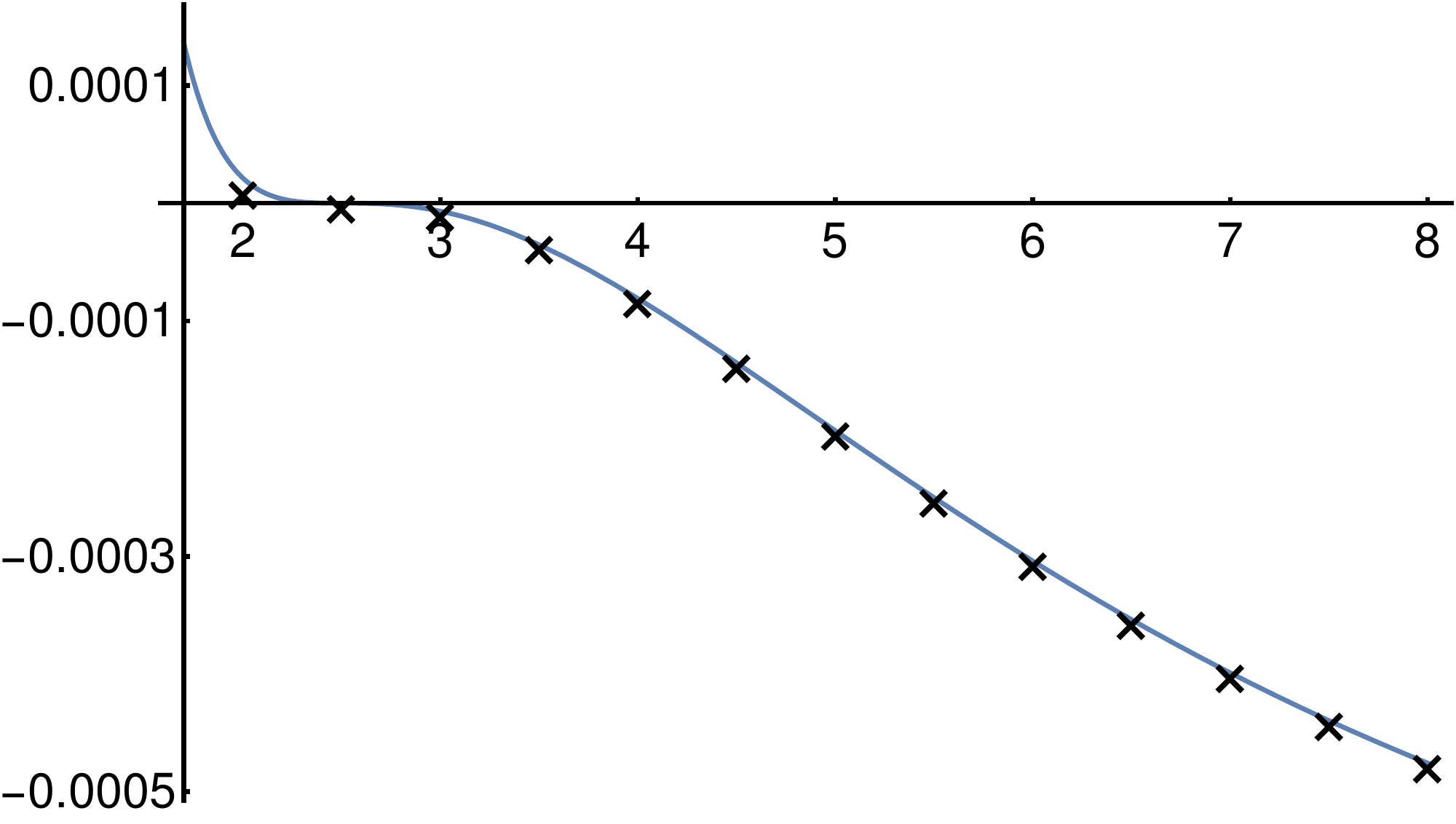}};
\node at (8.5,0) {\includegraphics[width=7cm]{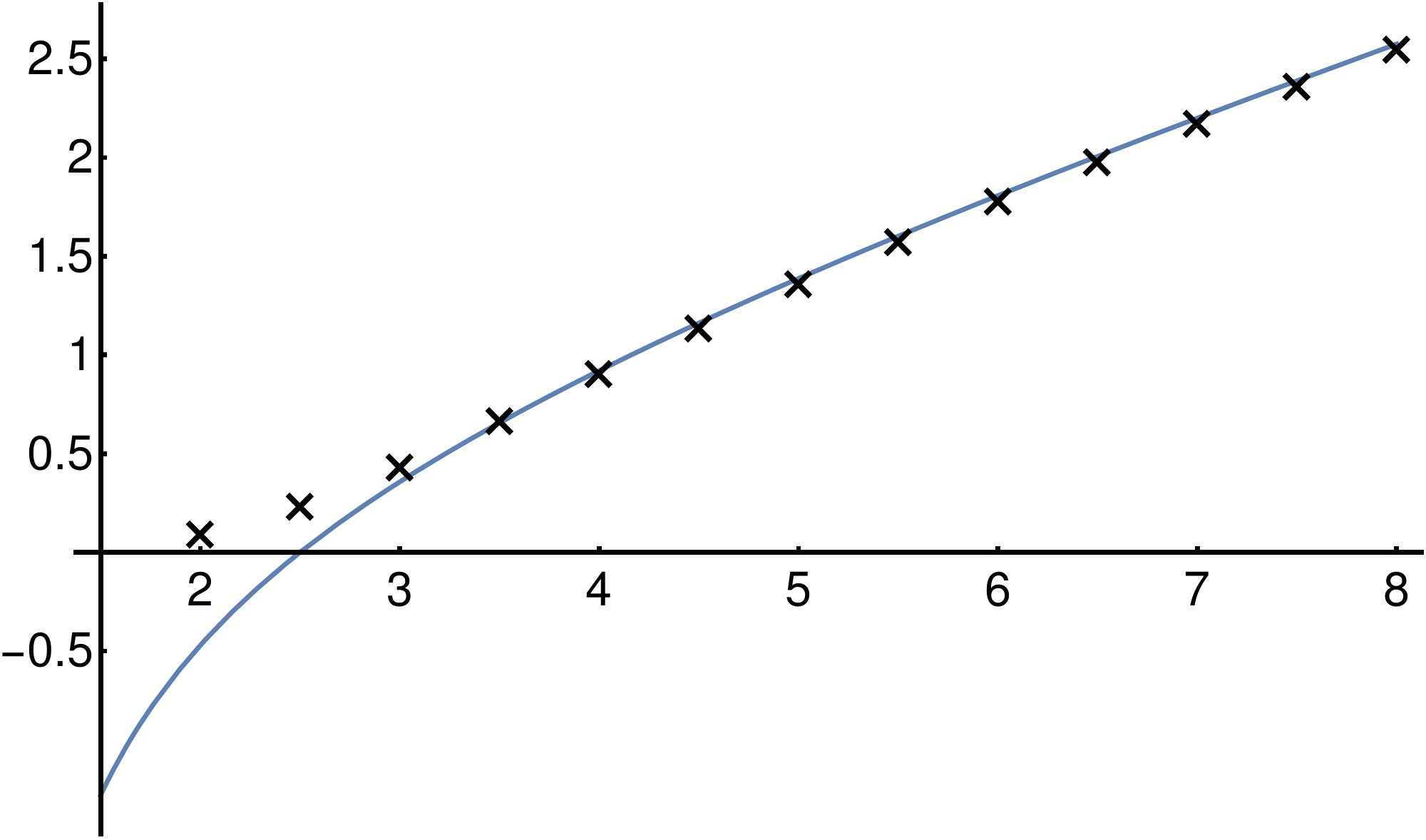}};
\node at (12.3,-0.65) {$n$};
\node at (5.5,2.3) {$P_3$};
\node at (-2.5,2.3) {\small $P_0/P_3$};
\node at (3.7,1) {$n$};
\end{tikzpicture}
\end{center}
\caption{\label{fig1}\small%
The crosses represent 
numerical data for the ratio $P_0/P_3$ (left panel) and $P_3$ (right panel) obtained through the 
study of the sum rule 
for the Bethe roots \eqref{aksjaa8732hdshg}\,-\,\eqref{aksjaa8732hdshgAAA}. The scaling limit is taken
for the spin chain with $r=5$, where the RG invariants are specified as in 
eqs.\,\eqref{ais78176326512assadas} and \eqref{saaaakju32hjsdB}
with $A=1$, $j=0$ and $\mu=2$. The values of $\mathfrak{a}_2$
and $\mathfrak{a}_4$ were fixed according to \eqref{iaiassausa}, $\mathfrak{a}_3$ was set to zero,
while the computation was performed for various values of $\mathfrak{a}_1$.
The solid line 
is a plot of the analytic formula given by \eqref{9023sdj*@} and \eqref{9023sdj*@A} for the left and
right panels, respectively. 
Note that there is a loss of accuracy as  $n \lesssim \frac{5}{2}$.}
\end{figure}

As an example, consider again the case with 
$r=5$, $A=1$, $(\mu,j)=(2,0)$. The RG invariants $\mathfrak{a}_2$ and $\mathfrak{a}_4$,
with corresponding exponents $d_2=\frac{2}{5}$ and $d_4=\frac{4}{5}$,
are taken as in \eqref{iaiassausa},  while the value of
\bea
{\mathfrak a}_{1}=\bigg(\frac{N}{5 N_0}\bigg)^{\frac{6}{5}}\  \frac{1}{5}\ \sum_{\ell=1}^5(\eta_\ell)^{-1}
\eea
is  kept free now. Then, the scaling limit of the Bethe roots for the
ground state would be described by the ODE
\bea
 \bigg[-\partial_y^2+p^2+ \re^{(n+5)y}+E^5\ \re^{5 y}-
 c\, E^2\,\re^{\frac{2}{5}(n+10)\, y}-\tilde{c}_1\,E\,\re^{\frac{1}{5}(n+10)y}
 \,\bigg]\,\psi=0\ .
\eea
Among other things, the differential equation implies
\be\label{aksjaa8732hdshg}
\bigg(\frac{N}{5N_0}\bigg)^{-\frac{2n}{5(n+5)}}\,
h_1^{(N,{\rm reg})}=\tilde{c}_1\,\frac{(n+5)^{\frac{2}{n+5}-\frac{8}{5}}}{%
\Gamma^2(\frac{4}{5}-\frac{1}{n+5})}\,
f_1\big(\tfrac{{\tt k}}{2},\tfrac{1}{5}+\tfrac{1}{n+5}\big)+o(1)\qquad\qquad(N\to\infty)\,,
\ee
where 
\bea
h_1^{(N,{\rm reg})}=\sum_{m=1}^{N/2} (\zeta_m)^{-1}+\frac{ N }{2r\cos(\frac{\pi}{r}+\frac{\pi}{n+r})}\ \sum_{\ell=1}^r(\eta_\ell)^{-1}
\eea
with $r=5$
and the function $f_1$ is defined in formula \eqref{sa8932jhdssaaaa} in Appendix \ref{AppA}.
The dependence of the l.h.s. on $c$, the twist parameter ${\tt k}$ and
the RG invariant $\mathfrak{a}_1$ can be investigated numerically. One finds that  
\be\label{aksjaa8732hdshgAAA}
\tilde{c}_1=P_0\,c^3+P_3\,\mathfrak{a}_1\qquad\qquad {\rm for}\qquad\qquad n>\tfrac{5}{2}
\ee
with certain $n$ dependent constants $P_0$ and $P_3$. 
Since $\tilde{c}_1$ must vanish as $\mathfrak{a}_1=-c^3 Z_1^{(\mu=2)}$, 
one obtains
\be\label{9023sdj*@}
P_0/P_3=\tfrac{\nu}{5}\, \big( \mathfrak{ g}_{6}+
\tfrac{1}{5}\, (\nu+1)\, \mathfrak{ g}_{2}\mathfrak{ g}_{4}+\tfrac{1}{150}\,
(\nu+1)(\nu-4)\, \mathfrak{ g}_{2}^3\big)\ .
\ee
Also notice that when  $c=0$, the RG invariants $\mathfrak{a}_2=\mathfrak{a}_4=0$ and one gets back the
case with $\mu=1$  discussed above. This yields the relation $P_3=1/Z_1^{(\mu=1)}$ or, explicitly,
\be\label{9023sdj*@A}
P_3=\frac{5\,\Gamma(\frac{5}{2n})}{\Gamma(\frac{1}{2}+\frac{5}{2n})}\,
\frac{\Gamma(\frac{7}{10}-\frac{1}{2n})}{\Gamma(\frac{1}{5}-\frac{1}{2n})}\ .
\ee
In fig.\,\ref{fig1}, numerical data
for $P_0$ and $P_3$ obtained from the study of the sum rule \eqref{aksjaa8732hdshg}
with $\tilde{c}_1$ as in
\eqref{aksjaa8732hdshgAAA}  is compared against the  predictions
\eqref{9023sdj*@},\,\eqref{9023sdj*@A}. One observes that there is apparent deviation between the
numerical results and the analytic expressions as $n$ becomes smaller than $\frac{5}{2}$. 
This is due to the slow decay of the remainder term $o(1)$ in eq.\,\eqref{aksjaa8732hdshg}. A simple
 fit of the form $b_1+b_2\,N^{-\delta}$ of the numerical data, which worked well  for large $n$,
yields a small value of the exponent $\delta$ and,
in fact, becomes highly unstable. As such, we were not able to determine
with confidence  the limiting value of the l.h.s. of  \eqref{aksjaa8732hdshg} as $n<\frac{5}{2}$. In all likelihood,
this means that  its large\,-\,$N$ behaviour can not be adequately described as $b_1+b_2\,N^{-\delta}$ 
containing  a single small exponent $\delta$.

\subsubsection{Example: $r=7$, $A=1$ and $(\mu,j)=(1,0)$\label{sec424}}
\begin{figure}
\begin{center}
\scalebox{1}{
\begin{tikzpicture}
\node at (-4.6,3.6) { $\delta$};
\node at (5.5,-2.8) {$n$};
\node at (0,0) {\includegraphics[width=10cm]{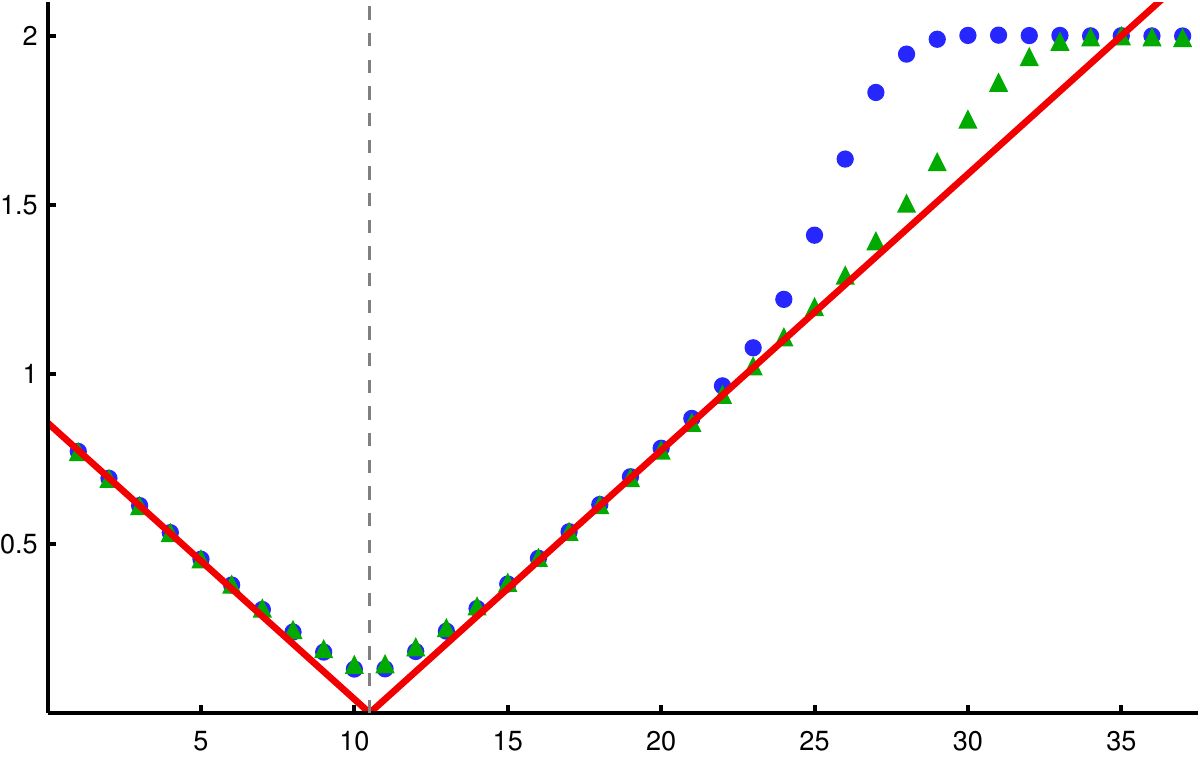}};
\end{tikzpicture}
}
\end{center}
\caption{\small%
The figure corresponds to the case $r=7$, $A=1$ and $(\mu,j)=(1,0)$.
In order to numerically analyze the correction terms in eqs.\,\eqref{90328721hajs} and \eqref{sa982h3jhdf}
the quantity $(\frac{N}{7N_0})^{-\frac{2n}{7(n+7)}}\,
h_1^{(N,{\rm reg})}$  was computed for increasing $N$. 
This was fitted  with $h_1^{(\infty)}+{\rm const}\times N^{-\delta}$, where $h_1^{(\infty)}$
stands for the known limiting value, i.e., given in  \eqref{90328721hajs} for  $n>\frac{21}{2}$ and
zero for $n<\frac{21}{2}$.
The  results for $\delta$
are represented by the blue dots. The green triangles depict  data for $\delta$ that was obtained by applying the similar
procedure to
$(\frac{N}{7N_0})^{-\frac{2n}{7(n+7)}}\,
(h_1^{(N,{\rm reg})}|_{{\tt k}=0.02}-h_1^{(N,{\rm reg})}|_{{\tt k}=0.04})$  instead.
The solid red line is a plot of
$\delta=|\frac{6}{7}-\frac{4n}{49}|$. Note that near $n\approx 10.5$  the value of the exponent $\delta$ becomes small and
the simple fit with a single exponent seems to be unreliable. For  $n>35$  one observes
$\delta=2$ with  good accuracy. Again, in the vicinity $n\lesssim 35$ the  fit with a single exponent
looks inadequate, which may explain the  significant spread in the numerical data obtained via the fitting procedures.
\label{fig4}}
\end{figure}

To understand better the problems that occurred for the cases with $r=5$  considered above, it
is instructive to turn to another example, where
 $r=7$, $A=1$ and $(\mu,j)=(1,0)$. Then there is only a single non-trivial RG invariant 
\bea\label{389hsdyu3a}
{\mathfrak a}_{1}=\bigg(\frac{N}{r N_0}\bigg)^{d_1}\  \frac{1}{r}\ \sum_{\ell=1}^r(\eta_\ell)^{-1}
\eea
with $r=7$ and $d_1=\frac{10}{7}$.
The scaling limit of the Bethe roots for the ground state is described
by the ODE
\bea\label{asjh7823hgdsaaa}
 \bigg[-\partial_y^2+p^2+ \re^{(n+7)y}+E^7\ \re^{7 y}-
 c\, E\,\re^{\frac{1}{7}(n+14)\, y}
 \,\bigg]\,\psi=0\,,
\eea
where $c$, therein, is related to the value of $\mathfrak{a}_1$ as
\be\label{as8932hjuysdty}
\mathfrak{a}_1= \frac{\Gamma(\frac{1}{2}+\frac{7}{2n})}{7\,\Gamma(\frac{7}{2n})}\,
\frac{\Gamma(\frac{1}{7}-\frac{3}{2n})}{
\Gamma(\frac{9}{14}-\frac{3}{2n})}\, c\ .
\ee
The proportionality coefficient is free of singularities and zeroes as 
$n>\frac{21}{2}$. It turns out that when the position of the first pole
in the $\mathfrak{a}_s$ --- $c$ relation occurs at sufficiently large values of $n$, one can obtain
more accurate numerical data in its vicinity.
In particular, we found with confidence that
\be\label{90328721hajs}
\bigg(\frac{N}{7N_0}\bigg)^{-\frac{2n}{7(n+7)}}\,
h_1^{(N,{\rm reg})}=c\ \frac{(n+7)^{\frac{2}{n+7}-\frac{12}{7}}}{%
\Gamma^2(\frac{6}{7}-\frac{1}{n+7})}\,
f_1\big(\tfrac{{\tt k}}{2},\tfrac{1}{7}+\tfrac{1}{n+7}\big)+o(1)\qquad\qquad (n>\tfrac{21}{2})\,,
\ee
where
\be\label{kja8923hjsd}
o(1)=\begin{cases}O\big(c\,N^{\frac{6}{7}-\frac{4n}{49}}\big) & \ \ \ {\rm for} \ \ \ \frac{21}{2}<n<35 \\[0.2cm]
O\big(c\,N^{-2}\big)& \ \ \ {\rm for} \ \ \ n>35
\end{cases}\,.
\ee
It follows from the relation 
\eqref{as8932hjuysdty} that
when $n$ approaches $\frac{21}{2}$ from above with the RG invariant $\mathfrak{a}_1$ kept fixed, $c$ must tend to zero.
Then, the differential equation \eqref{asjh7823hgdsaaa} becomes the one corresponding to the ${\cal Z}_7$ invariant
case at $n=\frac{21}{2}$.  We also performed a numerical study in the domain $0<n\le\frac{21}{2}$.
It was found that if $\mathfrak{a}_1$ is still specified as in \eqref{389hsdyu3a} with $d_1=\frac{10}{7}$,
the scaling limit of the Bethe roots does not depend on the 
value of this RG invariant and is the same as for $\mathfrak{a}_1=0$. However, it impacts the rate of convergence,
in particular,
\be\label{sa982h3jhdf}
\bigg(\frac{N}{7N_0}\bigg)^{-\frac{2n}{7(n+7)}}\,
h_1^{(N,{\rm reg})}=O\big(c\,N^{-\frac{6}{7}+\frac{4n}{49}}\big)\qquad\qquad (0<n<\tfrac{21}{2})\, .
\ee
In fig.\,\ref{fig4} the numerical data  in support of  eqs.\,\eqref{kja8923hjsd} and \eqref{sa982h3jhdf} is presented.
\medskip

We explored
the  possibility  of achieving a non-trivial scaling limit by making the exponent  in \eqref{389hsdyu3a}
$n$\,-\,dependent
such that $0<d_1< \frac{10}{7}$. It was found that if   $d_1$ is replaced by $\tilde{d}_1$ where
\be
\tilde{d}_1=\frac{4}{7}+\frac{4n}{49}\qquad\qquad (0<n<\tfrac{21}{2})
\ee
then the limits
\be
h_i^{(\infty)}=\lim_{N\to\infty}\bigg(\frac{N}{7N_0}\bigg)^{-\frac{2in}{7(n+7)}}\,
h_i^{(N,{\rm reg})}\qquad\qquad\qquad (i=1,2,\ldots)
\ee
exist and are non-vanishing for generic values of the RG invariant $\mathfrak{a}_1$ in \eqref{389hsdyu3a}.
This suggests that the Bethe roots for the ground
state exhibit the scaling behaviour. 
\medskip

\subsubsection{The case $\frac{rj}{A}< \mu<\frac{r}{A+1}\,(j+\frac{1}{2})$\label{sec425}}
For  $A=1$, and $(\mu,j)=(1,0)$  it is possible to give  a uniform description for
any $r\ge 5$, which incorporates the case $r=7$ just discussed.  With regards to this,
it should be pointed out that $\mathfrak{a}_1$ has exponent  $d_1=\frac{2}{r}(r-2)$,
while formula \eqref{saaaakju32hjsdB} implies that for all the other RG invariants
$d_s>2$. As was already mentioned, their values do not affect the scaling limit
and they can be set to zero. This way,  the scaling of the Bethe roots depends only on the 
single RG invariant 
$\mathfrak{a}_1$ \eqref{389hsdyu3a}. Similar to what was discussed for $r=7$
we confirmed that, in
order to achieve a non-trivial scaling limit, the exponent $d_1$ should be taken as
\be\label{ask9812132}
d_1=\begin{cases}\frac{2}{r}\, (r-2)\ \ & {\rm for}\ \ \ n>\frac{1}{2}\,r\,(r-4)\\[0.2cm]
\frac{4}{r^2}\,  (n+r)\ \ & {\rm for}\ \ \ 0<n<\frac{1}{2}\,r\,(r-4)
\end{cases}\, .
\ee
For $n>\frac{1}{2}\,r\,(r-4)$ the scaled Bethe roots are described by the differential equation
\bea\label{3092Jjhasjqwoioiew}
 \bigg[-\partial_y^2+p^2+ \re^{(n+r)y}+E^r\ \re^{r y}-
 c\, E\,\re^{\frac{1}{r}(n+2r)\, y}
 \,\bigg]\,\psi=0\,.
\eea
We believe that  formulae
 \eqref{ask9812132} and \eqref{3092Jjhasjqwoioiew} are applicable for  $A=1$, $(\mu,j)=(1,0)$ and any $r\ge 5$,
despite that for $r=5$  they are difficult to confirm numerically in the domain $0<n<\frac{5}{2}$.
\medskip

If one accepts that all the RG invariants with   $d_s\ge 2$  may be ignored,  an interesting prediction is reached
for the case of generic values of $r$, $A$ and $j$ with $\mu$ restricted by the condition
\be
\frac{rj}{A}<\mu<
\frac{  r}{A+1}\  
\Big(j+\frac{1}{2}\Big) \, .
\ee
The latter  implies  that the only non-trivial RG invariant is $\mathfrak{a}_\mu$ with 
the corresponding exponent  $d_\mu=\frac{2}{r}\,\big((j+1)r-(A+1)\,\mu\big)$ such that
 $1<d_\mu<2$. It is expected that
a scaling limit, which is different than the one for the ${\cal Z}_r$ invariant case, can be achieved provided
\be\label{askj8921aaaaaaa}
d_\mu=\begin{cases}\frac{2M}{r}\ \ \ \ \ \ \ \ & {\rm for}\ \ \ n>n_{\rm min}\\[0.2cm]
\frac{4Ln}{r^2}+\frac{2}{r}(r-M)\ \ \ \ \ \ \ & {\rm for}\ \ \ 0<n<n_{\rm min}
\end{cases}\,  ,
\ee
where 
\be\label{ask8923jaaa21}
n_{\rm min}=\tfrac{r}{L}\,(M-\tfrac{r}{2})\qquad\qquad (L> 0)
\ee
and  we use the integers  $M$  and $L$ defined in eqs.\,\eqref{3902jds892hjs} and \eqref{jh7823iujhsdghas},
respectively.
For $n>n_{\rm min}$, the differential equation appearing in the scaling limit would be \eqref{3892jh12ssasa}.
If in the domain $0<n<n_{\rm min}$ the exponent $d_\mu$ is not assigned an $n$ dependence but kept fixed as
$d_\mu=\frac{2M}{r}$,   the scaled Bethe roots would be the same as for 
the ${\cal Z}_r$ invariant case.  
When the exponent is modified according to the second line of \eqref{askj8921aaaaaaa} 
we observed in many examples that the Bethe roots exhibit non-trivial scaling behaviour.
Describing the corresponding ODE for $0<n<n_{\rm min}$ is beyond the scope of this work.
\medskip

It should be emphasized that formula
  \eqref{ask8923jaaa21}  does not make sense for $L=0$. This  is again a signal that the case 
$A\mu=rj$
requires special attention.

\medskip
Let us give a short summary of this section.
For sufficiently large $n>n_{\rm min}$,
the scaling limit can be organized such that the
scaled Bethe roots are described by the differential equation \eqref{3892jh12ssasa},\,\eqref{askjas8712jhsa}.
The specialization of the RG invariants
is given by eqs.\,\eqref{ais78176326512assadas},\,\eqref{askj7823h21*asas} with exponents \eqref{saaaakju32hjsdB}.
The values of the non-vanishing $\mathfrak{a}_s$ are related to the parameters of the differential equation as in 
\eqref{sajk8937hj21hg}\,-\,\eqref{3902jdsjh}. The values of some of the RG invariants with $d_s>1$ turn out to have no
effect on the scaling limit and can be taken  to be zero. 
For given  $A$ and $(\mu,j)$ obeying the inequalities \eqref{askjas8712jhsa}, the
  minimal sets of non-trivial $\mathfrak{a}_s$, along
with their exponents, are listed in the tables contained in Appendix \ref{AppD2} for $3\le r\le 10$. Therein, we  also quote
 the corresponding values of the lower bound $n_{\rm min}$.

\medskip

\section{The case  $A\mu=rj$  with $A=1,\ldots,r-2$\label{sec5}}
A formal specification (ignoring the constraint \eqref{askjas8712jhsa}) of the
ODE \eqref{3892jh12ssasa} to the case at hand yields
\bea\label{askj8aaaa723hd98jh21}
 \Big[-\partial_y^2+p^2+ \re^{(n+r)y}-(-1)^A\,E^r\ \re^{ry}-
c\, E^{\mu}\ \re^{\mu y}
 \,\Big]\,\psi=0\ .
\eea
In view of the condition $A\mu=rj$, 
$\mu$  is not an arbitrary  integer from the segment $[1,r-1]$. Instead, if $H$
is the greatest common divisor of $r$ and $A$,
\be
H={\rm gcd}(r,A)\,,
\ee
then 
\bea\label{asj2893jhds}
\mu=\frac{r J}{H}\ \ \ \ \  \ \ (J=1,\ldots, H-1)\ .
\eea
It turns out that the case $A\mu=rj$ with $A=1,\ldots,r-2$ 
falls outside of the scope of the  discussion in the previous section and the ODE \eqref{askj8aaaa723hd98jh21}
does not actually appear in the scaling limit of the spin chain (except  when both $r,A$  are even and $\mu=\frac{r}{2}$).

\medskip
\subsection{Special features of the scaling limit\label{sec51}}
Let's first consider the case when the positive integer $\mu<\frac{r}{2}$ or, equivalently, $J<\frac{H}{2}$.  Similarly to what was discussed in sec.\,\ref{sec425} one may expect that all of the RG invariants have no effect on the scaling limit apart from
$\mathfrak{a}_\mu$,
whose  corresponding exponent $d_\mu=\frac{2}{H}\,(H-J)$ 
is such that $1<d_\mu <2$. In fact, a numerical study shows 
that the scaling behaviour does not depend on the value of 
$\mathfrak{a}_\mu$  as well, and the scaled Bethe roots coincide with those
for the ${\cal Z}_r$ invariant case.\footnote{%
A non-trivial scaling behaviour for the Bethe roots for the ground state can be achieved by swapping 
$d_\mu= \frac{2}{H}\,(H-J)$ for $\tilde{d}_\mu=\frac{2 J}{H}<1$ in the definition of the RG invariant $\mathfrak{a}_\mu$ 
\eqref{ais78176326512assadas}. This follows from the relation between the case $A\mu=r j>0$ and $A=0$ described in
sec.\,\ref{askj9812a9a8sjh122}
together with the results of sec.\,\ref{sasau}.}
\medskip

For $r$  even,  $\mu=\frac{r}{2}$ is among the admissible values \eqref{asj2893jhds}.  Then
it is sufficient to keep only one RG invariant $\mathfrak{a}_{\frac{r}{2}}$ to be non-vanishing with 
exponent $d_{\frac{r}{2}}=1$. However 
formulae \eqref{saj32jhghsaaaaa}-\eqref{3902jdsjh} for
the coefficient $Z_{\frac{r}{2}}$, which relates the value of
$\mathfrak{a}_{\frac{r}{2}}$ with the parameter $c$ in the differential equation,
give infinity.
This  suggests to modify the definition of the RG invariant  as
\bea\label{ask8923jjh21hga}
{\mathfrak a}_{\frac{r}{2}}\equiv  \frac{2}{r^2}\ \frac{1}{\log\big(\frac{N}{rN_0}\big)}\ \bigg(\frac{N}{rN_0}\bigg)\ \sum_{\ell=1}^r(\eta_\ell)^{-\frac{r}{2}}\ .
\eea
With such a setup, the investigation of the scaling behaviour becomes difficult to carry out numerically.
Even  $N\sim 5000$ turns out to be  insufficient to   reliably observe 
the existence of a scaling limit. Nevertheless, in Appendix \ref{AppE} we present a numerical scheme,
which dramatically improves convergence and allows one to confirm that the differential equation occurring in the scaling limit
is  given by \eqref{askj8aaaa723hd98jh21} with $\mu=\frac{r}{2}$. The parameter $c$ is related to the value of the
RG invariant as
\bea\label{kjas893aaa2187213}
{\mathfrak a}_{\frac{r}{2}}=(-1)^{\frac{1}{2}(r-A+2)}\ 
\frac{2}{r\sqrt{\pi}}\ \frac{\Gamma(\frac{1}{2}+\frac{r}{2n})}{\Gamma(1+\frac{r}{2n})}
\ c\ .
\eea
 Note that the argument of the logarithm in  \eqref{ask8923jjh21hga} involves the 
constant $rN_0$. It can be replaced by any positive number without affecting the scaling limit.
\medskip

In the case $\mu>\frac{r}{2}$  the integer $J$ in formula \eqref{asj2893jhds}
is restricted as $\frac{H}{2}< J\le H-1$. It turns out that the minimal set of non-trivial RG invariants
consists of
\bea\label{askaaaj8923jd}
\mathfrak{a}_s :\ \ s=\tfrac{rJ }{H}\,i_s\ \ \ \ ({\rm mod}\ r)\,,\qquad  \qquad i_s=1,2,\ldots,\big[\tfrac{H}{2(H-J)}\big]\, .
\eea
The corresponding exponents are given by
\bea\label{skjaaaa8932j12hasl}
d_{s}= \frac{2}{H}\,(H-J)\, i_s
\eea
and are all less than or equal to one.
A numerical study shows that the Bethe roots for the ground state develop a scaling behaviour, which differs from the
one that occurs in the ${\cal Z}_r$ invariant case. 
However, the scaling limit  is not described by the ODE \eqref{askj8aaaa723hd98jh21}. For instance,
when $r=9$, $A=3$ ($H=3$) and $\mu=6$ 
 there is only one non-trivial 
RG invariant $\mathfrak{a}_6$ with exponent $d_6=\frac{2}{3}$.
The differential equation yields the prediction that  the limits \eqref{akjs871323221hg} with $s=1,2,\ldots,8$ vanish except for
\be\label{asjk98321h12jjh6A}
h_6^{(\infty)}=\lim_{N\to\infty}\bigg(\frac{9N_0}{N}\bigg)^{\frac{4n}{3(n+9)}}\,h_6^{(N,{\rm reg})}\ .
\ee
However, we found that together with $h_6^{(\infty)}$, the limit
\be\label{asjk98321h12jjh6B}
h_3^{(\infty)}=\lim_{N\to\infty}\bigg(\frac{9N_0}{N}\bigg)^{\frac{2n}{3(n+9)}}\,h_3^{(N,{\rm reg})}
\ee
exists and is  not zero. 
\medskip

The first idea that may come to mind for describing the scaling limit for $\mu=\frac{r J}{H}>\frac{r}{2}$
 is to replace   \eqref{askj8aaaa723hd98jh21} by the more general differential equation of the form 
\be\label{aksju1873762121aaaa}
 \Big[-\partial_y^2+p^2+ \re^{(n+r)y}-(-1)^A\,E^r\ \re^{ry}-c\, E^{\mu}\ \re^{\mu y}-
\sum_{i}{b}_{i}\, E^{\mu_i}\ \re^{\mu_i y}
 \,\Big]\,\psi=0\,,
\ee
where the summation  runs over the  positive  integer $i=\big[\tfrac{H}{2(H-J)}\big]+1,
\,\ldots,\,\big[\tfrac{H-1}{H-J}\big]$ and
\bea\label{sakuyaaaa12hs78ashg}
\mu_i=i\, \mu - (i-1)\, r \ \ \ \ \ \ \ \ \ \ \ \  \ \ \ \ \ \ \ \ \ \big(\mu=\tfrac{r J}{H}\big)\ .
\eea
Numerical data supports this conjecture, where the values of the RG invariants
$\mathfrak{a}_s$   \eqref{askaaaj8923jd}
are related to $c$ as in eqs.\,\eqref{sajk8937hj21hg}-\eqref{3902jdsjh}. However, while the coefficients $b_i$
depend simply on $c$, they
turn out to be complicated functions of  $p=\frac{1}{2}(n+r)\,{\tt k}$ and $n$:
\be
b_i= c^i\  {\hat b}_i(p,n)\, .
\ee
For instance,  the differential equation  \eqref{aksju1873762121aaaa}, specialized to the case $r=9$, $A=3$ ($H=3$) and $\mu=6$,
leads to the prediction
\bea\label{askj8923jdsf}
h^{(\infty)}_{3}&=& b_1\, \frac{(n+9)^{-\frac{2 (n+6)}{n+9}}}{\Gamma^2\big(\frac{n+6}{n+9}\big)}\, f_1\big(\tfrac{{\tt k}}{2},\tfrac{3}{n+9}\big)\\
h^{(\infty)}_{6}&=& c\, \frac{(n+9)^{-\frac{2 (n+3)}{n+9}}}{\Gamma^2\big(\frac{n+3}{n+9}\big)}\, f_1\big(\tfrac{{\tt k}}{2},\tfrac{6}{n+9}\big)+b^2_1\  \frac{(n+9)^{-\frac{4 (n+6)}{n+9}}}{\Gamma^4\big(\frac{n+6}{n+9}\big)}\, f_2\big(\tfrac{{\tt k}}{2},\tfrac{3}{n+9}\big)\,.\nonumber
\eea
Eliminating $b_1$,  one obtains
\bea\label{sai9823h21090921}
h^{(\infty)}_{6}&=& c\ \frac{(n+9)^{-\frac{2 (n+3)}{n+9}}}{\Gamma^2\big(\frac{n+3}{n+9}\big)}\, f_1\big(\tfrac{{\tt k}}{2},\tfrac{6}{n+9}\big)+\Bigg(\frac{h^{(\infty)}_{3}}{f_1\big(\tfrac{{\tt k}}{2},\tfrac{3}{n+9}\big)} \Bigg)^2\, f_2\big(\tfrac{{\tt k}}{2},\tfrac{3}{n+9}\big)\label{ghtn6}\ .
\eea
Here the functions $f_1$ and $f_2$ are given by eqs.\,\eqref{sa8932jhdssaaaa}-\eqref{jassusau}
 in the Appendix \ref{AppA}, while the $\mathfrak{a}_s$ --- $c$  relation
reads explicitly as  
\bea\label{as8932hjuysdtyy}
\mathfrak{a}_6=-\frac{\Gamma \big(\frac{3}{2 n}\big) \Gamma \big(\frac{3 }{2 }+\frac{9}{2 n}\big)}{(n+9) \Gamma \big(1+\frac{9}{2 n}\big) \Gamma \big(\frac{1}{2}+\frac{3}{2 n}\big)}\ c\ .
\eea
Numerical data in support of \eqref{sai9823h21090921}  is presented in  tab.\,\ref{tab0}.
\medskip

In tab.\,\ref{tab2a} we list the cases where $A\mu=rj$ and $\mu>\frac{r}{2}$ for all possible values of $r\le 10$. 
Together with the integers $s$ labeling the non-trivial RG invariants and the corresponding exponents $d_s$, presented are
the values
of the integers $\mu_i$ entering into the differential equation \eqref{aksju1873762121aaaa}. Note that the two cases $A=3$ and $A=6$
with $r=9$ lead to the same ODE. In fact, it is easy to check that the two sets of Bethe
roots for the ground state are related as: $\{\zeta_j\}|_{A=3}=\{\re^{\frac{\ri\pi}{3}}\,\zeta_j\}|_{A=6}$.

\begin{table}
	\begin{center}
		\begin{tabular}{|l|c|c||c|c|}
\cline{2-5}
\multicolumn{1}{c|}{} &  \multicolumn{2}{c||}{} & \multicolumn{2}{c|}{} \\[-0.4cm]
\multicolumn{1}{c|}{} & \multicolumn{2}{c||}{${\tt k}=0.02$} & \multicolumn{2}{c|}{${\tt k}=0.04$} \\[0.05cm]
\cline{2-5}
\hline
 &  & & &\\[-0.4cm]
$n$ & $h_6^{(\infty)}$ from \eqref{asjk98321h12jjh6A} & $h_6^{(\infty)}$ from \eqref{sai9823h21090921}  &$h_6^{(\infty)}$ from \eqref{asjk98321h12jjh6A}   &
$h_6^{(\infty)}$ from \eqref{sai9823h21090921} \\[0.05cm]
\hline
 &  & & &\\[-0.4cm]
10 &  0.189683 & 0.189499  & 0.182812   &  0.182641 \\[0.05cm]
\hline
 &  & & &\\[-0.4cm]
20 & 0.096347  &  0.096342  & 0.089699  &  0.089695  \\[0.05cm]
\hline
 &  & & &\\[-0.4cm]
30 & 0.070086   &  0.070084  & 0.063443 &  0.063442 \\[0.05cm]
\hline
 &  & & &\\[-0.4cm]
40  &  0.057235  & 0.057232  &  0.050550  & 0.050548  \\[0.05cm]
\hline
		\end{tabular}
	\end{center}
	\caption{\label{tab0}\small%
Compared is numerical data obtained for $h_6^{(\infty)}$ in two different ways
for $r=9$, $A=3$ and various values of ${\tt k}$ and $n$.
In the first case, $h_6^{(\infty)}$ was calculated from the solution of the Bethe Ansatz equations via formulae \eqref{asjk98321h12jjh6A},\,\eqref{asosaiisa}
and  \eqref{sa982321uuy}.  In the second, it was found from the r.h.s. of \eqref{sai9823h21090921}, where  $h_3^{(\infty)}$ was calculated numerically using eq.\,\eqref{asjk98321h12jjh6B}.
The value of the single non-trivial  RG invariant $\mathfrak{a}_{6}$, with exponent $d_6=\frac{2}{3}$, 
was fixed according to \eqref{as8932hjuysdtyy}, where $c=1$.
Computations were performed for spin chains with $N\lesssim 5000$. }
\end{table}
\begin{table}
\begin{center}
\begin{tabular}{|c|c|l|l|l|l|} 
\cline{3-6}
\multicolumn{1}{c}{} & &  & &  &  \\[-0.37cm]
\multicolumn{1}{c}{} & & $\mu$ & $s$ & $d_s$ & $\mu_i$\\[0.03cm]
\hline
& & & &  &  \\[-0.37cm]
$\ r=6$ & $\ A=3\ $ & 4 & $\{4\}$ & $d_4=\frac{2}{3}$  & $\mu_2=2$\\[0.03cm]
\hline
\hline
& & & &  &  \\[-0.37cm]
$\ r=8$ & $\ A=4\ $ & 6 & $\{6,4\}$ & $d_6=\frac{1}{2},\, d_4=1$  & $\mu_3=2$\\[0.03cm]
\hline
\hline
& & & &  &  \\[-0.37cm]
\multirow{3}{*}{\vspace{0.35cm} $r=9$} & $\ A=3 \ $ & 6 & $\{6\}$ & $d_6=\frac{2}{3}$  & $\mu_2=3$\\[0.03cm]
\cline{2-6}
& & & &  &  \\[-0.37cm]
&  $\ A=6\ $ & 6 & $\{6\}$ & $d_6=\frac{2}{3}$  & $\mu_2=3$\\[0.03cm]
\hline
\hline
& & & &  &  \\[-0.37cm]
\multirow{3}{*}{\vspace{0.35cm} $r=10$} &  \multirow{3}{*}{\vspace{0.35cm}  $\!\! A=5$} & 6 & $\{6\}$ & $d_6=\frac{4}{5}$  & $\mu_2=2$\\[0.03cm]
\cline{3-6}
& & & &  &  \\[-0.37cm]
 &  & 8 & $\{8,6\}$ & $d_8=\frac{2}{5},\,d_6=\frac{4}{5}$  & $\mu_3=4,\,\mu_4=2$\\[0.03cm]
\hline
\end{tabular}
\caption{\label{tab2a}%
\small
Listed are all possible cases where $A\mu=rj$ with $\mu>\frac{r}{2}$ up to $r=10$. Apart from
the integers $s$ and exponents $d_s$, 
which specify the RG invariants $\mathfrak{a}_s$ belonging to the set \eqref{askaaaj8923jd}, 
the values of the integers $\mu_i$ \eqref{sakuyaaaa12hs78ashg} entering into the 
differential equation \eqref{aksju1873762121aaaa}  are given in the last column.}
\end{center}
\end{table}

\medskip
In sec.\,\ref{sec41}  a  subtlety was mentioned that occurs when both $r$, $A$ are even  with  
$\mu$ and $j$  given by \eqref{nash1278as}. Setting $K=1$ into the formula $\mu=\frac{r}{2K}\, (2K-1)$
one gets back the case $\mu=\frac{r}{2}$ discussed above. If $K>1$ 
then $\mu>\frac{r}{2}$. For $r$ even we expect that the RG invariant
$\mathfrak{a}_{\frac{r}{2}}$ should be specified as in \eqref{ask8923jjh21hga}. Moreover, it seems likely
that \eqref{kjas893aaa2187213} is generalized as
\bea
{\mathfrak a}_{\frac{r}{2}}=
(-1)^{\frac{1}{2}(r-A+2)}\ \
\frac{2\Gamma(K-\frac{1}{2})}{r\pi K!}\ \frac{\Gamma(\frac{1}{2}+\frac{r}{2n})}{\Gamma(1+\frac{r}{2n})}\ 
\ c^K\ .
\eea
Unfortunately,  we were not able to numerically verify this relation for $K>1$.
Notice that the first such case corresponds to $r=8$, $A=4$ and $\mu=6$, see  tab.\,\ref{tab2a}.
\medskip

\subsection{Relation to $A=0$\label{askj9812a9a8sjh122}}
It turns out that the cases $A\mu=rj$ with $A=1,2,\ldots,r-2$  and $A=0$
are related.  The link already occurs for the Bethe roots corresponding to the
ground state of the spin chain
at finite lattice size $N$.  
It allows one
to carry over the discussion of the case $A=0$ contained in  sec.\,\ref{sasau} which,
among other things, gives a relation between the coefficients $b_i$ in the ODE 
\eqref{aksju1873762121aaaa} with the eigenvalues of the so-called quasi-shift operators.
\medskip

Recall that the integer $\mu$ is used to specify the minimal set of non-trivial RG invariants
$\mathfrak{a}_s$. In view of  \eqref{askj7823h21*asas}, when
performing the scaling limit, one may take
\be\label{askj893h21998aaaaaksjd}
\sum_{\ell=1}^r(\eta_\ell)^{-s}=0\qquad\qquad {\rm for}\qquad\qquad s\ne \sigma,2\sigma,\ldots,r\ ,
\ee
where  $\sigma>1$ stands for the greatest common divisor of
$\mu$ and $r$:
\be
\sigma={\rm gcd}(r,\mu)\,.
\ee
The condition \eqref{askj893h21998aaaaaksjd} may be fulfilled if one sets
\be\label{as989asdasdasdasd831}
\eta_{\ell+\frac{r}{\sigma}}=\re^{\frac{2\pi\ri}{\sigma}}\,\eta_\ell \ .
\ee
When $s$ is  a multiple of $\sigma$, i.e., $s=\sigma m$ with  $m=1,2,\ldots, r/\sigma-1$, then
instead of \eqref{askj893h21998aaaaaksjd} one has
\be\label{ksajs873h21hgasa}
\frac{1}{rm}\sum_{\ell=1}^{r/\sigma}(\eta_\ell)^{-\sigma m}=
\begin{cases} \big(\frac{r N_0}{N}\big)^{d_{\sigma m}}\,\mathfrak{a}_{\sigma m} \qquad\qquad & 
 {\rm if}\qquad \sigma m\ne\frac{r}{2}
\\[0.2cm]
 \frac{1}{\log(\frac{N}{rN_0})}\,
\big(\frac{r N_0}{N}\big)\,\mathfrak{a}_{\frac{r}{2}}
\qquad\qquad & {\rm if}\qquad  \sigma m=\frac{r}{2}
\end{cases}\,,
\ee
while the normalization condition \eqref{iaassauasas} becomes
\be
\frac{\sigma}{r}\ \sum_{\ell=1}^{r/\sigma}\,(\eta_\ell)^{-r}=(-1)^{r-1}\,.
\ee
For given values of $\mathfrak{a}_{\sigma m}$ and $N$, this may be treated    as a system
of $r/\sigma$ equations for determining the set
 $\{\eta_1,\ldots,\eta_{r/\sigma}\}$ and hence, in view of \eqref{as989asdasdasdasd831}, 
all the inhomogeneities.
\medskip

The  $N/2$ Bethe roots for the ground state ($S^z=0$)
 are split into groups with roughly equal phases, which can be labelled by the integer
$a=1,2,\ldots,r$ such that 
 $\arg(\zeta_m^{(a)})\approx \frac{\pi}{r}\,(2a-2-A) \mod 2\pi$.
If \eqref{as989asdasdasdasd831}  is obeyed and
\be
\zeta_m^{(a+\frac{r}{\sigma})}=\re^{\frac{2\pi\ri}{\sigma}}\,\zeta_m^{(a)}\,,
\ee
then the algebraic system of $N/2$ Bethe Ansatz equations  \eqref{baekasdba}    reduces to a system of
$\tilde{N}/2$ equations with
\be\label{sa9832jbsdb12}
\tilde{N}=N/\sigma\ .
\ee
Namely,
\be\label{ask8932jh21d}
\Bigg(\prod_{\ell=1}^{\tilde{r}}
\frac{\tilde{\eta}_\ell+\tilde{q}^{+1} \,\,\tilde{\zeta}_j}
{\tilde{\eta}_\ell+\tilde{q}^{-1}\,\tilde{\zeta}_j }\Bigg)^{\frac{\tilde{N}}{\tilde{r}}}
=-\omega^2\, 
\prod_{i=1}^{\tilde{N}/2}\,
\frac{\tilde{\zeta}_i-\tilde{q}^{+2}\,\tilde{\zeta}_j }
{\tilde{\zeta}_i-\tilde{q}^{-2}\,\tilde{\zeta}_j }
\,\qquad\qquad \big(j=1,2,\ldots,\tfrac{\tilde{N}}{2}\big)\,,
\ee
where 
\be\label{askj3h2178has}
\{\tilde{\zeta}_j\}_{j=1}^{\tilde{N}/2}=
\bigcup_{a=1}^{\tilde{r}}\,\Big\{(-1)^{I}\,\big( \zeta_j^{(a)}\big)^\sigma\Big\}_{j=1}^{N/(2r)}
\ee
and we also use 
\be\label{jas8913hhgez}
\tilde{r}=r/\sigma\,,\qquad\qquad\quad
\tilde{q}= (-1)^{I}\,q^\sigma\,,\qquad\qquad\quad
\tilde{\eta}_\ell=(-1)^{\sigma-1}\,(\eta_\ell)^\sigma\qquad \quad(\ell=1,2,\ldots,\tilde{r})\,. 
\ee
The above formulae involve the sign factor  $(-1)^I$ with
$
I=\frac{A\sigma}{r}
$, which is an integer due to the condition $A\mu=rj$.
\medskip

From the definition \eqref{jas8913hhgez}, it follows that
\be
\tilde{q}=\exp\Big({\frac{\ri\pi}{\tilde{n}+\tilde{r}}}\Big)\qquad{\rm with}\qquad \tilde{n}=n/\sigma\,,
\ee
i.e., the argument of $\tilde{q}$ lies in the interval $(0,\frac{\pi}{\tilde{r}})$. 
Thus the algebraic system \eqref{ask8932jh21d}
coincides with the Bethe Ansatz equations for the $\tilde{r}$\,-\,site periodic spin chain with the set of
inhomogeneities $\{\tilde{\eta}_\ell\}_{\ell=1}^{\tilde{r}}$
in the regime $\tilde{A}=0$.  In view of \eqref{ksajs873h21hgasa}, the RG invariants $\tilde{\mathfrak{a}}_s$ are expressed 
in terms of the original ones as
\be
\tilde{\mathfrak{a}}_s=(-1)^{(\sigma-1)s}\,
\sigma^{1-d_{\sigma s}}\,\mathfrak{a}_{\sigma s}\qquad\qquad {\rm and}\qquad\qquad \tilde{d}_s=d_{\sigma s}\ .
\ee
It turns out that if $\{\zeta_j\}$  is the solution to the Bethe Ansatz equations 
\eqref{baekasdba} corresponding to the ground state, then $\tilde{\zeta}_j$ \eqref{askj3h2178has} are the Bethe roots 
for the ground
state of the spin chain with  $(N,r,q,\eta_\ell)$ swapped for $(\tilde{N},\tilde{r},\tilde{q},\tilde{\eta}_\ell)$
defined in eqs.\,\eqref{sa9832jbsdb12} and \eqref{jas8913hhgez}. 

\medskip

\section{General differential equation for $A=1,2,\ldots,r-2$}

Consider  the equation
\bea\label{sa9823uy721}
 \bigg[-\partial_y^2+p^2+ \re^{(n+r)y}-(-1)^A\,E^r\ \re^{ry}-\sum_{(\mu,j)\in\,\bm{\Xi}_{r,A}}
 c_{\mu,j}\, E^{\mu}\ \re^{\big(  (A\mu-rj)\,\frac{n+r}{r}+\mu\big) y}\,\bigg]\,\psi=0\ ,
\eea
where  the sum goes over pairs of non-negative integers  belonging to the set
 \bea\label{skaj8923h1287}
\bm{\Xi}_{r,A}&=&\big\{(\mu,j)\, :\ \  
\tfrac{rj}{A}<\mu<
\tfrac{  r}{A+1}\  
(j+1)  \ \ \& \ \    j\ge 0\big\}\ .
 \eea
We have excluded $(\mu,j)$  for which the 
condition
$A\mu=rj$ is satisfied for the reasons discussed in the previous section.
It turns out that for any $r\ge 3$ and
$A=1,2,\ldots,r-2$
the number of admissible pairs obeys the condition
\be
|\bm{\Xi}_{r,A}|=|\bm{\Xi}_{r,r-1-A}|
\ee
and
is bounded as\footnote{%
We checked  \eqref{sa8923hjsassaa} up to $r=200$. 
If $r$ is a prime number then $|\bm{\Xi}_{r,A}|=\frac{r-1}{2}$ for any $A=1,2,\ldots,r-2$,
while as $A=\frac{r}{2}-1$ and $A=\frac{r}{2}$ ($r$ even)  one has $|\bm{\Xi}_{r,A}|=[\tfrac{r}{4}]$.}
\be\label{sa8923hjsassaa}
[\tfrac{r+1}{4}]\le |\bm{\Xi}_{r,A}|\le [\tfrac{r-1}{2}]\ .
\ee
This way, the amount of terms in the sum in \eqref{sa9823uy721} is less than the number of functionally 
independent RG invariants. Thus one arrives at the hypothesis that for sufficiently large $n$ and
given  $c_{\mu,j}$,  it is always possible to organize the scaling limit
such that the scaled Bethe roots are described by the ODE \eqref{sa9823uy721}. 
\medskip

The specification of the RG invariants $\mathfrak{a}_s$
and their relation to the coefficients  $c_{\mu,j}$ 
  can be deduced, in principle, 
for any given value of $r$ and $A=1,2,\ldots,r-2$ similar to how it was done previously.  
Unfortunately, we did not find a convenient  form for these relations that works in general.
For odd  $r$, $A=\frac{r-1}{2}$  and $n>0$ they are given by eqs.\,\eqref{hassy},\eqref{sakj87321} and \eqref{sakj87321AAA}.
Below we present the explicit formulae for even $r$ and $A=\frac{r}{2}-1,\,\frac{r}{2}$ as well as
arbitrary $r$ with $A=1,r-2$.
\medskip

\subsection{The case $A=1$}

The definition \eqref{skaj8923h1287} for the set of admissible values of $(\mu,j)$, specialized to $A=1$,  yields
\be
\bm{\Xi}_{r,1}=\big\{(\mu,j)\, :\ \ \mu=1,\ldots \,, \big[\tfrac{r-1}{2}\big]\ \ \&\ \ j=0\, \big\}\ ,
\ee
so that the ODE  \eqref{sa9823uy721} takes the form
\bea\label{hasysd}
 \Bigg[-\partial_y^2+p^2+ \re^{(n+r)y}+E^r\ \re^{r y}-\sum_{\mu=1}^{[\frac{r-1}{2}]}
 c_{\mu}\, E^{\mu}\ \re^{\frac{n+2r}{r}\, \mu y}\,\Bigg]\,\psi=0\ .
\eea
One should distinguish between the case when $r$ is an odd or even integer. 
For the former,  a computation  shows that the
exponents for the RG invariants $\mathfrak{a}_s$   are given by
\bea
 d_{s}=\frac{2}{r}\, o_s\ \ \ \ \ \ \ \ \ \ \ (r-{\rm odd})\ ,
\eea
where
\bea
o_s\equiv \begin{cases}
r-2s\ \ \ \ \ & {\rm for}\ \ \ s=1,2,\ldots, \frac{r-1}{2}\\[0.1cm]
2r-2s\ \ \ \ \  & {\rm for}\ \ \ s= \frac{r+1}{2},\,\frac{r+3}{2},\,\ldots, r-1
\end{cases}\ .
\eea
The values of ${\mathfrak a}_{s}$ are related to the coefficients  of  the ODE \eqref{hasysd} as
\bea
{\mathfrak a}_{s}=  (-1)^{\frac{(r+1)}{2} o_s}\ \ \frac{1}{s}\ \sum_{k=1}^{ o_s}(-1)^k\ \frac{\Gamma(s\nu+k)}{ k!\Gamma(s\nu) }\ \sum_{j_1+j_2\ldots +j_k=o_s-k\atop
 j_1,\ldots, j_k\geq 0 }R_{2j_1+1}\ldots R_{2j_k+1}
\eea
with $\nu=\frac{2n}{n+r}$ and we use the same notation as in \eqref{sakj87321}. Namely
\be\label{jasususa}
R_{2j+1}=\frac{1}{r}\sum_{
\alpha_1+2\alpha_2+\ldots+(j+1)\alpha_{j+1}=j+1\atop
\alpha_1,\ldots,\,\alpha_{j+1}\geq 0
}
\frac{(-1)^{\alpha_1+\ldots+\alpha_{j+1}}}{\alpha_1! \alpha_2!\ldots \alpha_{j+1}!}\ 
\frac{\Gamma(\alpha_1+\ldots+\alpha_{j+1}-\frac{2j+1}{r})}{\Gamma(1-\frac{2j+1}{r})}\ 
G_2^{\alpha_1}G_4^{\alpha_2}\dots G_{2j+2}^{\alpha_{j+1}}
\ee
where now
\bea
G_{2m}=\frac{\Gamma(\frac{3}{2}+\frac{r}{2n}) }{\Gamma(\frac{r}{2n})}\ \sum_{k=1}^m \re^{\frac{\ri\pi}{2}(m-k)}\ \frac{
\Gamma\big(\frac{k}{2}-
(\frac{1}{2r}+\frac{1}{n})\, m+\frac{r}{2n}\big )
}{k!\Gamma\big(\frac{3}{2}-\frac{k}{2}
- (\frac{1}{2r}+\frac{1}{n})\, m+\frac{r}{2n}\big )}
 \sum_{
\mu_{1}+\ldots +\mu_{k}=\frac{1}{2}( kr-m)\atop
1\leq \mu_1,\ldots,\mu_k\leq \frac{r-1}{2}}
c_{\mu_{1}}\cdots c_{\mu_{k}}\, .
\eea
Our numerical work shows that the value of the
 RG invariant $\mathfrak{a}_{s}$ with $s=\frac{r+1}{2}$, whose exponent is
$d_{s}=2-\frac{2}{r}$, has no effect on the scaling limit. Thus it can be set to  zero
and the  number of non-trivial RG invariants is equal to $r-2$.
\medskip

In the case of even $r$ the analogous relations read as follows.  The exponents corresponding to   ${\mathfrak a}_{s}$ are given by
\bea
d_s=\frac{4}{r}\, e_s \ \ \ \ \ \ \ \ \ \ \ (r-{\rm even})\ ,
\eea
where we use
\bea
e_s\equiv \begin{cases}
\frac{r}{2}-s\ \ \ \ \ &{\rm for}\ \ \ s=1,\ldots\frac{r}{2}-1\\[0.1cm]
r-s\ \ \ \ \ &{\rm for}\ \ \ s=\frac{r}{2},\ldots,r-1
\end{cases} \ .
\eea
As for the $\mathfrak{a}_s$ --- $c_\mu$ relation, one has
\bea\label{askj893aaa2jh21}
{\mathfrak a}_{s}=(-1)^{e_s-1}\ \ \frac{1}{s}\times \begin{cases}
 \frac{1}{2\ri }\, \  
 \big( F^{(+)}_{s}-F^{(-)}_{s}\big)\ 
 \ \ \ & {\rm for}\ \ \ \ \ \  \  s=1,\ldots,\tfrac{r}{2}-1\\[0.1cm]
\frac{1}{2}\, \big( F^{(+)}_{s}+F^{(-)}_{s}\big)\ \ \ \ \ &  {\rm for}\ \ \ \ \ \ \  s=\frac{r}{2},\ldots,r-1
 \end{cases}\ .
\eea
Here
\bea\label{aiiasaasu}
F^{(\pm )}_{s}=
\sum_{k=1}^{e_s}(-1)^k\ \frac{\Gamma(\frac{s\nu}{2}+k) }{ k!\Gamma(\frac{s\nu}{2})  }\ \sum_{j_1+j_2\ldots +j_k=e_s-k\atop
j_1,\ldots, j_k\geq 0 }R^{(\pm)}_{2j_1+1}\ldots R^{(\pm)}_{2j_k+1}
\eea
with
\bea\label{iasisausau}
{ R}^{(\pm)}_{2j+1}&=&\frac{2}{r}\sum_{
\alpha_1+2\alpha_2+\ldots+(j+1)\alpha_{j+1}=j+1
\atop
\alpha_1,\ldots,\,\alpha_{ j+1}\geq 0
}
\frac{(-1)^{\alpha_1+\ldots+\alpha_{j+1}}}{\alpha_1! \alpha_2!\ldots \alpha_{j+1}!}\ 
\frac{\Gamma(\alpha_1+\ldots+\alpha_{j+1}-\frac{2}{r}\, (2j+1))}{\Gamma(1-\frac{2}{r}\, (2j+1))}\\[0.1in]
&\times&
\big(G^{(\pm)}_2\big)^{\alpha_1}
\big(G^{(\pm)}_4\big)^{\alpha_2}
\dots \big(G^{(\pm)}_{2j+2}\big)^{\alpha_{j+1}}
\eea
and
\bea
G^{(\pm)}_{2m}=\frac{\Gamma(\frac{3}{2}+\frac{r}{2n})}{\Gamma(\frac{r}{2n})}\ \sum_{k=1}^m\re^{\frac{\ri\pi }{2 }(r\pm 1) k} \ \frac{
\Gamma\big(\frac{k}{2}-(\frac{1}{r}+\frac{2}{n})\, m+\frac{r }{2n}\big)}{k!\Gamma\big(\frac{3}{2}-\frac{k}{2}-(\frac{1}{r}+\frac{2}{n})\, m+\frac{r}{2n}\big )}
 \sum_{
\mu_{1}+\ldots +\mu_{k}=\frac{r}{2} k- m\atop
1\leq \mu_1,\ldots,\mu_k\leq\frac{r}{2}-1}
c_{\mu_{1}}\cdots c_{\mu_{k}}\ .
\eea
\medskip

The following comment is in order. Formula \eqref{askj893aaa2jh21}
yields that $\mathfrak{a}_{r-1}\equiv 0$. Moreover, the RG invariant $\mathfrak{a}_{\frac{r}{2}}$
has exponent   $d_{\frac{r}{2}}=2$ and its value does not affect the scaling limit.
In fact, for even $r$ the set of $r-3$ RG invariants turns out to not be minimal in general. For example,
we found that in the cases with $r\le 10$ the scaled Bethe roots are also insensitive to the
value of $\mathfrak{a}_{\frac{r}{2}+1}$, in spite that the corresponding exponent $d_{\frac{r}{2}+1}<2$.
This is similar to what happens for the RG invariant $\mathfrak{a}_{\frac{r-1}{2}}$ in the case of 
 odd $r$. 
\medskip

Finally, 
  the above formulae for both  odd and even $r$  are literally applicable  provided
\be
n>\frac{1}{2}\,r\,(r-4)\ .
\ee

\medskip
\subsection{The case even $r$ and $A=\frac{r}{2}-1$}
Since
\be\label{kasu}
\bm{\Xi}_{r,\frac{r}{2}-1}=\big\{(\mu,j)\, :\ \ \mu=2j+1\ \ \&\ \ j=0,1,\ldots,\big[\tfrac{r}{4}\big]-1\,\big\}\ ,
\ee
the corresponding 
ODE reads explicitly as
\bea\label{caaa}
 \Bigg[-\partial_y^2+p^2+ \re^{(n+r)y}+(-1)^\frac{r}{2}\,E^r\ \re^{ry}-\sum_{j=0}^{ [\frac{r}{4}]-1}
 c_{2j+1}\, E^{2j+1}\ \re^{\big(  \frac{n+r}{2}-(2j+1)\,\frac{n}{r}\big) y}\,\Bigg]\,\psi=0\ .
\eea
Based on an analysis of the differential equation and  a numerical study of the scaling limit,
we expect that there are only the  $[\frac{r}{4}]$ non-trivial RG invariants:
\bea\label{aiiswwsuusa}
{\mathfrak a}_{2j+1}=\frac{1}{2j+1}\ \bigg(\frac{N}{r N_0}\bigg)\  \frac{1}{r}\ \sum_{\ell=1}^r(\eta_\ell)^{-2j-1}\ \ \ \ \ \ \ \ \ \ \ \ \ \ \ \ \ 
\big( 0\leq j\leq  \big[\tfrac{r}{4}\big]-1\big)\ .
\eea
The relation between their values and the coefficients $c_{2j+1}$ is especially simple
\bea
{\mathfrak a}_{2j+1}
=(-1)^j \ \frac{\Gamma(\frac{1}{2}+\frac{r}{2n})}{r\Gamma(\frac{r}{2n})}\ 
\frac{\Gamma\big(\frac{1}{2}-\frac{2j+1}{r}\big)}{ \Gamma\big(1-\frac{2j+1}{r}\big )}\ c_{2j+1}\ .
\eea
The differential equation \eqref{caaa} describes the scaling limit of the ground state Bethe roots
for any $n>0$.

\medskip
\subsection{The case even $r$ and $A=\frac{r}{2}$}

The set $\bm{\Xi}_{r,\frac{r}{2}}$ coincides with $\bm{\Xi}_{r,\frac{r}{2}-1}$ \eqref{kasu}:
\be
\bm{\Xi}_{r,\frac{r}{2}}=\big\{(\mu,j)\, :\ \ \mu=2j+1\ \ \&\ \ j=0,1,\ldots,\big[\tfrac{r}{4}\big]-1\,\big\}
\ee
and the corresponding 
ODE takes the form
\bea\label{caaadd}
 \Bigg[-\partial_y^2+p^2+ \re^{(n+r)y}-(-1)^\frac{r}{2}\,E^r\ \re^{ry}-\sum_{j=0}^{ [\frac{r}{4}]-1}
 c_{2j+1}\, E^{2j+1}\ \re^{\big(  \frac{n+r}{2}+2j+1\big) y}\,\Bigg]\,\psi=0\ .
\eea
The specification of the RG invariants and the $\mathfrak{a}_s$ --- $c_{2j+1}$ relation
looks similar to those for the case $A=1$ and we'll use the analogous notations.
However, now one should distinguish between even and odd  $\frac{r}{2}$. 
For even $\frac{r}{2}$  the
exponents for the RG invariants $\mathfrak{a}_s$   are given by
\bea
 d_{s}=\frac{2}{r}\, o_s\ \ \ \ \ \ \ \ \ \ \ \big(r/2-{\rm even}\big)\ ,
\eea
where
\bea
o_s\equiv \begin{cases}
\frac{r}{2}-s\ \ \ \ \ & {\rm for}\ \ \ s=1,3,\ldots, \frac{r}{2}-1\\[0.1cm]
r-s\ \ \ \ \  & {\rm for}\ \ \ s= \frac{r}{2},\,\frac{r}{2}+2,\,\ldots, r-2
\end{cases}\ .
\eea
The values of ${\mathfrak a}_{s}$ are related to the coefficients  of  the ODE \eqref{caaadd} as
\bea
{\mathfrak a}_{s}=   (-1)^{\scriptscriptstyle{[\frac{o_s}{2}]}-1}\ \ \frac{1}{s}\, \sum_{k=1}^{ o_s}(-1)^{k}
\ \frac{\Gamma(s\nu+k)}{ k!\Gamma(s\nu) }\ \sum_{j_1+j_2\ldots +j_k=o_s-k\atop
 j_1,\ldots, j_k\geq 0 }R_{2j_1+1}\ldots R_{2j_k+1}
\eea
with $\nu=\frac{2n}{n+r}$ and  we use the same notation \eqref{jasususa}
where now
\bea
G_{2m}=\frac{\Gamma(\frac{3}{2}+\frac{r}{2n})}{\Gamma(\frac{r}{2n})}\ \sum_{k=1} ^m\re^{-\frac{\ri\pi }{4} rk}\frac{
\Gamma\big(\frac{k}{2}-\frac{m}{n}+\frac{r}{2n}\big)}{k!\Gamma\big(\frac{3}{2}-\frac{k}{2}-\frac{m}{n}+\frac{r}{2n}\big )}\ \sum_{
{j_1+\ldots +j_k=\frac{r-2}{4}\,k-\frac{m}{2} }\atop
0\leq j_1,\ldots, j_k\leq\frac{r}{4}-1
}
c_{2j_1+1}\cdots c_{2j_k+1}\ .
\eea

In the case of odd $\frac{r}{2}$ the  exponents are given by
\bea
d_s=\frac{4}{r}\, e_s \ \ \ \ \ \ \ \ \ \ \ \big(r/2-{\rm odd}\big)\ ,
\eea
where 
\bea
e_s\equiv \begin{cases}
\frac{1}{4}\,(r-2s)\ \ \ \ \ &{\rm for}\ \ \ s=1,\,3,\,\ldots,\,\frac{r}{2}-2\\[0.1cm]
\frac{1}{2}\,(r-s)\ \ \ \ \ &{\rm for}\ \ \ s=\frac{r}{2}+1,\,\frac{r}{2}+3,\, \ldots,\,r-4
\end{cases}\  .
\eea
The  $\mathfrak{a}_s$ --- $c_{2j+1}$ relation reads as
\bea
{\mathfrak a}_{s}=(-1)^{e_s-1}\ \ \frac{1}{s}\times \begin{cases}
 \frac{1}{2\ri }\,  
 \big( F^{(-)}_{s}-F^{(+)}_{s}\big)\ 
 \ \ \ & {\rm for}\ \ \ \ \ \  \   s=1,\,3,\,\ldots,\,\frac{r}{2}-2\\[0.1cm]
\frac{1}{2}\, \big( F^{(+)}_{s}+F^{(-)}_{s}\big)\ \ \ \ \ &  {\rm for}\ \ \ \ \ \ \  s=\frac{r}{2}+1,\,\frac{r}{2}+3,\, \ldots,\,r-4
 \end{cases}\ .
\eea
Here the functions $F^{(\pm )}_{s}$ are defined by  eqs.\eqref{aiiasaasu} and \eqref{iasisausau} but now 
\bea
G^{(\pm)}_{2m}=\frac{\Gamma(\frac{3}{2}+\frac{r}{2n})}{\Gamma(\frac{r}{2n})}\ \sum_{k=1}^m 
\re^{\frac{\ri\pi }{2}(\frac{r-2}{2}\pm 1) k}\
\frac{\Gamma\big(\frac{k}{2}-\frac{2m}{n}+\frac{r}{2n}\big)}{k!\Gamma\big(\frac{3}{2}-\frac{k}{2}-\frac{2m}{n}+\frac{r}{2n}\big )}\ 
\sum_{
{j_1+\ldots +j_k= \frac{r-2}{4}\, k-m }\atop
0\leq j_1,\ldots j_k\leq \frac{r-2}{4}-1
}
c_{2j_1+1}\cdots c_{2j_k+1}\ .
\eea
The differential equation \eqref{caaadd} describes the scaling limit of the ground state Bethe roots
for any $n>0$. 

\smallskip

\subsection{The case $A=r-2$\label{aaaaasak932kjdsjj12}}
The set of admissible values of $(\mu,j)$  \eqref{skaj8923h1287}  are  given by
\be
\bm{\Xi}_{r, r-2}=\big\{(\mu,j)\, :\ \ \mu=j+1\ \ \&\ \ j=0,1,\ldots,\big[\tfrac{r-1}{2}\big]-1\big\}
\ee
and the ODE  becomes
\bea\label{hasytsatddsssd}
 \Bigg[-\partial_y^2+p^2+ \re^{(n+r)y}-(-E)^r\ \re^{r y}-\sum_{\mu=1}^{[\frac{r-1}{2}]}
 c_{\mu}\, E^{\mu}\ \re^{\big(  n+r- \frac{2n+r}{r}\, \mu\big) y}\,\Bigg]\,\psi=0\ .
\eea
We have found that for any $n>0$ the differential equation describes the scaling limit of the Bethe roots for the ground state
if the minimal set of non-trivial RG invariants consists of the $[\frac{r}{2}]$ members
\bea\label{aiisssssuuswa}
{\mathfrak a}_{s}=\frac{1}{s}\ \bigg(\frac{N}{r N_0}\bigg)^{\frac{2s}{r}}\  \frac{1}{r}\ \sum_{\ell=1}^r(\eta_\ell)^{-s}\ \ \ \ \ \ \ \ \ \ \  \ \ \ \big( s=1,\ldots, \big[\tfrac{r}{2}\big]\big)\ .
\eea
Their values are related to the coefficients of eq.\eqref{hasytsatddsssd} as
\bea
{\mathfrak a}_s 
=\frac{1}{s}\ \sum_{k=1}^s(-1)^k\ \frac{\Gamma(s\nu +k)}{k! \Gamma(s\nu )}\ 
\sum_{ 
j_1+j_2\ldots +j_k=s-k
\atop j_1,\ldots, j_k\geq 0}
R_{2j_1+1}\ldots R_{2j_k+1}\ .
\eea
Here  $\nu=\frac{2n}{n+r}$ and again we use the notation  $R_{2j+1}$ \eqref{jasususa}  with
\bea
 G_{2m}= \frac{\Gamma(\frac{3}{2}+\frac{r}{2n})}{\Gamma(\frac{r}{2n})} \ \sum_{k=1}^m
 \frac{
\Gamma\big(k- (\frac{2}{r}+\frac{1}{n})\, m+\frac{r}{2n}\big)}{k!\, \Gamma\big(\frac{3}{2} - (\frac{2}{r}+\frac{1}{n})\, m+\frac{r}{2n}\big )}
 \ \sum_{
 \mu_1+\ldots +\mu_k=m
 \atop
1\le \mu_1,\ldots, \mu_k\le [\frac{r-1}{2}]
 }
c_{\mu_1}\cdots c_{\mu_k}\ .
\eea

\section{\label{sasau} ODEs for $A=0$}
As $A=0$ the integer $\mu$ may take any value from $1$ to $r-1$, while $j=0$ (see \eqref{as9812aaaasy32hsd}). 
Then the  ODE \eqref{a873wehg12},\,\eqref{akjsaussauy}
becomes
\bea\label{askj3jhdsjhjaaasa}
 \bigg[-\partial_y^2+p^2+ \re^{(n+r)y}-E^r\,\re^{ry}-\sum_{\mu=1}^{r-1}
 c_{\mu}\, E^{\mu}\, \re^{\mu y}\,\bigg]\,\psi=0\ .
\eea
This is a more general version of equation \eqref{aksju1873762121aaaa}, which 
describes the scaling limit of the Bethe roots for the ground state when
$A\mu=rj> 0$. The latter is  connected to the case $A=0$ since, as was explained in 
sec.\,\ref{askj9812a9a8sjh122},  the ground state Bethe roots 
for $A\mu=rj$ and $A=0$ are simply related.  In view of this, 
one may expect   that  in performing the scaling limit in the regime $\gamma\in(0,\frac{\pi}{r})$ one should take
 the  RG invariants $\mathfrak{a}_s$  with $ \frac{r}{2}\le s \le r-1$
to be non-vanishing with the corresponding exponents given by
\be
d_s=\tfrac{2}{r}\,(r-s)\qquad\qquad\qquad \big(s=\big[\tfrac{r+1}{2}\big],\ldots,r-1\big)\, .
\ee
The  $\mathfrak{a}_s$ --- $c_\mu$  relation is obtained along the similar lines as  before yielding that
\bea\label{sakj90d12jh32aaaaa}
{\mathfrak a}_s 
= (-1)^{s-1}\ \ \frac{1}{s}\ \sum_{k=1}^{r-s}(-1)^k\ \frac{\Gamma(s\nu +k)}{k! \Gamma(s\nu )}\ 
\sum_{ 
j_1+j_2\ldots +j_k=r-s-k
\atop j_1,\ldots, j_k\geq 0}
R_{2j_1+1}\ldots R_{2j_k+1}\ \ \ \ \ \big(\tfrac{r}{2}<s\leq r-1\big)\, ,
\eea
where  $\nu=\frac{2n}{n+r}$ and we use the notation  $R_{2j+1}$ \eqref{jasususa}  with
\bea
G_{2m}=\frac{\Gamma(\frac{3}{2}+\frac{r}{2n})}{\Gamma(\frac{r}{2n})}
 \sum_{k=1}^m \frac{
  \Gamma\big(\frac{r-2 m}{2n}\big)}{k!\Gamma\big(\frac{3}{2}-k+\frac{r-2 m}{2n}\big )}
\ 
 \sum_{\mu_{1}+\ldots +\mu_{k}=r k-m}
c_{\mu_{1}}\cdots c_{\mu_{k}}\ .
\eea
These formulae are literally applicable for $\frac{r}{2}<s\le r-1$. If $r$ is even   and $s=\frac{r}{2}$
the summand in the r.h.s. of \eqref{sakj90d12jh32aaaaa} at $k=1$ coincides with $\nu R_{r-1}$,
 which contains a divergent term $G_r$. As was discussed, this
 signals that the definition of the 
 RG invariant $\mathfrak{a}_{\frac{r}{2}}$  needs to be modified:
\be
{\mathfrak a}_{\frac{r}{2}}\equiv  \frac{2}{r^2}\ \frac{1}{\log\big(\frac{N}{rN_0}\big)}\ \bigg(\frac{N}{rN_0}\bigg)\ \sum_{\ell=1}^r(\eta_\ell)^{-\frac{r}{2}}
\ee
and, similar to \eqref{kjas893aaa2187213}, we expect that its value
is expressed through the coefficients of the ODE as 
\be\label{sakj90d12jh32aaaaab}
{\mathfrak a}_{\frac{r}{2}}=\frac{2\Gamma(\frac{1}{2}+\frac{r}{2n})}{\sqrt{\pi}r\Gamma(1+\frac{r}{2n})}\ 
 \sum_{k=1}^{\frac{r}{2}}(-1)^{\frac{r}{2}-k} \ \frac{\Gamma\big(k-\frac{1}{2}\big )}
 {\sqrt{\pi} k!}\ 
 \sum_{\mu_{1}+\ldots +\mu_{k}=r (k-\frac{1}{2})}
c_{\mu_{1}}\cdots c_{\mu_{k}}\ .
\ee
\medskip

In the case under consideration, in taking the scaling limit, the
$[\frac{r}{2}]$ non-trivial
RG invariants  can be treated as
 the independent parameters.
A quick inspection shows that formulae \eqref{sakj90d12jh32aaaaa}\,-\,\eqref{sakj90d12jh32aaaaab}
involve the
$[\frac{r}{2}]$  coefficients $c_\mu$, where $\mu=[\frac{r+1}{2}],[\frac{r+1}{2}]+1,\ldots, r-1$
only. These relations can be inverted to express $c_{[\frac{r+1}{2}]},\ldots,c_{r-1}$ in terms  of
$\mathfrak{a}_{[\frac{r+1}{2}]},\ldots,\mathfrak{a}_{r-1}$.
The remaining coefficients of the ODE $c_\mu$ with $\mu=1,\ldots, [\frac{r-1}{2}]$ depend on the RG invariants
and the anisotropy parameter $\gamma=\frac{\pi}{n+r}$. Contrary to $c_\mu$ with $\mu\ge \frac{r}{2}$, 
they also turn out to  be complicated functions of
$p=\frac{1}{2}\,(n+r)\,{\tt k}$, where ${\tt k}$ is the twist parameter.
While there are no simple analytical expressions for these $c_\mu$, in principle, they can  be 
determined numerically.
One way of doing  so is by studying the scaling behaviour of
the sums  of inverse powers of the  Bethe roots, i.e., the sum rules. In practice,  however,
obtaining $c_\mu$ becomes a difficult computational task for 
$r\ge 6$. In this regard, we found it also useful to
consider the products 
\be\label{sjk89aaaa3jh21}
{\cal K}^{(\ell)}=\re^{\ri \pi{\tt k}}\,q^{-\frac{N}{2}}\,
\prod_{m=1}^{M}
\frac{\zeta_m+\eta_{\ell}\,q^{+1}}{\zeta_m+\eta_{\ell}\,q^{-1}}\qquad\qquad (\ell=1,2,\ldots,r)\,,
\ee
where $M=N/2-S^z$ so that  in the case of the ground state $M=N/2$.
Such products are the eigenvalues of the so-called quasi-shift operators, which
are members of the commuting family  that includes the spin chain Hamiltonian
(for details see, e.g., sec. 6.2 of ref.\cite{Bazhanov:2020new}). Note that $\prod_{\ell=1}^r{\cal K}^{(\ell)}$
coincides with the eigenvalue of the lattice translation operator \eqref{asj21gh}.
Having at hand the numerical values of
the Bethe roots $\{\zeta_m\}_{m=1}^{N/2}$ corresponding to the ground state 
and computing \eqref{sjk89aaaa3jh21} for increasing numbers of lattice sites, we observed that
the following limits exist
\bea\label{1902jhahasbqbvwpzl}
 {b}_\mu=\lim_{N\to\infty}\Bigg[
 \bigg(\frac{N}{rN_0}\bigg)^{\frac{r-2\mu}{r}}
 \frac{1}{2\pi \ri r  }\ \sum_{\ell=1}^r\re^{\frac{\ri\pi}{r} \mu (r+1-2\ell)}\ \log\big({\cal K}^{(\ell)}\big)\Bigg]
\eea
(here the branches of the logarithms are chosen in such a way that $\log({\cal K}^{(\ell)})$
vary continuously with the values of the RG invariants and vanish when all $\mathfrak{a}_s=0$).
Moreover, it was checked for $r=3,4,5$ that a relation of the form
\bea\label{asaaak98122131212}
c_\mu=\sum_{k=1}^{r-\mu}
\sum_{
\mu_{1}+\ldots +\mu_{k}=(k-1) r+\mu
\atop
1\le \mu_1,\ldots, \mu_k\le r-1 }
B^{(\mu)}_{\mu_1\ldots\mu_k}\, {b}_{\mu_1}\cdots {b}_{\mu_k}\ \ \ \ \ \ \ \ \ \ \ \ \ \ \ (\mu=1,\ldots,r-1)
\eea
is satisfied by the  $r-1$ numbers ${b}_\mu$ and
the coefficients $c_\mu$. Here
 $B_{\mu_1\ldots\mu_k}^{(\mu)}$ depends on $n$ and $r$, but not on the twist parameter ${\tt k}$.
\medskip

Let's illustrate the above on the simplest example with $r=3$. In this case, the value of the RG invariant 
$\mathfrak{a}_1$ may be set to zero.  The coefficient $c_2$
 in the differential equation
\bea\label{ask12JHasgg12asqw12}
 \bigg[-\partial_y^2+p^2+ \re^{(n+3)y}-E^3\,\re^{3y}-
 c_{1}\, E\, \re^{ y}- c_{2}\, E^2\, \re^{2y}\,\bigg]\,\psi=0
\eea
is related to $\mathfrak{a}_2$ as
\be
{\mathfrak a}_2=-  \frac{\Gamma(\frac{1}{2}+\frac{3}{2n})}{3\Gamma(\frac{3}{2n})}\ \frac{\Gamma(\frac{1}{2n})}{\Gamma(\frac{1}{2}+\frac{1}{2n})}\ c_2\ .
\ee
Formula \eqref{asaaak98122131212}, specialized to the case at hand, becomes
\bea\label{sakj8912aaahj21asass}
c_1=B_1^{(1)}\, b_1+B_{22}^{(1)}\, (b_2)^2\ ,\ \ \ \ \ \ \ \ \qquad \qquad c_2=B_2^{(2)}\, b_2\ .
\eea
Numerical values of   $B_{22}^{(1)}$  and $B_2^{(2)}$ for different $n$ are presented in
 tab.\,\ref{Btab}.
As for $B_1^{(1)}$, we found that the analytical expression
\bea
B^{(1)}_1=\frac{\sqrt{\pi}\,\Gamma(\frac{1}{2n})}{\Gamma(\frac{1}{2}+\frac{1}{2n})}
\eea
is in agreement 
with the data within the accuracy of our computations (at least $\sim 10^{-6}$ relative error). 
\medskip

\begin{table}
\begin{center}
\begin{tabular}{|l|l|l||l|l|l|}
\hline
 & &  & &  &  \\[-0.37cm]
$n$ & $B_{22}^{(1)}$ & $B_2^{(2)}$  & $n$ & $B_{22}^{(1)}$ & $B_2^{(2)}$   \\[0.03cm]
\hline
\hline
 & &  & &  &  \\[-0.37cm]
$0.5$ & $-24.31181$  &$6.335556$  &   $5.5$  & $-214.2092$   & $17.56490$   \\[0.03cm]
\hline
 & &  & &  &  \\[-0.37cm]
$1.0$ & $-33.51001$ &  $8.000000$ &   $6.0$  &  $-244.4040$ &  $18.57161$  \\[0.03cm]
\hline
 & &  & &  &  \\[-0.37cm]
$1.5$ & $-45.26519$ & $9.248695$  &  $6.5$  & $-276.6034$   & $19.57701$   \\[0.03cm]
\hline
 & &  & &  &  \\[-0.37cm]
$2.0$ &$-59.22168$  & $10.37020$    &  $7.0$ & $-310.8065$   & $20.58141$   \\[0.03cm]
\hline
 & &  & &  &  \\[-0.37cm]
$2.5$ & $-75.27314$ &  $11.43904$  &  $7.5$ &  $-347.0127$  &   $21.58504$  \\[0.03cm]
\hline
 & &  & &  &  \\[-0.37cm]
$3.0$ & $-93.37681$  &  $12.48197$   & $8.0$  &$-385.2213$   &  $22.58808$  \\[0.03cm]
\hline
 & &  & &  &  \\[-0.37cm]
$3.5$ & $-113.5123$  &  $13.51060$  &  $8.5$ & $-425.4320$  &   $23.59065$  \\[0.03cm]
\hline
 & &  & &  &  \\[-0.37cm]
$4.0$ &$-135.6686$  &  $14.53070$  &  $9.0$ & $-467.6444$   &  $24.59284$  \\[0.03cm]
\hline
 & &  & &  &  \\[-0.37cm]
$4.5$ & $-159.8393$  & $15.54535$   & $9.5$ & $-511.8583$  & $25.59473$   \\[0.03cm]
\hline
 & &  & &  &  \\[-0.37cm]
$5.0$ & $-186.0204$  &  $16.55638$  & $10.0$ & $-558.0735$  & $26.59636$   \\[0.03cm]
\hline
\end{tabular}
\end{center}
\caption{\label{Btab}\small%
Listed are the numerical values of $B_{22}^{(1)}$ and $B_2^{(2)}$ from the relation \eqref{sakj8912aaahj21asass},
which expresses  the coefficients $c_1$, $c_2$ in the differential equation \eqref{ask12JHasgg12asqw12} in terms
of  $b_1$, $b_2$ defined by \eqref{sjk89aaaa3jh21},\,\eqref{1902jhahasbqbvwpzl} with $r=3$.}
\end{table}

The numerical computation of the limits $b_\mu$ \eqref{1902jhahasbqbvwpzl} is straightforward for any $r$.
Hence, knowledge of  $B^{(\mu)}_{\mu_1\ldots\mu_k}$ in the relation \eqref{asaaak98122131212}
would provide an effective way of determining all
 $r-1$ coefficients  $c_\mu$ in the differential equation \eqref{askj3jhdsjhjaaasa}.  
Unfortunately, at the current moment 
the  analytical formula for $B^{(\mu)}_{\mu_1\ldots\mu_k}$ is not known.

\bigskip
The following comment is in order here. 
In sec.\,\ref{sec24}, devoted to the case  $A=\frac{r-1}{2}$ with odd $r$, it was mentioned that the
Bethe roots for a low energy state develop a scaling behaviour not only in the vicinity of $\zeta=0$, but also $\zeta=\infty$.
This is manifest in that the limits
\be
\bar{E}_{m}^{(a)}=\lim_{N\to\infty\atop m-{\rm fixed}}\,\bigg(\frac{N}{r N_0}\bigg)^{\frac{2n}{r(n+r)}}
\Big(\zeta_{M_a-m}^{(a)}\Big)^{-1}
\ee 
exist and are non-vanishing (for the ground state $S^z=0$ and  $M_a=N/(2r)$). In the case $A=0$, 
we found that if the scaling limit is performed, where the set of  non-trivial RG invariants is taken to be
 $\{\mathfrak{a}_s\}_{s=[\frac{r+1}{2}]}^{r-1}$ 
with corresponding exponents  $d_s=\tfrac{2}{r}\,(r-s)$, then
the scaled Bethe roots $\bar{E}_{m}^{(a)}$ are described in terms of the ODE of the form
\bea\label{askj3jhdaaasaaaassa}
 \Bigg[-\partial_{\bar y}^2+{\bar p}^2+ \re^{(n+r){\bar y}}-{\bar E}^r\, \re^{r{\bar y}}-\sum_{\mu=1}^{[\frac{r}{2}]}
 {\bar c}_{\mu}\, {\bar E}^{\mu}\, \re^{\mu {\bar y}}\,\Bigg]\,{\bar \psi}=0\ .
\eea
The coefficients $\bar{c}_\mu$, entering therein, depend on  the values of the RG invariants, 
the anisotropy $n$ and $\bar{p}=-\frac{1}{2}\,(n+r)\,{\tt k}$. 
\medskip

For given values of $\mathfrak{a}_s$ and $N$
the relations  \eqref{ais78176326512assadas}, along with the normalization condition \eqref{iaassauasas}, may be treated as a system of
algebraic equations, which determines the set of inhomogeneities $\{\eta_\ell\}_{\ell=1}^r$.
 For large $N$ it is not difficult to show that
\bea\label{asj9ajahsg1v21v}
\frac{1}{r}\,\sum_{\ell=1}^r(\eta_\ell)^r=(-1)^{r-1}+O(N^{-2})\ ,
\eea
while
\bea\label{aksj12haajgshd}
\frac{1}{s}\ \bigg(\frac{N}{rN_0}\bigg)^{{\bar d}_s} \
\frac{1}{r}\, \sum_{\ell=1}^r(\eta_\ell)^s={\bar {\mathfrak a}}_s+O(N^{-2})\ \ \ \ \ \ \ \ \ \ \ (s=1,\ldots,r-1)\ .
\eea
Here
\bea
{\bar d}_s=\frac{2 s}{r}
\eea
and the limiting values ${\bar {\mathfrak a}}_s$ are certain polynomials of the RG invariants ${\mathfrak a}_1,\ldots,{\mathfrak a}_{r-1}$. In particular,
\bea\label{slaksaau3aa2aaaa9a1a2aa98}
{\bar {\mathfrak a}}_1&=&(-1)^r\, {\mathfrak a}_{r-1}\nonumber\\
{\bar {\mathfrak a}}_2&=&(-1)^r\, {\mathfrak a}_{r-2}+\frac{r}{2}\ ({\mathfrak a}_{r-1})^2\\
{\bar {\mathfrak a}}_3&=&(-1)^r\,{\mathfrak a}_{r-3}+r\, {\mathfrak a}_{r-2}\,{\mathfrak a}_{r-1}+(-1)^r\ \frac{r^2}{3}\   ({\mathfrak a}_{r-1})^3\ \ \ \ \ \ \ \ (r>3)\ .
\nonumber
\eea
The scaling limit can be performed with the set of RG invariants chosen 
to be  $\{\bar{\mathfrak{a}}_s\}$. The latter would be defined by
formula \eqref{aksj12haajgshd} with the  term $O(N^{-2})$ ignored, and
are kept independent of $N$. 
Also, one may impose the normalization condition \eqref{asj9ajahsg1v21v}, again, without the remainder term.
The two  schemes based on the different sets of RG invariants  lead to the same limiting 
values   $E_m^{(a)}$ and $\bar{E}_m^{(a)}$, provided that ${\mathfrak{a}}_s$ and 
 $\bar{\mathfrak{a}}_s$  are related as in 
\eqref{slaksaau3aa2aaaa9a1a2aa98}.
The scaled Bethe roots would be  described 
by the ODEs
\eqref{askj3jhdsjhjaaasa} and \eqref{askj3jhdaaasaaaassa}.
Since the Bethe Ansatz equations \eqref{baekasdba}
are invariant w.r.t. the simultaneous inversion
$\zeta_j\mapsto \zeta_j^{-1}$, $\eta_\ell\mapsto (\eta_\ell)^{-1}$ and
$\omega\mapsto \omega^{-1}$ (${\tt k}\mapsto -{\tt k}$), the results for 
 $E_m^{(a)}$  carry over to $\bar{E}_m^{(a)}$ with a change in nomenclature only.

\smallskip

\subsection*{Comment on the ${\cal Z}_r$ invariant case with $\eta_\ell=(-1)^r\,\re^{\frac{\ri\pi}{r}\,(2\ell-1)}$}
In the work \cite{Kotousov:2023zps}, the scaling limit for the low energy states of the
${\cal Z}_r$ invariant spin chain in the regime $\gamma\in(0,\frac{\pi}{r})$ was considered. 
It was found that such states can be labelled by $S^z$, the winding number ${\tt w}\in\mathbb{Z}$,
the pair of non-negative integers $({\tt L},\bar{\tt L})$; and also, for   $r$ even, $s$ which may take any
real value.
The corresponding eigenvalues of the CFT Hamiltonian, defined according to eqs.\,\eqref{uassaysa},\,\eqref{kjas8923hjds},
are given by
\be\label{asjk1892hjdaaa}
{\cal E}_{\rm CFT}=\frac{p^2+\bar{p}^2}{n+r}
-\frac{r}{12}\,+{\tt L}+\bar{{\tt L}}+\begin{cases}\dfrac{s^2}{2n}& \ \ \ {\rm for}\ \ \ r\ {\rm even}
\\[0.2cm]
0 & \ \ \ {\rm for}\ \ \ r\ {\rm odd}
\end{cases}\, ,
\ee
where
\bea
p=\frac{S^z}{2}+\frac{n+r}{2}\ ({\tt k}+{\tt w})\ ,\ \ \ \ \ \ \ {\bar p}=\frac{S^z}{2}-\frac{n+r}{2}\  ({\tt k}+{\tt w})\ .
\eea
For any low energy Bethe state, 
the differential equations were  proposed, which describe the scaling limit of the Bethe roots.
In particular, the ODEs for the
so-called ``primary'' Bethe states, for which the integers ${\tt L}=\bar{\tt L}=0$, take the form
\begin{subequations}\label{kkjj2198asdasdasdasdasd}
\bea
&&\Bigg[-\partial^2_y+p^2+\re^{(n+r)y}-E^r\,\re^{ry}+
\sum_{\mu=1}^{[\frac{r}{2}]} (-1)^\mu\,s_\mu\, E^\mu\,\re^{\mu y}\,\Bigg]\,\psi=0\\[-0.05cm]
&&\qquad\qquad\qquad\qquad\qquad\qquad\qquad\qquad\qquad\qquad\qquad\qquad\qquad \quad  .\nonumber
\\[-0.05cm]
&&\Bigg[-\partial_{\bar y}^2+{\bar p}^2+ \re^{(n+r){\bar y}}-{\bar E}^r\, \re^{r{\bar y}}-\sum_{\mu=1}^{[\frac{r}{2}]}
 (-1)^\mu\,{\bar s}_\mu\, {\bar E}^{\mu}\, \re^{\mu {\bar y}}\,\Bigg]\,{\bar \psi}=0
\eea
\end{subequations}
The coefficients $s_\mu$, $\bar{s}_\mu$ 
are expressed in terms of the eigenvalues of the quasi-shift operators \eqref{sjk89aaaa3jh21}:
\begin{subequations}
\bea
s_\mu&=&+{B_\mu}\ \lim_{N\to\infty}\Bigg[
 \bigg(\frac{N}{rN_0}\bigg)^{\frac{r-2\mu}{r}}
 \frac{1}{2\pi \ri r  }\ \sum_{\ell=1}^r\re^{+\frac{\ri\pi}{r} \mu (r+1-2\ell)}\ \log\big({\cal K}^{(\ell)}\big)\Bigg]\\[0.0cm]
&&\qquad\qquad\qquad\qquad\qquad\qquad\qquad\qquad\qquad\qquad\qquad\qquad\qquad\ \    ,\nonumber\\[0.0cm]
 {\bar s}_\mu&=&-{B_\mu}\ \lim_{N\to\infty}\Bigg[
 \bigg(\frac{N}{rN_0}\bigg)^{\frac{r-2\mu}{r}}
 \frac{1}{2\pi \ri r  }\ \sum_{\ell=1}^r\re^{-\frac{\ri\pi}{r} \mu (r+1-2\ell)}\ \log\big({\cal K}^{(\ell)}\big)\Bigg]
\eea
\end{subequations}
where
\bea
B_\mu=(-1)^{(r-1)\mu}\ 2n\ 
 \frac {\sqrt{\pi}\,\Gamma(1+\frac{r-2\mu}{2n})}
 {\Gamma(\frac{1}{2}+\frac{r-2\mu}{2n})}
 \qquad\qquad \qquad\qquad \big(\,\mu=1,\ldots,[\tfrac{r}{2}]\,\big)\ .
\eea
Note that when $r$ is even, $s_{\frac{r}{2}}=\bar{s}_{\frac{r}{2}}$ and coincides with $s$, which
 appears in formula \eqref{asjk1892hjdaaa}.
\medskip

The pair \eqref{kkjj2198asdasdasdasdasd}
 resembles the differential equations  \eqref{askj3jhdsjhjaaasa} and \eqref{askj3jhdaaasaaaassa}.
However, it describes the scaling limit of a class of low energy excited states, not
just the ground state. The values of $s_\mu$ and $\bar{s}_\mu$ specify the primary Bethe state and
 are not arbitrary. For  odd $r$ they belong to a certain discrete set, which 
is determined through a so-called ``quantization condition''. In the case of even $r$,
the parameter  $s\equiv s_{\frac{r}{2}}=\bar{s}_{\frac{r}{2}}$ may be any real number, but once it is fixed,
the remaining $s_\mu$, $\bar{s}_\mu$ with $\mu<\frac{r}{2}$  take a discrete set of admissible values, which is also defined
 by the
quantization condition. In contrast, the pair of ODEs \eqref{askj3jhdsjhjaaasa} and \eqref{askj3jhdaaasaaaassa}
correspond to the ground state.  The coefficients $c_\mu$ with $\mu\ge \frac{r}{2}$
are related to the values of the RG invariants through eqs.\,\eqref{sakj90d12jh32aaaaa}-\eqref{sakj90d12jh32aaaaab}.
We expect that the values of the remaining  $c_\mu$ with $\mu<\frac{r}{2}$ as well as $\bar{c}_\mu$  in
 \eqref{askj3jhdaaasaaaassa}
are a particular solution of a certain quantization condition. Determining the latter would be crucial for describing
the scaling limit of the low energy spectrum for the spin chain with softly broken ${\cal Z}_r$ symmetry
in the regime $\gamma\in(0,\frac{\pi}{r})$. However, this problem
lies outside the scope of the paper. 

\smallskip
\section{Concluding remarks}
In this work, we  described within the ODE/IQFT approach the scaling behaviour of the Bethe roots for just the 
ground state of the  spin\,-\,$\frac{1}{2}$ chain associated with the inhomogeneous six-vertex model.
Even so, the results indicate the presence of a remarkable variety
of multiparametric integrable structures in the underlying field theories.
Further development of the ODE/IQFT correspondence would require
obtaining the class of differential equations that describe the scaling limit of all the 
low energy Bethe states.  In the ${\cal Z}_r$ invariant case with 
anisotropy parameter $\gamma\in(0,\frac{\pi}{r})$ this was done in ref.\cite{Kotousov:2023zps}.
For the more general setup discussed in the present work, a preliminary analysis shows that  a
creative
application of the original ideas from ref.\cite{Bazhanov:2003ni} is required.
An important physical
question is uncovering the field theories governing the critical behaviours
for $0<\gamma<\pi$. It is relatively well understood now in the case $r=2$ 
and for the domain $\pi(1-\frac{1}{r})<\gamma<\pi$  with arbitrary $r$ due to the works
\cite{Jacobsen:2005xz,Ikhlef:2008zz,IJS2,Ikhlef:2011ay,Candu:2013fva,Frahm:2013cma,Bazhanov:2019xvyA} and
\cite{Kotousov:2021vih}, respectively. 
\medskip

Among the most interesting directions of  research in the area
of the ODE/IQFT correspondence is its extension to the spin chain associated with the
higher spin\,-\,$j$ generalization of the 
inhomogeneous six-vertex model and especially  the  $j\to\infty$ limit 
for the isotropic (XXX) spin chain, i.e., at the boundary of the domain of criticality,
$\gamma=0$ and $\gamma=\pi$.
 This may be useful for exploring the 
 isotropic spin chain
built from (infinite dimensional)
 unitary representations of $\mathfrak{sl}(2,\mathbb{C})$. The latter is relevant to high energy
physics since, as was  discovered in the pioneering paper of Lipatov \cite{Lipatov:1993yb},
 this spin chain (in a certain setup)
describes the wave functions of compound states of reggeized gluons
in multicolour QCD ($N_c\to\infty$) in the generalized leading logarithmic approximation.
The idea was further developed by  Faddeev and Korchemsky in  \cite{Faddeev:1994zg}.
In the works \cite{Derkachov:2001,Derkachov:2002wz} the spectrum of the spin chain was studied for
small lattice sizes $N$, which is interpreted as the number of gluons. The analysis becomes complicated
for increasing $N$. This is where, one may hope, that the ODE/IQFT approach would be useful.

\section*{Acknowledgments}
The authors acknowledge discussions with Andreas Kl{\"u}mper and thank him for providing 
us with a copy of the paper \cite{Kluemper} before it was published.
SG and GK are grateful to Holger Frahm and M\'{a}rcio Martins for stimulating conversations.
\smallskip

\noindent
The research of SG   is funded
by the Deutsche Forschungsgemeinschaft under  grant
No.\,Fr 737/9-2.
Part of this work was carried out during GK's visits to the NHETC at Rutgers University. 
He is grateful for the support and hospitality he received during the stays.
\smallskip

\noindent
The research of SL is supported by the NSF under grant number NSF-PHY-2210187.

\newpage

\appendix 
\section{Auxiliary functions appearing in the sum rules\label{AppA}}
Here we present the explicit formulae for $f_1$, $f_2$ and $f_3$ entering into eq.\eqref{aissuaus}. We use the same notation
as in ref.\cite{Bazhanov:2019xvyA}, where these functions have previously appeared.

\medskip

The function $f_1$ is defined as
\bea\label{sa8932jhdssaaaa}
f_1(h,g)&=&\frac{\pi\Gamma(1-2g)}{\sin(\pi g)}\ \frac{\Gamma(g+2h)}{\Gamma(1-g+2 h)}\ .
\eea
As for $f_2$ and $f_3$, the expression for these is more complicated.
In particular, $f_2$ is given by the integral
\be\label{jassusau000}
f_2(h, g)=2^{1-4g}\,\frac{\Gamma^2(1-g)}{\Gamma^2(\frac{1}{2}+g)}\
\frac{\Gamma(2g+2 h)}{\Gamma(1-2g+2 h)}\,\int_{ -\infty}^\infty\frac{\rd x}{2\pi}\,\frac{S_1(x)}{x+\ri h}\ \ \ \ \ \  
\big(0<g<\tfrac{1}{2},\ \Re e(h)>0\big)
\ee
with
\be\label{Sdef1a}
S_1(x)=\sinh(2\pi x)\,\Gamma(1-2g+2\ri x)\,\Gamma(1-2g-2\ri x)\,\big(\Gamma(g+2\ri x)\Gamma(g-2\ri x)\big)^2\ .
\ee
The above is applicable for $0<g<\frac{1}{2}$.
Its analytic continuation to the interval $\tfrac{1}{2}<g<1$, yields
\bea\label{jassusau}
f_2(h,g)&=&2^{1-4 g}\,\frac{\Gamma^2(1-g)}{\Gamma^2(\frac{1}{2}+g)}\
\frac{\Gamma(2g+2 h)}{\Gamma(1-2g+2h)}\,\Bigg(\int_{-\infty}^\infty\frac{\rd x}{2\pi}\,
\frac{S_1(x)}{x+\ri h}  \\[0.2cm]
&-&
\frac{\sin(2\pi g)\Gamma(3-4 g)\Gamma^2(1-g)\Gamma^2(3g-1)}{(2h+1-2g)
(2h-1+2 g)}\Bigg)\qquad
 \qquad \big(\,\tfrac{1}{2}<g<1, \ \Re e(h)>0\, \big)\ .\nonumber
\eea

\bigskip
In the case of $f_3$, the following expression is valid  for
$0<g<\tfrac{1}{2}$ and $\Re e(h)>0$:
\bea\label{ausususay}
f_3(h,g)&=& 2^{2-6 g}\, \sqrt{\pi}\ \frac{\Gamma^3(1-g)}{
\Gamma^3(\frac{1}{2}+
g)}\ 
\frac{\Gamma(3g-1+2 h)}{
 \Gamma(2-3g +2 h)}\\
&\times&\Bigg(\,
-{\sin(4\pi g)\over \pi^2}\, 
\int_{-\infty}^{+\infty}\frac{\rd y}{2 \pi}
\int_{-\infty}^{+\infty}\frac{\rd x}{ 2 \pi}\,
\frac{S_2(x,y)}{ (y+\ri  h )(x-y-\ri 0)}\, 
+\frac{1}{3}\, \int_{-\infty}^{+\infty}\frac{\rd x}{ 2 \pi} 
\,\frac{S_3(x)}{ x+\ri h}\, \Bigg),\nonumber
\eea
where  the functions $S_2$ and $S_3$ read as
\bea
S_2(x,y)&=&\sinh(2 \pi y)\, \sinh(2 \pi x)\, \Gamma(g+2 \ri y)\,
\Gamma(g-2 \ri y)\,
\Gamma(2g+2 \ri y)\, \Gamma(2g-2 \ri y)\nonumber\\[0.2cm]
&\times&\Gamma(2-3g+2 \ri y)\, \Gamma(2-3g-2 \ri y)\,
\Gamma(1-2g+2 \ri x)\, \Gamma(1-2g-2 \ri x)\nonumber\\[0.2cm]
&\times&
\big(\Gamma(g+2 \ri x)\, \Gamma(g-2 \ri x)\big)^2\\[0.2cm]
S_3(x)&=&\sinh(2 \pi x)\, \Gamma(2-3 g+2 \ri x)
\, \Gamma(2-3 g-2 \ri x)\, 
\big(\Gamma(g+2 \ri x)\, \Gamma(g-2 \ri x)\big)^3\nonumber\\[0.2cm]
&\times & 
\frac{\sin(4\ri\pi x+2\pi g)-2 \sin(2\pi g)}{
\sin(2\ri\pi x+2\pi g)}\ .\nonumber
\eea
For $\tfrac{1}{2}<g<\tfrac{2}{3}$ and $\Re e(h)>0$ one has
\bea
f_3(h,g)&=& 2^{2-6 g}\, \sqrt{\pi}\ \frac{\Gamma^3(1-g)}{
	\Gamma^3(\frac{1}{2}+
	g)}\ 
\frac{\Gamma(3g-1+2 h)}{
	\Gamma(2-3g +2 h)}\\
&\times&\Bigg[\,
-{\sin(4\pi g)\over \pi^2}\, 
\int_{-\infty}^{+\infty}\frac{\rd y}{2 \pi}
\int_{-\infty}^{+\infty}\frac{\rd x}{ 2 \pi}\,
\frac{S_2(x,y)}{ (y+\ri  h )(x-y-\ri 0)}\, 
+\frac{1}{3}\, \int_{-\infty}^{+\infty}\frac{\rd x}{ 2 \pi} 
\,\frac{\tilde{S}_3(x)}{ x+\ri h}\, \Bigg],\nonumber
\eea
where
\bea
\tilde{S}_3(x)&=&S_3(x)-3\pi^{-2}\sin(4\pi g)\sin (2 \pi  g) \sinh (2 \pi  x) \Gamma (3-4 g) \Gamma^2 (1-g) \Gamma^2 (3 g-1) \nonumber\\
&\times& \Gamma (2-3 g-2 \ri x)  \Gamma (2-3 g+2 \ri x)\\
&\times& \Gamma (g+2 \ri x) \Gamma (g-2 \ri x)\Gamma (2 g+2 \ri x-1) \Gamma (2 g-2 \ri x-1)\nonumber\,.
\eea
Finally if $\tfrac{2}{3}<g<1$ and $\Re e(h)>0$, the analytic continuation of \eqref{ausususay} gives
\bea
f_3(h,g)&=& 2^{2-6 g}\, \sqrt{\pi}\ \frac{\Gamma^3(1-g)}{
	\Gamma^3(\frac{1}{2}+
	g)}\ 
\frac{\Gamma(3g-1+2 h)}{
	\Gamma(2-3g +2 h)}\\
&\times&\Bigg[\,
-{\sin(4\pi g)\over \pi^2}\, 
\int_{-\infty}^{+\infty}\frac{\rd y}{2 \pi}
\int_{-\infty}^{+\infty}\frac{\rd x}{ 2 \pi}\,
\frac{S_2(x,y)}{ (y+\ri  h )(x-y-\ri 0)}\, 
+\frac{1}{3}\, \int_{-\infty}^{+\infty}\frac{\rd x}{ 2 \pi} 
\,\frac{\tilde{S}_3(x)}{ x+\ri h}\,+S_4 \Bigg],\nonumber
\eea
where $S_4$ is defined as 
\bea
S_4&=&\frac{\sin (2 \pi  g) \sin (3 \pi  g) \Gamma (5-6 g) \Gamma (2-2 g) \Gamma^3 (1-g) \Gamma^2 (3 g-1) \Gamma (5 g-3)}{\pi  \big((2-3 g)^2-4 h^2\big)}\nonumber\\[0.2cm]
&+&\frac{4 \sin (3 \pi  g) \cos (\pi  g)\Gamma (4-6 g) \Gamma^3 (2-2 g) \Gamma^3 (4 g-2)}{
3 \big((2-3 g)^2-4 h^2\big)\,\big(2 \cos (2 \pi  g)+2 \cos (4 \pi  g)+1\big)} \\[0.2cm]
&\times&\big(3 g-6 h-2+(15 g-6 h-10) \cos (2 \pi  g)+(9 g-6 h-6) \cos (4 \pi  g)+2 (3 g-2) \cos (6 \pi  g)\big)\ .\nonumber
\eea

\section{Lagrange formula\label{AppB}}
Consider the algebraic equation
\bea
X^r+
 \sum_{m=1}^{r-1}  G_{m} X^{r-m}={ Y}^r\ .
\eea
Let $X(Y)$  be the solution such that
\be
X(Y)\to Y\qquad\qquad {\rm as} \qquad\qquad Y\to\infty\ .
\ee
In fact, $X(Y)$ is a multivalued function of $Y$ and this condition specifies
its branch for sufficiently large $Y$. It was found by Lagrange 
\cite{Lagrange}, see also \cite{Belardinelli}, that $X(Y)$ admits
 a convergent power series expansion of the form
\bea
X(Y)=Y+\sum_{k=1\atop 
k\not=0\,{\rm mod}\, r}^{\infty} R_k\ Y^{-k}\ ,
\eea
where
\bea
R_k=\frac{1}{r}\sum_{\alpha_1,\ldots\alpha_{n-1}\geq 0
\atop
\alpha_1+2\alpha_2+\ldots+(r-1)\alpha_{r-1}=k+1}
\frac{(-1)^{\alpha_1+\ldots+\alpha_{r-1}}}{\alpha_1!\alpha_2!\ldots\alpha_{r-1}!}\ 
\frac{\Gamma(\alpha_1+\ldots+\alpha_{r-1}-\frac{k}{r})}{\Gamma(1-\frac{k}{r})}\ 
G_1^{\alpha_1}G_2^{\alpha_2}\dots G_{r-1}^{\alpha_{r-1}}\ .
\eea

\section{Relations  between $\mathfrak{a}_{2j+1}$ and $c_{2j+1}$ for $r=3,5,7$ and $A=\frac{r-1}{2}$\label{AppC}}
For $r$ odd and $A=\frac{r-1}{2}$ the formula \eqref{hassy} was obtained in the main body of the text
 that expresses the RG invariants $\mathfrak{a}_{2j+1}$ 
\eqref{sajk89732hjsdA}
with $j=1,2,\ldots,\frac{r-1}{2}$   in terms of the coefficients $c_{2j+1}$ of the ODE \eqref{hasytsat}.
The inversion of this relation leads to eq.\,\eqref{aks322091}.  Here we write it out explicitly for 
$r=3,5,7$.
\medskip

In the case $r=3$ there is only a single non-trivial RG invariant $\mathfrak{a}_1$. The coefficient
$c_1$ entering into the differential equation is simply proportional to it:
\bea
c_1=C_0^{(0)}\,  {\mathfrak a}_1\ ,
\eea
where we use the notation  $C_j^{(j)}$ from eq.\eqref{as983hjds21aaaa}, namely,
\be
C_j^{(j)}=(-1)^j\,\frac{r\Gamma\big(
\frac{r}{2n}\big)
\Gamma\big(1-\frac{(2j+1)(n-r)}{2r n}\big)}{
\Gamma\big(\frac{1}{2}+
\frac{r}{2n}\big)\Gamma\big(\frac{1}{2}-
\frac{(2j+1)(n-r)}{2r n}\big)}\ .
\ee
For $r=5$ the two coefficients  $c_1$ and $c_3$ are expressed in terms of the
 two RG invariants $\mathfrak{a}_1$ and $\mathfrak{a}_3$ as:
\bea\label{aisoassiao}
c_{1}&=&C_0^{(0)}\, {\mathfrak a}_1+\frac{n-5}{20n}\ \Big(\big(C_1^{(1)}\big)^2-5C_0^{(0)}\Big)\, {\mathfrak a}^2_3
\nonumber\\[0.1in]
c_{3}&=&C_1^{(1)}\,{\mathfrak a}_{3}\ .
\eea
For $r=7$, one has
\bea\label{iaisaisaiasiasias}
c_1&=&C_0^{(0)}\, {\mathfrak a}_1- \frac{n-7}{14 n}\   \Big( 7 C_0^{(0)}+ C_1^{(1)} C_2^{(2)}\Big)\ 
 {\mathfrak a}_3 {\mathfrak a}_5
\nonumber\\[0.2in]
&+&
\frac{n-7}{n^2}\ \bigg(\,  \frac{15 n-7}{24}\ C_0^{(0)}
+ \frac{3 (n-7)}{56}\  C_1^{(1)}\, C_2^{(2)}-\frac{3 n+28}{588}\ \big(C_2^{(2)}\big)^3 \bigg)
\ {\mathfrak a}_5^3\nonumber\\[0.2in]
c_3&=&  C_1^{(1)}\,{\mathfrak a}_3-\frac{3 (n-7)}{28 n}\  \Big(7 C_1^{(1)} +
\big( C_2^{(2)}\big)^2\Big)\ {\mathfrak a}_5^2\\[0.2in]
c_5&=&C_2^{(2)} \, {\mathfrak a}_5\ .\nonumber
\eea

\section{Scaling limit at the free fermion point with $\eta_\ell=(-1)^r\,\re^{\frac{\ri\pi}{r}\,(2\ell-1)}$ \label{AppD}}
Here we focus on the case with  $r$ odd and $q=\ri$. In addition, the inhomogeneities are set to be
$\eta_\ell=(-1)^r\,\re^{\frac{\ri\pi}{r}\,(2\ell-1)}$ for which
the spin chain possesses ${\cal Z}_r$ invariance. In this case the Bethe Ansatz equations
take the especially simple form
\be
\Bigg(
\frac{1+(\ri\zeta_j)^r}
{1-(\ri\zeta_j)^r}\Bigg)^{\frac{N}{r}}
=-\re^{\ri\pi (2{\tt k}+S^z)}\ .
\ee
\medskip

 In sec.\,\ref{sec21}  we described the
scaling limit of the eigenvalue of the $Q$\,-\,operator for the ground state of the spin chain.
Formula \eqref{asj9823hjs} remains valid in the case of any low energy Bethe state provided $n>r$.
However, at $n=r$ 
it should be modified as
\be\label{asj9823hjsaaa}
\slim_{N\to\infty} \bigg(\frac{N}{2r}\bigg)^{\frac{1}{2r} (-1)^AE^r}\, A_+\Big(\big(2N/\pi\big)^{-\frac{1}{r}}\,E\Big)=D_+(E)
\qquad\qquad \qquad\qquad (n=r)\,, 
\ee
where
recall that the symbol ``$\slim$'' is there as a reminder that the formula holds true only for the low energy states.
The necessity of the $N$\,-\,dependent factor can be illustrated on the ground state eigenvalue. 
The scaling limit of the Bethe roots for the ground state, where $S^z=0$,
yields
\bea\label{asiu81eiowq12}
E^{(a)}_m=\re^{\frac{\ri \pi}{r}(2a-\frac{r-1}{2})}\ 
\big((2m_a-1+2{\tt k})\, r\big)^{\frac{1}{r}}\ \ \ \ \  \qquad\qquad \ (a=1,\ldots, r)
\eea
with 
\be\label{asiu81eiowq12AA}
m_a=m=1,2,\ldots\ . 
\ee
Then, it is easy to see that the limit in the l.h.s. of \eqref{asj9823hjsaaa}
exists and leads to 
\be
D_+^{({\rm vac})}=   \frac{\Gamma(\frac{1}{2}+{\tt k})}{\Gamma\big(\frac{1}{2}+{\tt k}+
\epsilon^r\big)}\ ,
\ee
where we use the short-cut notation
 \bea
 \epsilon=\re^{-\frac{\ri \pi}{2r} (r+1)}\,(2r)^{-\frac{1}{r}}\ E \ .
 \eea
\medskip

Staying in the sector $S^z=0$, we now consider the scaling limit 
of $A_+(\zeta)$ for 
a low energy excited Bethe state. The chiral state $|\alpha\rangle$ appearing in eq.\,\eqref{askj8723jh2} can be 
characterized by the location of a finite number of ``holes'' at $n^-_{j,a}$ in the vacuum distribution of the integers 
$m_a$ given by \eqref{asiu81eiowq12AA}   as well as ``particles'' $n^+_{j,a}$. 
The positive integers $n^\pm_{j,a}$ are ordered as
  \bea
  1\leq n_{1,a}^{\pm}<  n_{2,a}^{\pm}<\ldots<   n_{M^{\pm}_a,a}^{\pm}\ .
  \eea
Note that for the particles, ${\tt k}-n^+_{j,a}+\frac{1}{2}$ is a negative number and we take
its $r$-th root to be $\re^{\frac{\ri\pi }{r}}\ (n^+_{j,a}-{\tt k}-\frac{1}{2})^{\frac{1}{r}}$ so that
the corresponding $E_m^{(a)}$  lies on the ray $\arg(E_m^{(a)})=\frac{\pi}{2r}\,(4a-r-1)$ 
(${\rm mod}\ 2\pi$). Then the scaling limit of $A_+(\zeta)$ yields 
\bea
D_+&=&   \frac{\Gamma(\frac{1}{2}+{\tt k})}{\Gamma\big(\frac{1}{2}+{\tt k}+
\epsilon^r\big)}
\ 
  \frac{\prod_{a=1}^r \prod_{j=1}^{M^-_a}\
 \re^{\frac{2\pi \ri }{r} a}  \big(n_{j,a}^--\frac{1}{2}+{\tt k}\big)^{\frac{1}{r}}}
 { \prod_{a=1}^r\prod_{j=1}^{M^+_a}\re^{\frac{ \pi\ri}{r} (2a+1)} \big(n_{j,a}^{+}-\frac{1}{2}-{\tt k}\big)^{\frac{1}{r}}}\nonumber\\[0.2in]
 &\times&
 \frac{\prod_{a=1}^r \prod_{j=1}^{M^+_a}\
 \Big(\re^{\frac{\pi\ri}{r} (2a+1)} \big(n_{j,a}^{+}-\frac{1}{2}-{\tt k}\big)^{\frac{1}{r}}+\epsilon\Big)}
 {\prod_{a=1}^r \prod_{j=1}^{M^{-}_a}\
 \Big(\re^{\frac{2\pi\ri}{r} a} \big(n_{j,a}^{-}-\frac{1}{2}+{\tt k}\big)^{\frac{1}{r}}+\epsilon\Big)}\ .
  \eea
As $\epsilon\to+\infty$, $D_+$ develops the asymptotic behaviour
 \bea\label{kjass873hjqsda}
 \log D_+\asymp\log ({\mathfrak{C}_+})-\epsilon^r\ \big(
 r \log(\epsilon)-1\big)-\big( {\tt M}+ r{\tt k}
 \big)\, \log(\epsilon)+\sum_{s=1}^\infty 
  \frac{(-1)^s}{s}\ D_{+,s}\ \epsilon^{-s}\, .
 \eea
Here ${\tt M}$ stands for the difference between the total number of holes and particles, 
 \bea
 {\tt M}= \sum_{a=1}^r\big(M_a^{-}-M_a^{+}\big)\,,
 \eea
while $\mathfrak{C}_+$ is given by
 \bea
{\mathfrak{C}_+}=  \frac{ \Gamma(\frac{1}{2}+{\tt k})}{\sqrt{2\pi}}
 \frac{\prod_{a=1}^r \prod_{j=1}^{M^-_a}\
 \re^{\frac{2\pi\ri}{r}a} \big(n_{j,a}^--\frac{1}{2}+{\tt k}\big)^{\frac{1}{r}}}
 { \prod_{a=1}^r\prod_{j=1}^{M^+_a}\re^{\frac{ \pi \ri}{r} (2a+1)} \big(n_{j,a}^{+}-\frac{1}{2}-{\tt k}\big)^{\frac{1}{r}}}
\ .
\eea
The coefficients $D_{+,s}$ entering into the sum read as
 \bea\label{as7823jhsd12}
 D_{+,s}&=&r\delta_{s,0\,({\rm mod}\, r)}\ \frac{B_{\frac{s}{r}+1}(\frac{1}{2}+{\tt k})}{\frac{s}{r}+1}\\[0.2in]
 &+& \sum_{a=1}^r \Bigg(\, \sum_{j=1}^{M_a^{-}}  \re^{\frac{ \ri \pi  }{r}\,  2as}\,
 \Big(n_{j,a}^{-}-\tfrac{1}{2}+{\tt k}\Big)^{\frac{s}{r}}\ 
 - 
 \sum_{j=1}^{M_a^{+}}  \re^{\frac{ \ri \pi  }{r} (2a+1)s}\,\Big(n_{j,a}^{+}-\tfrac{1}{2}-{\tt k}\Big)^{\frac{s}{r}} 
\, \Bigg)\, ,\nonumber
 \eea
 where $B_{m+1}(z)$  are the Bernoulli polynomials. 
\medskip

Within the usual ODE/IQFT interpretation \cite{Bazhanov:1996dr},
the coefficients $D_{+,s}$ in the expansion \eqref{kjass873hjqsda} would be the eigenvalues of
certain integrals of motion.  In particular, $D_{+,r}=I_1$ --- the eigenvalue of the first local integral of motion,
\be
\hat{I}_1=\int_0^{2\pi}\frac{{\rm d} x}{2\pi}\, T\,,
\ee
which is 
the chiral component of the CFT Hamiltonian \eqref{932jd8932jds}.
Formula \eqref{as7823jhsd12}, specialized to $s=r$ yields
 \be
 I_1= \frac{r}{2}\ \bigg({\tt k}+\frac{{\tt M}}{r}\bigg)^2-\frac{{\tt M}^2}{2r}-\frac{r}{24}+
  \sum_{a=1}^r\bigg( \sum_{j=1}^{M_a^{-}}\big(n_{j,a}^{-}-\tfrac{1}{2}\big)+
\sum_{j=1}^{M_a^{+}} \big(n_{j,a}^{+}-\tfrac{1}{2}\big)\,\bigg)\ .
 \ee
Let's write the integer ${\tt M}$ as 
\bea
{\tt M}={\tt m}+r{\tt w}
 \eea
with 
\be
{\tt w}=0,\pm1,\pm2,\ldots\,,\qquad\qquad {\tt m}=0,1,2,\ldots,r-1\, .
\ee
Then the  state with the lowest value of $I_1$ with given ${\tt M}\ge 0$ would have
all $M_a^+=0$, while
 ${\tt m}$ members of the set $\{M_a^-\}_{a=1}^r$ would be equal to ${\tt w}+1$ and the rest
would coincide with ${\tt w}$. The location of the holes is described by
$n_{j,a}^-=j$ for $j\le M_{a}^-$.
For ${\tt M}< 0$ all $M_a^-=0$, while the set   $\{M_a^+\}_{a=1}^r$  
is comprised of $r-{\tt m}$ integers $|{\tt w}|$  and ${\tt m}$ integers $|{\tt w}+1|$. 
Then one can re-write $I_1$ in the form which is applicable for any  ${\tt M}\in\mathbb{Z}$,
 \be
 I_1= \frac{r}{2}\ \bigg({\tt k}+{\tt w}+\frac{\tt m}{r}\bigg)^2+\frac{{\tt m}\,(r-{\tt m})}{2r}-\frac{r}{24}+{\tt L}\,,
 \ee
where ${\tt L}$ is a non-negative integer given by
\be
{\tt L}=  \sum_{a=1}^r\bigg( \sum_{j=1}^{M_a^{-}}\big(n_{j,a}^{-}-\tfrac{1}{2}\big)+
\sum_{j=1}^{M_a^{+}} \big(n_{j,a}^{+}-\tfrac{1}{2}\big)\,\bigg)-
{\tt m}\sum_{j=1}^{|{\tt w}+1|}\,(j-\tfrac{1}{2})-(r-{\tt m})\sum_{j=1}^{|{\tt w}|}\,(j-\tfrac{1}{2})\ .
\ee
Note that 
the states with ${\tt L}=0$, having the lowest value of $I_1$, are $\frac{r!}{{\tt m}!(r-{\tt m})!}$ times degenerate.
Analogous computations can be repeated
 for the chiral state $|\bar{\alpha}\rangle$. One should keep in mind that
in the sector with fixed $S^z$, the total number of (left and right) holes 
should coincide with the total number of particles, i.e.,
\be
\sum_{a=1}^r\big(M^-_a+\bar{M}^-_a\big)=\sum_{a=1}^r\big(M^+_a+\bar{M}^+_a\big)\, .
\ee
\medskip

In the above considerations, we focused on the sector $S^z=0$ to make the discussion as simple as possible.
This can be extended to the case of  $S^z$ being any (half-)integer.
The final expressions for $I_1$ and $\bar{I}_1$ are given by formulae 
\eqref{asjhyghdfvAAAA}-\eqref{asjhyghdfvBBBB}  where one should set $n=r$ therein. 
Then, the eigenvalues of the Hamiltonian $\hat{H}_{\rm CFT}$ \eqref{kjas8923hjds} coincide with $I_1+\bar{I}_1$, while for the
$r$\,-\,site lattice translation operator they are described by \eqref{asjhyghdfvCCCC}.
This result motivated 
the conjecture contained in sec.\ref{sec251}.
The latter was verified numerically on some specific examples away from the free fermion point.

\medskip

\section{Scaling limit for $r$, $A$ even and $(\mu,j)=(\frac{r}{2},\frac{A}{2})$\label{AppE}}
Here we present the numerical scheme mentioned in  sec.\,\ref{sec51}
which allows one to confirm that the scaling limit of the Bethe roots
with the single non-trivial RG invariant taken as in \eqref{ask8923jjh21hga}
is described by the differential equation \eqref{askj8aaaa723hd98jh21} with $\mu=\frac{r}{2}$.
\medskip

As was explained in sec.\,\ref{askj9812a9a8sjh122}, in the case  $\mu=\frac{r}{2}$
the ground state Bethe roots, where the inhomogeneities obey the $r$\,-\,site periodicity condition
$\eta_{J+r}=\eta_J$, are simply related to those for the spin chain with $\tilde{\eta}_{J+2}=\tilde{\eta}_J$.
The parameters of the reduced system $(\tilde{N},\tilde{q},\tilde{\eta}_\ell)$ are expressed in terms of the original
one $(N,q,\eta_\ell)$ as
in eqs.\,\eqref{sa9832jbsdb12} and \eqref{jas8913hhgez} with $\sigma=\frac{r}{2}$. In particular, 
\be\label{ashh1298sdhj21}
\tilde{N}=2N/r\,,\qquad\qquad
\tilde{q}=\exp\Big({\frac{\ri\pi}{\tilde{n}+2}}\Big)\,,\qquad\qquad {\rm where}\qquad\qquad \tilde{n}=2n/r\,.
\ee
From the definition of the RG invariant
 \eqref{ask8923jjh21hga} and the normalization condition \eqref{sajk89732hjsdC} one has
\bea\label{askj8912jhasasssasa}
\frac{1}{\tilde{\eta}_1}+\frac{1}{\tilde{\eta}_2}&=&
(-1)^{\frac{r}{2}-1}\ {\mathfrak a}_{\frac{r}{2}} \
 \bigg(\frac{4\tilde{N_0}}{\tilde{N}}\bigg)\ \log\bigg(\frac{r\tilde{N}}{4\tilde{N_0}}\bigg)\,
\nonumber\\[0.2cm]
\frac{1}{\tilde{\eta}_1^2}+\frac{1}{\tilde{\eta}_2^2}&=&-2\, ,
\eea
where  $\tilde{N}_0$ is given by formula \eqref{asm8932hg21} with $r$ and $n$ replaced by
$2$ and $\tilde{n}$, respectively. In what follows it would be convenient to use the parameter $\alpha$, defined as
\be\label{as238972813}
\alpha=\arg(\tilde{\eta}_1)=-\arg(\tilde{\eta}_2)\,,\qquad\qquad\qquad\qquad\alpha\in(0,\pi)\, .
\ee
Then the system of equations \eqref{askj8912jhasasssasa}  determining the inhomogeneities  implies
\be\label{sakj892jhsad2}
\alpha= \frac{\pi}{2}+(-1)^{\frac{r}{2}}\ {\mathfrak a}_{\frac{r}{2}} \ 
\bigg(\frac{2\tilde{N_0}}{\tilde{N}}\bigg)\ \log\bigg(\frac{r\tilde{N}}{4\tilde{N_0}}\bigg)\,
\ +O\big((\log \tilde{N})^3/\tilde{N}^{3}\big)\ .
\ee
\medskip

The scaling limit of the 2\,-\,site periodic spin chain in the regime with $\arg(\tilde{q})\in(0,\frac{\pi}{2})$ 
has attracted a lot of attention in the literature, see
\cite{Jacobsen:2005xz,Ikhlef:2008zz,Ikhlef:2011ay,Candu:2013fva,Frahm:2013cma}
and the more recent papers \cite{Bazhanov:2019xvy,Bazhanov:2019xvyA}. 
The interest originated following the work of Jacobsen and Saleur \cite{Jacobsen:2005xz}, which was devoted to the case
when the parameter $\alpha$ in \eqref{as238972813} is equal to $\frac{\pi}{2}$.
In the subsequent papers \cite{Ikhlef:2008zz} and \cite{Ikhlef:2011ay}, the  energy spectrum for ---
in the terminology used in sec.\,\ref{sasau} --- primary Bethe states was studied. These were originally defined
in \cite{Ikhlef:2008zz}
using the logarithmic form of the Bethe Ansatz equations. The corresponding Bethe roots 
 turn out to be all real and the states can be characterized by the integer ${\tt m}$,
which coincides with the difference between the number of  positive  $(\tilde{\zeta}_m>0)$ and
negative  $(\tilde{\zeta}_m<0)$ roots. 
In ref.\cite{Ikhlef:2011ay} an important relation was found between ${\tt m}$ and  the quantity
\be\label{jkas78132hj21hgas}
s=\frac{\tilde{n}}{2\pi}\log\big({\cal K}^{(2)}\,/\,{\cal K}^{(1)}\big)\,,
\ee
which is expressed in terms of the eigenvalues of the quasi-shift operators
\be
{\cal K}^{(\ell)}=\re^{\ri\pi{\tt k}}\,{\tilde{q}}^{-\frac{\tilde{N}}{2}}\,\prod_{m=1}^{\frac{\tilde{N}}{2}-\tilde{S}^z}\,
\frac{\tilde{\zeta}_m+\tilde{\eta}_\ell\,\tilde{q}^{+1}}{%
\tilde{\zeta}_m+\tilde{\eta}_\ell\,\tilde{q}^{-1}}\qquad\qquad\qquad (\ell=1,2)\, .
\ee
In the case of the primary Bethe states there is a natural way of assigning them a dependence on the lattice size  so that
$s$ becomes a function of $\tilde{N}$. Then  it was observed from the numerical data that, with a properly
chosen branch of the logarithm in the definition \eqref{jkas78132hj21hgas}, 
the leading large\,-\,$\tilde{N}$ behaviour of $s$ for fixed ${\tt m}$ is given by
\be\label{sakj821jhsdasassa}
s\sim\frac{\pi{\tt m}}{2\log(\tilde{N})}\qquad\qquad\qquad (\tilde{N}\to\infty)\ .
\ee
Moreover, the so-called ``quantization condition'' was proposed
that provides a refinement to this leading asymptotic. 
The more precise version of the relation, 
which is also applicable to the case of quasi-periodic boundary conditions, appeared in ref.\cite{Bazhanov:2019xvy}
and reads explicitly as
\be\label{askjh217832jhd}
4s\log\bigg(\frac{\tilde{N}}{2\tilde{N}_0}\bigg)-
\delta(s)-2\pi \,{\tt m}=O\big((\log \tilde{N})^{-\infty}\big)\ ,
\ee
where
\bea\label{asnmbqwhyxbw}
\delta(s)=-2\ri\, \log\Bigg(2^{-\frac{2\ri s}{\tilde{n}}(\tilde{n}+2)}\,\frac{\Gamma(\frac{1}{2}+\tilde{p}+\frac{\ri s}{2})}{%
\Gamma(\frac{1}{2}+\tilde{p}-\frac{\ri s}{2})}
\frac{\Gamma(\frac{1}{2}+\bar{\tilde{p}}+\frac{\ri s}{2})}{\Gamma(\frac{1}{2}+\bar{\tilde{p}}-\frac{\ri s}{2})}
\Bigg)
\eea
and
\be 
\tilde{p}=\frac{\tilde{S}^z}{2}+\frac{\tilde{n}+2}{2}\ \big({\tt k}+{\tt w}\big)\,,\qquad\qquad 
\bar{\tilde{p}}=\frac{\tilde{S}^z}{2}-\frac{\tilde{n}+2}{2}\ \big({\tt k}+{\tt w}\big)\ . 
\ee
The quantization condition holds
true up to power law corrections in $\tilde{N}$, as indicated by the r.h.s. of \eqref{askjh217832jhd}.
For given   ${\tt m}$, it gives a remarkably accurate description for the value of
$s$ at large $\tilde{N}$ \cite{Kluemper}. 
It follows that  $s(\tilde{N})$ tends to zero as $\tilde{N}\to\infty$
if the integer ${\tt m}$ is left unchanged. A non-trivial scaling limit can be achieved by increasing $|{\tt m}|$ simultaneously
with $\tilde{N}$ in such a way that $s$ is kept fixed, i.e., treated as the RG invariant.
Then  \eqref{askjh217832jhd} is interpreted as an equation
specifying the $\tilde{N}$\,-\,dependence of ${\tt m}$. This way, the states would be characterized by $s$, which
may take any real value. The corresponding eigenvalue of the CFT Hamiltonian \eqref{kjas8923hjds}
is given by \eqref{asjk1892hjdaaa} with $r=2$. 
In the  work \cite{Bazhanov:2019xvy} it was proposed that
the differential equation
\be\label{asnui12banaab}
\Big[-\partial^2_{\tilde{y}}+
\tilde{p}^2+\re^{(\tilde{n}+2)\tilde{y}}-\tilde{E}^2\,\re^{2\tilde{y}}-s\, \tilde{E}\,\re^{\tilde{y}}\Big]\psi=0
\ee
describes the scaled Bethe roots for the primary Bethe states when the 
scaling limit is taken with $s$ being treated as an RG invariant.

\medskip

The quantization condition was originally found for the case when $\alpha$ from \eqref{as238972813}
is equal to $\frac{\pi}{2}$. It was generalized by
 Frahm and Seel in ref.\cite{Frahm:2013cma}  to  
$\alpha\in(\gamma,\pi-\gamma)$ being a fixed number independent of the lattice size.
For our purposes, it is sufficient to use the results of that paper specialized to the ground state, i.e.,
the state of the spin chain Hamiltonian with the lowest possible energy.
 The important difference between $\alpha=\frac{\pi}{2}$ and  $\alpha\ne \frac{\pi}{2}$
is that the value of $s$ defined as  in \eqref{jkas78132hj21hgas}
for the ground state in the former case vanishes, while for the latter it
is no longer zero. Moreover, as $\tilde{N}\to\infty$, 
\be\label{ashj12hgaab}
s_{\scriptscriptstyle\rm GS}(\tilde{N})=\tilde{N}b_\infty+O(1)\ ,
\ee
where $s_{\scriptscriptstyle\rm GS}$ denotes the value of $s$ computed for the ground state and
\bea
b_\infty=-\frac{n}{\pi}\int_{0}^{\infty}\frac{\rd t}{t}\
\frac{\sinh(t)\,
\sinh\big((\tilde{n}+1)t\big)}{\sinh({\tilde n}t)
\,\sinh\big((\tilde{n}+2)t\big)}\, \sinh\big((1-\tfrac{2\alpha}{\pi})\,(\tilde{n}+2) t\big)\ .
\eea
The remainder term in \eqref{ashj12hgaab} admits a more accurate description. Namely, denoting it as
\be\label{asjyu12hgsd87hg12}
s_\alpha\equiv s_{\scriptscriptstyle\rm GS}(\tilde{N})-\tilde{N}b_\infty\ ,
\ee
the following quantization condition holds true
\be\label{aksjh12hghgds}
4s_\alpha\, \log\bigg(\frac{\tilde{N}}{2\tilde{N}_0}\bigg)
-\delta_{\tt k}(s_\alpha)=2\pi{\tt m}_{\alpha}
+\frac{\tilde{n}+2}{\tilde{n}}\,(\pi-2\alpha)\,\tilde{N}+O\big((\log \tilde{N})^{-\infty}\big)\,.
\ee
Here $\delta_{\tt k}(s)$ is given by \eqref{asnmbqwhyxbw} with 
${\tilde p}=-\bar{{\tilde p}}=\frac{1}{2}(\tilde{n}+2)\,{\tt k}$ and
 ${\tt m}_{\alpha}$ is the nearest integer to 
$-\frac{\tilde{n}+2}{2\pi \tilde{n}}\,(\pi-2\alpha)\,\tilde{N}$.
Despite the similarity of the above with \eqref{askjh217832jhd} one should keep in mind that
the  latter relation  gives the $\tilde{N}$ dependence of $s$ \eqref{jkas78132hj21hgas} for the primary Bethe states,
while \eqref{aksjh12hghgds} describes how $s_\alpha$, defined in \eqref{asjyu12hgsd87hg12} depends on $\tilde{N}$
for the ground state of the spin chain only.
\bigskip

In the relation \eqref{aksjh12hghgds}  the parameter $\alpha$ is assumed to be fixed.
Nevertheless, it inspired us to introduce the formal variable $s$
as the solution of the equation
\be\label{san8aaa7hj2bvs}
4s\, \log\bigg(\frac{\tilde{N}}{2\tilde{N}_0}\bigg)
-\delta_{\tt k}(s)=
(-1)^{\frac{r}{2}-1}\ \mathfrak{a}_{\frac{r}{2}}\ 
\frac{4\,(\tilde{n}+2)}{\tilde{n}}\,\tilde{N}_0\,\log\bigg(\frac{r\tilde{N}}{4\tilde{N}_0}\bigg)
\ ,
\ee
which is obtained from \eqref{aksjh12hghgds} by setting ${\tt m}_\alpha=0$;
substituting $\alpha$ for the $\tilde{N}$ dependent  expression in the r.h.s. of \eqref{sakj892jhsad2};
 and ignoring all correction terms. 
In the original scheme for performing the scaling limit $\mathfrak{a}_{\frac{r}{2}}$
 was treated 
as the RG invariant. We now switch the schemes and suppose that $s$
does not depend on the number of lattice sites, whereas the $\tilde{N}$ dependence of  $\mathfrak{a}_{\frac{r}{2}}$  
is dictated by \eqref{san8aaa7hj2bvs}.
It turns out that in the new scheme the sums 
over the Bethe roots 
$$
\bigg(\frac{2\tilde{N}_0}{\tilde{N}}\bigg)^{\frac{\tilde{n}j}{\tilde{n}+2}}\ 
\frac{1}{j}\ \sum_{m=1}^{\tilde{N}/2}(\tilde{\zeta}_m)^{-j}
$$
tend to their limiting values with corrections that decay as a power law in $\tilde{N}$. In contrast, if 
$\mathfrak{a}_{\frac{r}{2}}$   were to be kept fixed, the corrections would  decay considerably more slowly --- as
the inverse power of $\log(\tilde{N})$. Thus, the new scheme allows one to observe the scaling behaviour
for $\tilde{N}=\frac{2N}{r}\sim 1000$. On the other hand the definition \eqref{san8aaa7hj2bvs} implies
\be\label{sakui12bbsvb21}
s+(-1)^{\frac{r}{2}}\,\mathfrak{a}_{\frac{r}{2}}\,\frac{\tilde{n}+2}{\tilde{n}}\,\tilde{N}_0\,
=O\big(1/\log(\tilde{N})\big)\ .
\ee
Hence, the change of scheme will not affect the scaling limit, but rather, the rate at which it is 
achieved as $\tilde{N}\to\infty$. In this regard, one may note
 that the choice of the constant $rN_0$ in the logarithm in   \eqref{ask8923jjh21hga} 
 for the RG invariant $\mathfrak{a}_{\frac{r}{2}}$
is not essential: the constant can be replaced by any positive number.
\medskip

This way, we were able to establish that the scaling limit of the Bethe roots 
is described by the ODE \eqref{asnui12banaab}. Taking into account the relations  \eqref{ashh1298sdhj21} 
between the parameters of the original and
reduced systems and performing the change of variables 
\be
\tilde{y}=\frac{r}{2}\,{y}-\frac{r}{n+r}\,\log\Big(\frac{r}{2}\Big)
\ee
and
\begin{subequations}
\bea
\tilde{p}&=&\frac{2}{r}\,p\,,\qquad\qquad\qquad
\tilde{E}=(-1)^{\frac{A}{2}}\,\Big(\frac{2}{r}\Big)^{\frac{n}{n+r}}\,E^{\frac{r}{2}}\\[0.2cm]
\label{sanbhjauiqwBBB}
s&=&(-1)^{\frac{A}{2}}\,\frac{2c}{r}
\eea
\end{subequations}
one arrives at the differential equation \eqref{askj8aaaa723hd98jh21} with $\mu=\frac{r}{2}$.
The $\mathfrak{a}_{\frac{r}{2}}$ --- $c$ relation \eqref{kjas893aaa2187213} follows by 
combining \eqref{sanbhjauiqwBBB} with \eqref{sakui12bbsvb21}.

\medskip

\section{Supplementary tables \label{AppD2}}
Here we give the specification of the RG invariants $\mathfrak{a}_s$ \eqref{saaaakju32hjsd},
such that the scaled Bethe roots  \eqref{asjk732j} are described by the ODE
\bea
 \bigg[-\partial_y^2+p^2+ \re^{(n+r)y}-(-1)^A\,E^r\ \re^{ry}-
c\, E^{\mu}\ \re^{\big(  (A\mu-rj)\,\frac{n+r}{r}+\mu\big) y}
 \,\bigg]\,\psi=0
\eea
with $(\mu,j)$ taken from the set \eqref{cv}.  
\medskip

For  $r=3,4,\ldots,10$, $A=1,2,\ldots,r-2$ and
all possible pairs
$(\mu,j)$, presented in the tables are  the values of the integer $s$
 corresponding to the minimal non-trivial set of RG invariants along with their
exponents $d_s$. All other sums
$
\sum_{\ell=1}^r(\eta_\ell)^{-s}
$
can be taken to be zero. Also, the normalization condition 
\be
\frac{1}{r}\,\sum_{\ell=1}^r (\eta_\ell)^{-r}=(-1)^{r-1}
\ee
is being assumed. The values of the non-vanishing RG invariants are expressed in terms of 
the coefficient $c$ of the ODE as in formulae \eqref{sajk8937hj21hg}\,-\,\eqref{3902jdsjh}.  Then,
the scaling limit of the ground state Bethe roots is described by the differential equation 
for any $n>n_{\rm min}$, where $n_{\rm min}$ is listed in the last column of the tables.

\begin{table}[h!]
\begin{center}
\scalebox{0.95}{
\begin{tabular}{|c|c|c|l|l|l|l|}
\cline{2-6}
 \multicolumn{1}{c|}{ }& & & & &   \\[-0.4cm]
 \multicolumn{1}{c|}{ }& $A$ & $(\mu,j)$  & $ s$ & $d_s$ & $n_{\rm min}$\\[0.0cm]
\hline
& & & & &  \\[-0.4cm]
%--------------------------------------------------------------------------------------r=3
\multirow{1.}{*}{$r=3$} 
%---------------------------------------------------------------------A=1
& \multirow{1}{*}{$\vspace{-0.0cm} 1$}  & $(1,0)$  & $\{1\}$ & $d_1=\frac{2}{3}$ & $0$  \\[0.05cm]
\hline
\hline
& & & & &  \\[-0.4cm]
%--------------------------------------------------------------------------------------r=4
\multirow{2.5}{*}{\vspace{0.1cm}$r=4$} 
%---------------------------------------------------------------------A=1
& \multirow{1}{*}{$\vspace{-0.0cm} 1$}  & $(1,0)$  & $\{1\}$ & $d_1=1$ & $0$  \\[0.05cm]
\cline{2-6}
& & & & & \\[-0.4cm]
%---------------------------------------------------------------------A=2
& \multirow{1}{*}{$\vspace{-0.0cm} 2$}  & $(1,0)$  & $\{1,2\}$ & $d_1=\frac{1}{2},\,d_2=1$ & $0$  \\[0.05cm]
\cline{2-6}
\hline
\hline
& & & & & \\[-0.4cm]
%--------------------------------------------------------------------------------------r=5
\multirow{7.5}{*}{\vspace{-0.1cm}$r=5$} 
%---------------------------------------------------------------------A=1
& \multirow{2}{*}{$\vspace{-0.2cm} 1$}  
& $(1,0)$  & $\{1\}$ & $d_1=\frac{6}{5}$ & $\frac{5}{2}$  \\[0.05cm]
\cline{3-6}
&  & & & &   \\[-0.35cm]
& & $(2,0)$   &$\{2,4,1\}$ &  $d_2=\frac{2}{5},\,d_4=\frac{4}{5},\,d_1=\frac{6}{5}$ & $\frac{5}{2}$  \\[0.05cm]
\cline{2-6}
 & & & & & \\[-0.4cm]
%---------------------------------------------------------------------A=2
& \multirow{2}{*}{$\vspace{-0.2cm} 2$}
 & $(1,0)$   &$\{1\}$   & $d_1=\frac{4}{5}$ & 0 \\[0.05cm]
\cline{3-6}
 & & & & &  \\[-0.35cm]
& & $(3,1)$   &$\{3,1\}$   & $d_3=\frac{2}{5},\,d_1=\frac{4}{5}$ & 0 \\[0.05cm]
\cline{2-6}
& & & & & \\[-0.4cm]
%---------------------------------------------------------------------A=3
& \multirow{2}{*}{$\vspace{-0.2cm}	3$}
& $(1,0)$   &$\{1,2\}$   & $d_1=\frac{2}{5},\,d_2=\frac{4}{5}$ & 0 \\[0.05cm]
\cline{3-6}
& & & & &  \\[-0.35cm]
& & $(2,1)$   &$\{2\}$   &  $d_2=\frac{4}{5}$& 0 \\[0.05cm]
\hline
\hline
& & & & &  \\[-0.4cm]
%--------------------------------------------------------------------------------------r=6
\multirow{8}{*}{\vspace{0.15cm}$r=6$} 
%---------------------------------------------------------------------A=1
& \multirow{2}{*}{$\vspace{-0.2cm} 1$}  
& $(1,0)$  & $\{1\}$ & $d_1=\frac{4}{3}$ & $6$  \\[0.05cm]
\cline{3-6}
&  & & & &   \\[-0.35cm]
& & $(2,0)$   &$\{2\}$ &  $d_2=\frac{2}{3}$ & $0$  \\[0.05cm]
\cline{2-6}
& & & & & \\[-0.4cm]
%---------------------------------------------------------------------A=2
& \multirow{1}{*}{$\vspace{-0.0cm} 2$}
& $(1,0)$   &$\{1\}$   & $d_1=1$ & 0 \\[0.05cm]
\cline{2-6}
& & & & & \\[-0.4cm]
%---------------------------------------------------------------------A=3
& \multirow{1}{*}{$\vspace{-0.0cm} 3$}
& $(1,0)$   &$\{1\}$   & $d_1=\frac{2}{3}$ & 0 \\[0.05cm]
\cline{2-6}
& & & & & \\[-0.4cm]
%---------------------------------------------------------------------A=4
& \multirow{2}{*}{$\vspace{-0.2cm}	4$}
& $(1,0)$   &$\{1,2,3\}$   & $d_1=\frac{1}{3},\,d_2=\frac{2}{3},\,d_3=1$ & 0 \\[0.05cm]
\cline{3-6}
& & & & &  \\[-0.35cm]
& & $(2,1)$   &$\{2\}$   &  $d_2=\frac{2}{3}$& 0 \\[0.05cm]
\hline
\hline
& & & & &  \\[-0.4cm]
%--------------------------------------------------------------------------------------r=7
\multirow{20}{*}{\vspace{0.1cm}$r=7$} 
%---------------------------------------------------------------------A=1
& \multirow{4}{*}{$\vspace{0.1cm} 1$}  
& $(1,0)$  & $\{1\}$ & $d_1=\frac{10}{7}$ & $\frac{21}{2}$  \\[0.05cm]
\cline{3-6}
&  & & & &   \\[-0.35cm]
& & $(2,0)$   &$\{2\}$ &  $d_2=\frac{6}{7}$ & $0$  \\[0.05cm]
\cline{3-6}
&  & & & &   \\[-0.35cm]
& & $(3,0)$   &$\{3,6,2,5,1\}$ &  $d_3=\frac{2}{7},\, d_6=\frac{4}{7},\, d_{2}=\frac{6}{7},\, d_5=\frac{8}{7},\, d_{1}=\frac{10}{7} $ & $\frac{21}{2}$  \\[0.05cm]
\cline{2-6}
& & & & & \\[-0.4cm]
%---------------------------------------------------------------------A=2
& \multirow{4}{*}{$\vspace{0.1cm} 2$} & $(1,0)$   &$\{1\}$   & $d_1=\frac{8}{7}$ & $\frac{7}{4}$ \\[0.05cm]
\cline{3-6}
& & & & & \\[-0.35cm]
& & $(2,0)$   &$\{2,4,6,1\}$   & $d_2=\frac{2}{7},\,d_4=\frac{4}{7},\,d_6=\frac{6}{7},\, d_1=\frac{8}{7}$ & $\frac{7}{4}$ \\[0.05cm]
\cline{3-6}
& & & & & \\[-0.35cm]
& & $(4,1)$   &$\{4,1\}$   & $d_4=\frac{4}{7},\,d_1=\frac{8}{7}$ & $\frac{7}{4}$ \\[0.05cm]
\cline{2-6}
& & & & & \\[-0.4cm]
%---------------------------------------------------------------------A=3
& \multirow{4}{*}{$\vspace{0.1cm}	3$} & $(1,0)$   &$\{1\}$   & $d_1=\frac{6}{7}$ & $0$  \\[0.05cm]
\cline{3-6}
& & & & &  \\[-0.35cm]
& & $(3,1)$   &$\{3\}$   & $d_3=\frac{4}{7}$ & $0$ \\[0.05cm]
\cline{3-6}
& & & & &  \\[-0.35cm]
& & $(5,2)$   &$\{5,3,1´\}$   &  $d_5=\frac{2}{7},\,d_3=\frac{4}{7},\,d_1=\frac{6}{7}$& $0$ \\[0.05cm]
\cline{2-6}
& & & & & \\[-0.4cm]
%---------------------------------------------------------------------A=4
& \multirow{4}{*}{$\vspace{0.1cm} 4$} & $(1,0)$   &$\{1,2\}$   & $d_1=\frac{4}{7},\, d_2=\frac{8}{7}$ & $\frac{7}{2}$ \\[0.05cm]
\cline{3-6}
& & & & & \\[-0.35cm]
& & $(2,1)$   &$\{2\}$   & $d_2=\frac{8}{7}$ & $\frac{7}{2}$ \\[0.05cm]
\cline{3-6}
& & & & & \\[-0.35cm]
& & $(4,2)$   &$\{4,1,5,2\}$   & $d_4=\frac{2}{7},\, d_1=\frac{4}{7},\, d_5=\frac{6}{7},\, d_2=\frac{8}{7}$ & $\frac{7}{2}$ \\[0.05cm]
\cline{2-6}
& & & & & \\[-0.4cm]
%---------------------------------------------------------------------A=5
& \multirow{4}{*}{$\vspace{0.1cm} 5$} & $(1,0)$   &$\{1,2,3\}$   & $d_1=\frac{2}{7},\, d_2=\frac{4}{7},\, d_3=\frac{6}{7}$ & $0$ \\[0.05cm]
\cline{3-6}
& & & & & \\[-0.35cm]
& & $(2,1)$   &$\{2\}$   & $d_2=\frac{4}{7}$ & $0$ \\[0.05cm]
\cline{3-6}
& & & & & \\[-0.35cm]
& & $(3,2)$   &$\{3\}$   & $d_3=\frac{6}{7}$ & $0$ \\[0.05cm]
\hline
\end{tabular}
}
\end{center}
\end{table}
\pagebreak

\begin{table}[h!]
	\begin{center}
\scalebox{0.95}{
		\begin{tabular}{|c|c|c|l|l|l|l|}
			\cline{2-6}
			\multicolumn{1}{c|}{ }& & & & &   \\[-0.4cm]
			\multicolumn{1}{c|}{ }& $A$ & $(\mu,j)$  & $ s$ & $d_s$ & $n_{\rm min}$\\[0.0cm]
			\hline
			& & & & &  \\[-0.4cm]
			%--------------------------------------------------------------------------------------r=8
			\multirow{21}{*}{$r=8$} 
			%---------------------------------------------------------------------A=1
			& \multirow{4}{*}{$\vspace{0.1cm} 1$}  
			& $(1,0)$  & $\{1\}$ & $d_1=\frac{3}{2}$ & $16$  \\[0.05cm]
			\cline{3-6}
			&  & & & &   \\[-0.35cm]
			& & $(2,0)$   &$\{2\}$ &  $d_2=1$ & $0$  \\[0.05cm]
			\cline{3-6}
			&  & & & &   \\[-0.35cm]
			& & $(3,0)$   &$\{3,6,1\}$ &  $d_3=\frac{1}{2},\, d_6=1,\, d_{1}=\frac{3}{2} $ & $16$ \\[0.05cm]
			\cline{2-6}
			& & & & & \\[-0.4cm]
			%---------------------------------------------------------------------A=2
			& \multirow{4}{*}{$\vspace{0.1cm} 2$} & $(1,0)$   &$\{1\}$   & $d_1=\frac{5}{4}$ & $4$ \\[0.05cm]
			\cline{3-6}
			& & & & & \\[-0.35cm]
			& & $(2,0)$   &$\{2,4\}$   & $d_2=\frac{1}{2},\,d_4=1$ & $0$ \\[0.05cm]
			\cline{3-6}
			& & & & & \\[-0.35cm]
			& & $(5,1)$   &$\{5,2,7,4\}$   &  $d_5=\frac{1}{4},\,d_2=\frac{1}{2},\,d_7=\frac{3}{4},\,d_4=1$& $0$ \\[0.05cm]
			\cline{2-6}
			& & & & & \\[-0.4cm]
			%---------------------------------------------------------------------A=3
			& \multirow{2.5}{*}{$\vspace{0.1cm}	3$} & $(1,0)$   &$\{1\}$   & $d_1=1$ & $0$ \\[0.05cm]
			\cline{3-6}
			& & & & &  \\[-0.35cm]
			& & $(3,1)$   &$\{3\}$   & $d_3=1$ & $0$ \\[0.05cm]
			\cline{2-6}
			& & & & & \\[-0.4cm]
			%---------------------------------------------------------------------A=4
			& \multirow{2.5}{*}{$\vspace{0.1cm} 4$} & $(1,0)$   &$\{1\}$   & $d_1=\frac{3}{4}$ & 0 \\[0.05cm]
			\cline{3-6}
			& & & & & \\[-0.35cm]
			& & $(3,1)$   &$\{3,6,1,4\}$   & $d_3=\frac{1}{4},\, d_6=\frac{1}{2},\, d_1=\frac{3}{4},\, d_4=1$ & $0$ \\[0.05cm]
			\cline{2-6}
			& & & & & \\[-0.4cm]
			%---------------------------------------------------------------------A=5
			& \multirow{4}{*}{$\vspace{0.1cm} 5$} & $(1,0)$   &$\{1,2\}$   & $d_1=\frac{1}{2},\, d_2=1$ & $0$ \\[0.05cm]
			\cline{3-6}
			& & & & & \\[-0.35cm]
			& & $(2,1)$   &$\{2\}$   & $d_2=1$ & $0$ \\[0.05cm]
			\cline{3-6}
			& & & & & \\[-0.35cm]
			& & $(5,3)$   &$\{5,2\}$   & $d_5=\frac{1}{2},\,d_2=1$ & $0$\\[0.05cm]
						\cline{2-6}
			& & & & & \\[-0.4cm]
			%---------------------------------------------------------------------A=6
			& \multirow{3.5}{*}{$\vspace{-0.1cm} 6$} & $(1,0)$   &$\{1,2,3,4\}$   & $d_1=\frac{1}{4},\, d_2=\frac{1}{2},\, d_3=\frac{3}{4},\, d_4=1$ & $0$ \\[0.05cm]
			\cline{3-6}
			& & & & & \\[-0.35cm]
			& & $(2,1)$   &$\{2,4\}$   & $d_2=\frac{1}{2},\, d_4=1$ & $0$ \\[0.05cm]
			\cline{3-6}
			& & & & & \\[-0.35cm]
			& & $(3,2)$   &$\{3\}$   & $d_3=\frac{3}{4}$ & $0$ \\[0.05cm]
			\hline
		\end{tabular}
}
	\end{center}
\end{table}
\pagebreak

\begin{table}[h!]
	\begin{center}
\scalebox{0.95}{
		\begin{tabular}{|c|c|c|l|l|l|l|}
			\cline{2-6}
			\multicolumn{1}{c|}{ }& & & & &   \\[-0.4cm]
			\multicolumn{1}{c|}{ }& $A$ & $(\mu,j)$  & $ s$ & $d_s$ & $n_{\rm min}$\\[0.0cm]
			\hline
			& & & & &  \\[-0.4cm]
			& & & & &  \\[-0.4cm]
			%--------------------------------------------------------------------------------------r=9
			\multirow{29}{*}{\vspace{-0.2cm}$r=9$} 
			%---------------------------------------------------------------------A=1
			& \multirow{6}{*}{$\vspace{-0.1cm} 1$}  
			& $(1,0)$  & $\{1\}$ & $d_1=\frac{14}{9}$ & $\frac{45}{2}$  \\[0.05cm]
			\cline{3-6}
			&  & & & &   \\[-0.35cm]
			& & $(2,0)$   &$\{2\}$ &  $d_2=\frac{10}{9}$ & $\frac{9}{4}$  \\[0.05cm]
			\cline{3-6}
			&  & & & &   \\[-0.35cm]
			& & $(3,0)$   &$\{3\}$ &  $d_3=\frac{2}{3} $ & $0$ \\[0.05cm]
			\cline{3-6}
			&  & & & &   \\[-0.35cm]
			& & \multirow{2}{*}{\vspace{-0.2cm}%
				$(4,0)$}   &
			\multirow{2}{*}{\vspace{-0.2cm}$\{4,8,3,7,2,6,1\}$} &  $d_4=\frac{2}{9},\,d_8=\frac{4}{9},\,d_3=\frac{2}{3},\,d_7=\frac{8}{9},$  & \multirow{2}{*}{\vspace{-0.2cm}$\frac{45}{2}$ }\\[0.05cm]
			&  & & & &   \\[-0.35cm]
			& &  & &  $d_2=\frac{10}{9},\, d_6=\frac{12}{9},\, \,d_1=\frac{14}{9}  $  &  \\[0.05cm]
			\cline{2-6}
			& & & & & \\[-0.4cm]
			%---------------------------------------------------------------------A=2
			& \multirow{4}{*}{$\vspace{0.125cm} 2$} & $(1,0)$   &$\{1\}$   & $d_1=\frac{4}{3}$ & $\frac{27}{4}$ \\[0.05cm]
			\cline{3-6}
			& & & & & \\[-0.35cm]
			& & $(2,0)$   &$\{2\}$   & $d_2=\frac{2}{3}$ & $0$ \\[0.05cm]
			\cline{3-6}
			& & & & & \\[-0.35cm]
			& & $(5,1)$   &$\{5,1\}$   &  $d_5=\frac{2}{3},\,d_1=\frac{4}{3}$& $\frac{27}{4}$  \\[0.05cm]
			\cline{2-6}
			& & & & & \\[-0.35cm]
			%---------------------------------------------------------------------A=3
			& \multirow{3.5}{*}{$\vspace{-0.1cm}	3$} & $(1,0)$   &$\{1\}$   & $d_1=\frac{10}{9}$ & $\frac{3}{2}$ \\[0.05cm]
			\cline{3-6}
			& & & & &  \\[-0.35cm]
			& & $(2,0)$   &$\{2,4,6,8,1\}$   & $d_2=\frac{2}{9},\, d_4=\frac{4}{9},\, d_6=\frac{2}{3},\, d_8=\frac{8}{9},\, d_1=\frac{10}{9}$ & $\frac{3}{2}$ \\[0.05cm]
			\cline{3-6}
			& & & & & \\[-0.35cm]
			& & $(4,1)$   &$\{4,8,3\}$   & $d_4=\frac{4}{9},\,d_8=\frac{8}{9}$ & 0 \\[0.05cm]
			\cline{2-6}
			& & & & & \\[-0.35cm]
			%---------------------------------------------------------------------A=4
			& \multirow{5}{*}{$\vspace{-0.1cm} 4$} & $(1,0)$   &$\{1\}$   & $d_1=\frac{8}{9}$ & 0 \\[0.05cm]
			\cline{3-6}
			& & & & & \\[-0.35cm]
			& & $(3,1)$   &$\{3\}$   & $d_3=\frac{2}{3}$ & $0$ \\[0.05cm]
			\cline{3-6}
			& & & & & \\[-0.35cm]
			& & $(5,2)$   &$\{5,1\}$   & $d_5=\frac{4}{9},\, d_1=\frac{8}{9}$ & $0$ \\[0.05cm]
			\cline{3-6}
			& & & & & \\[-0.35cm]
			& & $(7,3)$   &$\{7,5,3,1\}$   & $d_7=\frac{2}{9},\,d_5=\frac{4}{9},\,d_3=\frac{2}{3},\,d_1=\frac{8}{9} $& $0$ \\[0.05cm]
			\cline{2-6}
			& & & & & \\[-0.35cm]
			%---------------------------------------------------------------------A=5
			& \multirow{4}{*}{$\vspace{0.1cm} 5$} & $(1,0)$   &$\{1\}$   & $d_1=\frac{2}{3}$& $0$ \\[0.05cm]
			\cline{3-6}
			& & & & & \\[-0.35cm]
			& & $(2,1)$   &$\{2\}$   & $d_2=\frac{4}{3}$ & $\frac{27}{2}$ \\[0.05cm]
			\cline{3-6}
			& & & & & \\[-0.35cm]
			& & $(4,2)$   &$\{4\}$   & $d_4=\frac{2}{3}$ & $0$\\[0.05cm]
			\cline{2-6}
			& & & & & \\[-0.35cm]
			%---------------------------------------------------------------------A=6
			& \multirow{4}{*}{$\vspace{0.115cm} 6$}  
			& $(1,0)$  & $\{1,2\}$ & $d_1=\frac{4}{9},\,d_2=\frac{8}{9}$ & $0$  \\[0.05cm]
			\cline{3-6}
			&  & & & &   \\[-0.35cm]
			& & $(2,1)$   &$\{2\}$ &  $d_2=\frac{8}{9}$ & $0$  \\[0.05cm]
			\cline{3-6}
			&  & & & &   \\[-0.35cm]
			& & $(5,3)$   &$\{5,1,6,2\}$ &  $d_5=\frac{2}{9},\,d_1=\frac{4}{9},\, d_6=\frac{2}{3},\, d_2=\frac{8}{9} $ & $0$ \\[0.05cm]
			\cline{2-6}
			& & & & & \\[-0.35cm]
			%---------------------------------------------------------------------A=7
			& \multirow{5}{*}{$\vspace{-0.0cm} 7$} & $(1,0)$   &$\{1,2,3,4\}$   & $d_1=\frac{2}{9},\,d_2=\frac{4}{9},\,d_3=\frac{2}{3},\,d_4=\frac{8}{9} $ & $0$ \\[0.05cm]
			\cline{3-6}
			& & & & & \\[-0.35cm]
			& & $(2,1)$   &$\{2,4\}$   & $d_2=\frac{4}{9},\, d_4=\frac{8}{9}$ & $0$ \\[0.05cm]
			\cline{3-6}
			& & & & & \\[-0.35cm]
			& & $(3,2)$   &$\{3\}$   &  $d_3=\frac{2}{3}$& $0$ \\[0.05cm]
			\cline{3-6}
			& & & & & \\[-0.35cm]
			& & $(4,3)$   &$\{4\}$   &  $d_4=\frac{8}{9}$& $0$ \\[0.05cm]
			\hline
		\end{tabular}
}
	\end{center}
\end{table}

\pagebreak

\begin{table}[h!]
	\begin{center}
\scalebox{0.95}{
		\begin{tabular}{|c|c|c|l|l|l|l|}
			\cline{2-6}
			\multicolumn{1}{c|}{ }& & & & &   \\[-0.35cm]
			\multicolumn{1}{c|}{ }& $A$ & $(\mu,j)$  & $ s$ & $d_s$ & $n_{\rm min}$\\[0.0cm]
			\hline
			& & & & &  \\[-0.35cm]
			%-----------------------------------------------------------------------------------r=10
			\multirow{35}{*}{\vspace{0.15cm}$r=10$} 
			%---------------------------------------------------------------------A=1
			& \multirow{5}{*}{$\vspace{-0.1cm}	1$} & $(1,0)$   &$\{1\}$   & $d_1=\frac{8}{5}$ & $30$ \\[0.05cm]
			\cline{3-6}
			& & & & &  \\[-0.35cm]
			& & $(2,0)$   &$\{2\}$   & $d_2=\frac{6}{5}$ & $5$ \\[0.05cm]
			\cline{3-6}
			& & & & & \\[-0.35cm]
			& & $(3,0)$   &$\{3\}$   & $d_3=\frac{4}{5}$ & $0$ \\[0.05cm]
			\cline{3-6}
			& & & & & \\[-0.35cm]
			& & $(4,0)$   &$\{4,8,2\}$   & $d_4=\frac{2}{5},\, d_8=\frac{4}{5},\, d_2=\frac{6}{5}$ & $5$ \\[0.05cm]
			\cline{2-6}
			& & & & & \\[-0.35cm]
			%---------------------------------------------------------------------A=2
			& \multirow{5}{*}{$\vspace{-0.1cm} 2$} & $(1,0)$   &$\{1\}$   & $d_1=\frac{7}{5}$ & $10$ \\[0.05cm]
			\cline{3-6}
			& & & & & \\[-0.35cm]
			& & $(2,0)$   &$\{2\}$   & $d_2=\frac{4}{5}$ & $0$ \\[0.05cm]
			\cline{3-6}
			& & & & & \\[-0.35cm]
			& & $(3,0)$   &$\{3,6,9,2,5,1\}$   & $d_3=\frac{1}{5},\,d_6=\frac{2}{5},\,d_9=\frac{3}{5},\,d_2=\frac{4}{5},\,d_5=1,\, d_1=\frac{7}{5}$ & $10$ \\[0.05cm]
			\cline{3-6}
			& & & & & \\[-0.35cm]
			& & $(6,1)$   &$\{6,2\}$   & $d_6=\frac{2}{5},\,d_2=\frac{4}{5} $& $0$ \\[0.05cm]
			\cline{2-6}
			& & & & & \\[-0.35cm]
			%---------------------------------------------------------------------A=3
			& \multirow{5}{*}{$\vspace{-0.1cm} 3$} & $(1,0)$   &$\{1\}$   & $d_1=\frac{6}{5}$& $\frac{10}{3}$ \\[0.05cm]
			\cline{3-6}
			& & & & & \\[-0.35cm]
			& & $(2,0)$   &$\{2,4\}$   & $d_2=\frac{2}{5},\,d_4=\frac{4}{5}$ &$0$\\[0.05cm]
			\cline{3-6}
			& & & & & \\[-0.35cm]
			& & $(4,1)$   &$\{4\}$   & $d_4=\frac{4}{5}$&0  \\[0.05cm]
			\cline{3-6}
			& & & & & \\[-0.35cm]
			& & $(7,2)$   &$\{7,4,1\}$   & $d_7=\frac{2}{5},\, d_4=\frac{4}{5},\, d_1=\frac{6}{5}$& $\frac{10}{3}$ \\[0.05cm]
			\cline{2-6}
			& & & & & \\[-0.35cm]
			%---------------------------------------------------------------------A=4
			& \multirow{2}{*}{$\vspace{-0.2cm} 4$} & $(1,0)$   &$\{1\}$   & $d_1=1$& $0$ \\[0.05cm]
			\cline{3-6}
			& & & & & \\[-0.35cm]
			& & $(3,1)$   &$\{3\}$   & $d_3=1$ &$0$\\[0.05cm]
			\cline{2-6}
			& & & & & \\[-0.35cm]
			%---------------------------------------------------------------------A=5
			& \multirow{2}{*}{$\vspace{-0.2cm} 5$} & $(1,0)$   &$\{1\}$   & $d_1=\frac{4}{5}$& $0$ \\[0.05cm]
			\cline{3-6}
			& & & & & \\[-0.35cm]
			& & $(3,1)$   &$\{3,6\}$   & $d_3=\frac{2}{5},\, d_6=\frac{4}{5}$ &$0$\\[0.05cm]
			\cline{2-6}
			& & & & & \\[-0.35cm]
			%---------------------------------------------------------------------A=6
			& \multirow{5}{*}{$\vspace{-0.0cm} 6$} & $(1,0)$   &$\{1,2\}$   & $d_1=\frac{3}{5},\, d_2=\frac{6}{5}$& $5$ \\[0.05cm]
			\cline{3-6}
			& & & & & \\[-0.35cm]
			& & $(2,1)$   &$\{2\}$   & $d_2=\frac{6}{5}$ &$5$\\[0.05cm]
			\cline{3-6}
			& & & & & \\[-0.35cm]
			& & $(4,2)$   &$\{4,8,2\}$   & $d_4=\frac{2}{5},\, d_8=\frac{4}{5},\, d_2=\frac{6}{5}$&  $5$ \\[0.05cm]
			\cline{3-6}
			& & & & & \\[-0.35cm]
			& & $(7,4)$   &$\{7,4,1,8,5\}$   & $d_7=\frac{1}{5},\, d_4=\frac{2}{5},\, d_1=\frac{3}{5},\,d_8=\frac{4}{5},\, d_5=1$& $0$  \\[0.05cm]
			\cline{2-6}
			& & & & & \\[-0.35cm]
			%---------------------------------------------------------------------A=7
			& \multirow{5}{*}{$\vspace{-0.0cm} 7$}  
			& $(1,0)$  & $\{1,2\}$ & $d_1=\frac{2}{5},\,d_2=\frac{4}{5}$ & $0$  \\[0.05cm]
			\cline{3-6}
			&  & & & &   \\[-0.35cm]
			& & $(2,1)$   &$\{2\}$ &  $d_2=\frac{4}{5}$ & $0$  \\[0.05cm]
			\cline{3-6}
			&  & & & &   \\[-0.35cm]
			& & $(3,2)$   &$\{3\}$ &  $d_3=\frac{6}{5} $ & $10$ \\[0.05cm]
			\cline{3-6}
			&  & & & &   \\[-0.35cm]
			& & $(6,4)$   &$\{6,2\}$ &  $d_6=\frac{2}{5},\,d_2=\frac{4}{5} $ & $0$ \\[0.05cm]
			\cline{2-6}
			& & & & & \\[-0.35cm]
			%---------------------------------------------------------------------A=8
			& \multirow{5}{*}{$\vspace{-0.05cm} 8$} & $(1,0)$   &$\{1,2,3,4,5\}$   & $d_1=\frac{1}{5},\,d_2=\frac{2}{5},\,d_3=\frac{3}{5},\,d_4=\frac{4}{5},\, d_5=1 $ & $0$ \\[0.05cm]
			\cline{3-6}
			& & & & & \\[-0.35cm]
			& & $(2,1)$   &$\{2,4\}$   & $d_2=\frac{2}{5},\, d_4=\frac{4}{5}$ & $0$ \\[0.05cm]
			\cline{3-6}
			& & & & & \\[-0.35cm]
			& & $(3,2)$   &$\{3\}$   &  $d_3=\frac{3}{5}$& $0$ \\[0.05cm]
			\cline{3-6}
			& & & & & \\[-0.35cm]
			& & $(4,3)$   &$\{4\}$   &  $d_4=\frac{4}{5}$& $0$ \\[0.05cm]
			\hline
		\end{tabular}
}
	\end{center}
\end{table}

\end{document}